\documentclass[12pt,english]{amsart}
\usepackage{ae,aecompl}
\usepackage[T1]{fontenc}
\usepackage[latin9]{inputenc}
\usepackage{geometry}
\usepackage{array}
\usepackage{float}
\usepackage{multirow}
\usepackage{amstext}
\usepackage{amsthm}
\usepackage{amssymb}
\usepackage{graphicx}
\usepackage{setspace}
\usepackage{esint}
\usepackage{graphicx,amsmath,amssymb,epsfig}
\usepackage{amsfonts}
\usepackage{geometry}
\usepackage{setspace}
\usepackage{lscape}
\usepackage{caption}
\usepackage{rotating}
\usepackage{xcolor}
\usepackage{amsfonts}
\usepackage{subfig}
\usepackage{babel}

\makeatletter
\floatstyle{ruled}
\newfloat{algorithm}{tbp}{loa}
\providecommand{\algorithmname}{Algorithm}
\floatname{algorithm}{\protect\algorithmname}
\numberwithin{equation}{section}
\numberwithin{figure}{section}
\theoremstyle{remark}
  
\newtheorem{rem}{\protect\remarkname}
\renewcommand{\Pr}{\mathbb{P}}

\newcommand{\X}{\mathcal{X}}

\newcommand{\Z}{\mathcal{Z}}

\newcommand{\T}{\mathcal{T}}

\newcommand{\Y}{\mathcal{Y}}

\theoremstyle{plain}
\newtheorem{theorem}{Theorem}

\newtheorem{condition}{Assumption}

\newtheorem{lemma}{Lemma}

\numberwithin{equation}{section}
\theoremstyle{definition}

\@ifundefined{showcaptionsetup}{}{ \PassOptionsToPackage{caption=false}{subfig}}
\makeatother
\providecommand{\remarkname}{Remark}

\input{tcilatex}

\begin{document}
\title{
Semiparametric Estimation of Structural Functions in Nonseparable Triangular Models}

\author{Victor Chernozhukov$^\dag$ \and Iv\'an Fern\'andez-Val$^\S$ \and Whitney Newey$^\ddag$ \and  \\ Sami Stouli$^\P$ \and Francis Vella$^{\vert}$}
\date{\today. We  gratefully acknowledge research support from the National Science Foundation.}

\thanks{\noindent $^\dag$ Department of Economics, MIT, 
vchern@mit.edu.}
\thanks{ $\S$ Department of Economics, Boston University, ivanf@bu.edu.}
\thanks{ $\ddag$ Department of Economics, MIT, 
wnewey@mit.edu.}
\thanks{ $\P$ Department of Economics, University of Bristol, 
s.stouli@bristol.ac.uk.}
\thanks{ $\vert$ Department of Economics, Georgetown University, 
Francis.Vella@georgetown.edu.}

\maketitle
\begin{abstract}
Triangular systems with nonadditively separable unobserved heterogeneity provide a theoretically appealing framework for the modelling of complex structural relationships.
However, they are not commonly used in practice due to the need for exogenous variables with large support for identification, the curse of dimensionality in estimation, and the lack of inferential tools. This paper introduces two classes of semiparametric nonseparable triangular models that address these limitations. They are based on distribution and quantile regression modelling of the reduced form conditional distributions of the endogenous variables. We show that average, distribution and quantile structural functions are identified in these systems through a control function approach that does not require a large support condition. We propose a computationally attractive three-stage procedure to estimate the structural functions where the first two stages consist of quantile or distribution regressions. We provide asymptotic theory and uniform inference methods for each stage. In particular, we derive functional central limit theorems and bootstrap functional central limit theorems for the distribution regression estimators of the structural functions. These results establish the validity of the bootstrap for three-stage estimators of structural functions, and lead to simple inference algorithms. We illustrate the implementation and applicability of all our methods with numerical simulations and an empirical application to demand analysis.

\end{abstract}



\section{Introduction}

Models with nonadditively separable disturbances provide an important
vehicle for incorporating heterogenous effects. However, accounting for
endogenous treatments in such a setting can be challenging. One
methodology which has been successfully employed in a wide range of models
with endogeneity is the use of control functions (see, for surveys, Imbens
and Wooldridge 2009, Wooldridge 2015 and Blundell, Newey and Vella 2017). 
The underlying logic of this approach is to account for the endogeneity by
including an appropriate control function in the conditioning variables.
This paper proposes some relatively simple control function procedures to
estimate objects of interest in a triangular model with nonseparable
disturbances. Our approach to circumventing the inherent difficulties in 
nonparametric estimation associated with the curse of dimensionality is to
build our models upon a semiparametric specification. This also alleviates 
the large support requirement on the exogenous instrument, 
or exclusion restriction, needed for nonparametric identification. 
Our goal is thus to provide
models and methods that are essentially parametric but still allow for
nonseparable disturbances in order to address strong data requirements that come with nonparametric formulations. 
These models can be interpreted as ``baseline"
models on which series approximations can be built by adding additional
terms.

We consider two kinds of baseline models, quantile regression and
distribution regression. These models allow the use of convenient and widely
available methods to estimate objects of interest including average, distribution and
quantile structural/treatment effects. 
A main feature of the baseline models is that interaction terms included
would not usually be present as leading terms in estimation. These included
terms are products of a transformation of the control function with the endogenous
treatment. Their presence is meant to allow for heterogeneity in the
coefficient of the endogenous variable. Such heterogenous coefficient linear
models are of interest in many settings, including demand analysis and 
estimation of returns to education, and provide a natural starting point
for more general models that allow for nonlinear effects of the endogenous
treatments.

We use these baseline models to construct estimators of the average,
distribution and quantile structural functions based on parametric quantile
and distribution regressions. 
These objects fully characterise the structural relationship between the endogenous treatment and the outcome of interest, 
and describe the average, distribution and quantiles of the outcome across treatment values, had the treatment been exogenous. 
We also show how these baseline models can be
expanded to include higher order terms, leading to more flexible structural function specifications. The estimation procedure consists of
three stages. First, we estimate the control function via quantile regression (QR) or
distribution regression (DR) of the endogenous treatment on the exogenous
covariates and exclusion restrictions. Second, we estimate the reduced form
distribution of the outcome conditional on the treatment, covariates and
estimated control function using DR or QR. Third, we construct estimators of the structural functions applying
suitable functionals to the reduced form estimator from the second stage.
We derive asymptotic theory for the estimators based on DR in all the stages
using a trimming device that avoids tail estimation in the construction of
the control function. We also establish the validity of the bootstrap for our inference 
on structural functions, which enables the formulation of convenient inference algorithms 
which we describe in detail. The modelling framework we propose thus allows us to address three key 
difficulties that have restricted the use of such models 
in empirical work -- the curse of dimensionality, the large support condition for 
identification and the lack of easily implementable inference methods -- while simultaneously retaining important features of the original nonparametric formulation. We give an
empirical application based on the estimation of Engel curves which illustrates how our approach leads to complete and flexible estimates of all structural functions and their confidence regions.

Our results for the average structural function in the linear random
coefficients model are similar to Garen (1984). Florens, Heckman, Meghir,
Vytlacil (2008) give identification and estimation results for a restricted
model with random coefficients for powers of the endogenous treatment.
Blundell and Powell (2003, 2004) introduce the average structural function,
and Imbens and Newey (2009) give general models and results for a variety of
objects of interest and control functions, including quantile structural
functions, under a large support condition on the exclusion restriction. This work also complements the literature on local identification
and estimation in triangular nonseparable models, as in Chesher (2003), Ma
and Koenker (2006), and Jun (2009), on global construction of structural functions (Stouli, 2012) 
and identification in the presence of an exclusion restriction with small support (Fevrier and d'Haultfoeuille, 2015; Torgovitsky, 2015). 
Chernozhukov, Fernandez-Val and Kowalski
(2015) developed a related two-stage quantile regression estimator for
triangular nonseparable models
. These papers did not consider structural
functions defined for nonseparable triangular models with multidimensional unobserved heterogeneity.

This paper makes four main contributions to the existing literature. First, we establish identification of structural functions in
both classes of baseline models, providing 
conditions that do not impose large support requirements on the exclusion restriction. Second,
we derive a functional central limit theorem and a bootstrap functional
central limit theorem for the two-stage DR estimators in the second stage.
These results are uniform over compact regions of values of the outcome. To
the best of our knowledge, this result is new. Chernozhukov, Fernandez-Val
and Kowalski (2015) derived similar results for two-stage quantile
regression estimators but their results are pointwise over quantile indexes.
Our analysis builds on Chernozhukov, Fernandez-Val, and Galichon (2010) and
Chernozhukov, Fernandez-Val, and Melly (2013), which established the
properties of the DR estimators that we use in the first stage. The theory
of the two-stage estimator, however, does not follow from these results
using standard techniques due to the dimensionality and entropy properties
of the first stage DR estimators. We follow the proof strategy proposed by
Chernozhukov, Fernandez-Val and Kowalski (2015) to deal with these issues.
Third, we derive functional central limit theorems and bootstrap functional
central limit theorems for plug-in estimators of functionals of the
distribution of the outcome conditional on the treatment, covariates and
control function via functional delta method. These functionals include all
the structural functions of interest. 
We also use a linear functional for the average structural function which
had not been previously considered. Fourth, we show that this linear operator
that relates the average of a random variable with its distribution is
Hadamard differentiable. Our modelling framework and theoretical results  
are also of interest for the study of nonseparable triangular models in various alternative settings\footnote{See Fernandez-Val et al. (2018) for an application to the analysis of nonseparable sample selection
models with censored selection rules.}, 
and will allow establishing the validity of bootstrap inference for the corresponding estimators.

The rest of the paper is organized as follows. Section \ref{sec:Model}
describes the baseline models and objects of interest. Section \ref%
{sec:Estimation} presents the estimation and inference methods. Section \ref%
{sec:Asymptotic-Theory} gives asymptotic theory. Section \ref{sec:num}
reports the results of an extensive 
empirical application to Engel curves.
Implementation algorithms and
proofs of the main result are given in the Appendix. The online Appendix
Chernozhukov et al. (2018) contains supplemental material, including results of 
numerical simulations calibrated to the application.

\section{Modelling Framework\label{sec:Model}}

We begin with a brief review of the triangular nonseparable model and some
inherent objects of interest. Let $Y$ denote an outcome variable of interest
that can be continuous, discrete or mixed continuous-discrete, $X$ a
continuous endogenous treatment, $Z$ a vector of exogenous variables, $%
\varepsilon $ a structural disturbance vector of unknown dimension, and $\eta$
a scalar reduced form disturbance\footnote{In our empirical application, we use household level data to study the structural relationship between the share of expenditure on either food or leisure, $Y$, and the log of total expenditure, $X$, with gross earnings of the head of household as the exclusion restriction $Z$. Additional examples and a general economic motivation of nonseparable triangular models are given in Chesher (2003) and Imbens and Newey (2009), for instance.}. The model is 
\begin{eqnarray*}
Y &=&g(X,\varepsilon ), \\
X &=&h(Z,\eta),\text{ \ }(\varepsilon ,\eta)\text{ indep of }Z,
\end{eqnarray*}%
where $\eta\mapsto h(z,\eta)$ is a one-to-one function for each $z$. This model
implies that $\varepsilon$ and $X$ are independent conditional on $\eta
$ and that $\eta$ is a one-to-one function of $V=F_{X}(X\mid Z)$, the
 cumulative distribution function (CDF) of $X$ conditional on $Z$
evaluated at the observed variables. Thus, $V$ is a control function.


Objects of interest in this model include the average structural function
(ASF), $\mu (x)$, quantile structural function (QSF), $Q(\tau ,x)$, and 
distribution structural function (DSF), $G(y,x)$,
where 
\begin{equation*}
\mu (x)=\int g(x,\varepsilon )F_{\varepsilon }(d\varepsilon ),
\end{equation*}%
and
\begin{equation*}
Q(\tau,x)=\tau ^{th}\text{ quantile of }g(x,\varepsilon ), \quad G(y,x)=\textrm{Pr}(g(x,\varepsilon)\leq y).
\end{equation*}%
Here $\mu (\tilde{x})-\mu (\bar{x})$ is like an average treatment effect, 
$Q(\tau ,\tilde{x})-Q(\tau ,\bar{x})$ is like a quantile treatment effect, and
$G(y ,\tilde{x})-G(y ,\bar{x})$ is like a distribution treatment effect
from the treatment effects literature. If the support of $V$ conditional on $%
X=x$ is the same as the marginal support of $V$ then these objects are
nonparametrically identified\footnote{Nonparametric identification thus requires the exclusion restriction $Z$ to have full support conditional on $X=x$; see Imbens and Newey (2009) for a detailed discussion.} by 
\begin{equation*}
\mu (x)=\int E[Y\mid X=x,V=v]F_{V}(dv),
\end{equation*}
and
\begin{equation*}Q(\tau ,x)=G^{\leftarrow }(\tau
,x),\quad G(y,x)=\int F_{Y}(y \mid X=x,V=v)F_{V}(dv),
\end{equation*}%
where 
$G^{\leftarrow }(\tau ,x)$ denotes the left-inverse of $y\mapsto G(y,x)$,
i.e. $G^{\leftarrow }(\tau ,x):=\inf \{y\in \mathbb{R}:G(y,x)\geq \tau \}$.

It is straightforward to extend this 
approach to allow for covariates in the model by further conditioning on or
integrating over them. Suppose that $Z_{1}\subset Z$ is included in the
structural equation, which is now $g(X,Z_{1},\varepsilon ).$ Under the
assumption that $\varepsilon $ and $V$ are jointly independent of $Z$, then $\varepsilon $ will be independent of $X$ and $Z_{1}$ 
conditional on $V$. Conditional on covariates and unconditional
average structural functions are identified by 
\begin{equation*}
\mu (x,z_{1})=\int E[Y \mid X=x,Z_{1}=z_{1},V=v]F_{V}(dv),
\end{equation*}%
and
\begin{equation*}
 \mu (x)=\int E[Y \mid X=x,Z_{1}=z_{1},V=v]F_{Z_1}(dz_{1})F_{V}(dv).
\end{equation*}%
Similarly, conditional on covariates and unconditional quantile and
distribution structural functions are identified by 
\begin{equation*}
Q(\tau ,x,z_{1})=G^{\leftarrow }(\tau ,x,z_{1}),\quad G(y,x,z_{1})=\int
F_{Y}(y\mid X=x,Z_{1}=z_{1},V=v)F_{V}(dv),
\end{equation*}%
and 
\begin{equation*}
Q(\tau ,x)=G^{\leftarrow }(\tau ,x),\quad G(y,x)=\int F_{Y}(y\mid
X=x,Z_{1}=z_{1},V=v)F_{Z_1}(dz_{1})F_{V}(dv),
\end{equation*}%
respectively.



Without functional form restrictions, the curse of dimensionality makes it
difficult to estimate the control function $V=F_{X}(X \mid Z)$, the conditional
mean $E[Y \mid X,Z_{1},V],$ and the conditional CDF $F_{Y}(Y \mid X,Z_{1},V)$, 
and the full support condition makes it difficult to achieve point identification of the structural functions.
These difficulties motivate our specification of baseline parametric
models in what follows. These baseline models provide good starting
points for nonparametric estimation and may be of interest in their
own right.

\subsection{Quantile Regression Baseline\label{sub:QRspec}}

We start with a simplified specification with one endogenous treatment $X$,
one exclusion restriction $Z$, and a continuous outcome $Y$. We show 
below how additional excluded variables and covariates can be included.

The baseline first stage is the QR 
model 
\begin{equation}
X=Q_{X}(V\mid Z)=\pi _{1}(V)+\pi _{2}(V)Z,\quad V\mid Z\sim U(0,1).\label{eq:QR} 
\end{equation}%
%
Note that $v\mapsto \pi _{1}(v)$ and $v\mapsto \pi _{2}(v)$ are infinite
dimensional parameters (functions). We can recover the control function $V$
from $V=F_{X}(X\mid Z)=Q_{X}^{-1}(X\mid Z)$ or equivalently from 
\begin{equation*}
V=F_{X}(X\mid Z)=\int_{0}^{1}1\{\pi _{1}(v)+\pi _{2}(v)Z\leq X\}dv.
\end{equation*}%
This generalized inverse representation of the CDF is convenient for
estimation because it does not require the conditional quantile function to
be strictly increasing to be well-defined. 

The baseline second stage has a reduced form: 
\begin{eqnarray}
Y &=& Q_{Y}(U\mid X,V),\quad U\mid X,V\sim U(0,1), \label{eq:QR2-1} \\
Q_{Y}(U\mid X,V)&=&\beta _{1}(U)+\beta _{2}(U)X+\beta _{3}(U)\Phi^{-1}(V)+\beta _{4}(U)X\Phi ^{-1}(V), \label{eq:QR2-2}
\end{eqnarray}%
where $\Phi ^{-1}$ is the standard normal inverse CDF. This transformation
is included to expand the support of $V$ and to encompass the normal system
of equations as a special case. An example of a structural model with this
reduced form is the random coefficient model 
\begin{equation*}
Y=g(X,\varepsilon )=\varepsilon _{1}+\varepsilon _{2}X,
\end{equation*}%
with the restrictions 
\begin{equation*}
\varepsilon _{j}=Q_{\varepsilon _{j}}(U\mid X,V)=\theta _{j}(U)+\gamma
_{j}(U)\Phi ^{-1}(V),\quad U\mid X,V\sim U(0,1),\quad j\in \{1,2\}.
\end{equation*}%
These restrictions include the control function assumption $\varepsilon
_{j}\perp \!\!\!\perp X\mid V$ and a joint functional form restriction,
where the unobservable $U$ is the same for $\varepsilon _{1}$ and $%
\varepsilon _{2}$. 
Substituting in the second stage equation, 
\begin{equation*}
Y=\theta _{1}(U)+\theta _{2}(U)X+\gamma _{1}(U)\Phi ^{-1}(V)+\gamma
_{2}(U)\Phi ^{-1}(V)X,\quad U\mid X,V\sim U(0,1),
\end{equation*}%
which has the form of \eqref{eq:QR2-1}-\eqref{eq:QR2-2}. 

The specification \eqref{eq:QR2-1}-\eqref{eq:QR2-2} is a baseline, or starting point, for a
more general series approximation to the quantiles of $Y$ conditional on $X$
and $V$ based on including additional functions of $X$ and $\Phi ^{-1}(V)$.
The baseline is unusual as it includes the interaction term $\Phi ^{-1}(V)X;$
it is more usual to take the starting point to be $(1,\Phi ^{-1}(V),X),$
which is linear in the regressors $X$ and $\Phi ^{-1}(V).$ The inclusion of
the interaction term is motivated by allowing the coefficient of $X$ to vary
with individuals, so that $\Phi ^{-1}(V)$ then interacts $X$ in the
conditional distribution of $\varepsilon _{2}$ given the control functions. All the parameters of model \eqref{eq:QR}-\eqref{eq:QR2-2} can be
estimated using the QR estimator (Koenker and Bassett, 1978).


The ASF of the baseline specification is: 
\begin{equation*}
\mu (x)=\int_{0}^{1}E[Y\mid X=x,V=v]dv=\beta _{1}+\beta _{2}x, 
\end{equation*}
where the second equality follows by $\int_{0}^{1}\Phi ^{-1}(v)dv=0$ and 
\begin{equation*}
E[Y \mid X,V]=\int_{0}^{1}Q_{Y}(u\mid X,V)du=\beta _{1}+\beta _{2}X+\beta
_{3}\Phi ^{-1}(V)+\beta _{4}X\Phi ^{-1}(V)
\end{equation*}%
with $\beta _{j}:=\int_{0}^{1}\beta _{j}(u)du$, $j\in \{1,\ldots ,4\}$. The
QSF does not appear to have a closed form expression. It is the solution to 
\begin{eqnarray*}
Q(\tau ,x) &=&G^{\leftarrow}(\tau ,x), \\
G(y,x) &=&\int_{0}^{1}\int_{0}^{1}1\{\beta _{1}(u)+\beta _{2}(u)x+\beta
_{3}(u)\Phi ^{-1}(v)+\beta _{4}(u)\Phi ^{-1}(v)x\leq y\}dudv.
\end{eqnarray*}%
%
A special case of the QR baseline is a heteroskedastic normal system of
equations. We use this specification in the numerical simulations of Section %
\ref{sec:num}.


\subsection{Distribution Regression Baseline\label{sub:DRspec}}

We start again with a simplified specification with one endogenous treatment 
$X$ and one excluded $Z$, but now the outcome $Y$ can be continuous,
discrete or mixed.

Let $\Gamma$ denote a strictly increasing continuous CDF such as the
standard normal or logistic CDF. The first stage equation is the
distribution regression model 
\begin{equation*}
\eta=\pi_{1}(X)+\pi_{2}(X)Z,\quad\eta \mid Z \sim \Gamma,
\end{equation*}
which corresponds to the specification of the control variable $V$ as\textbf{%
\ } 
\begin{equation}
V=F_{X}(X \mid Z)=\Gamma(\pi_{1}(X)+\pi_{2}(X)Z).  \label{eq:DR}
\end{equation}
\begin{sloppy}While the first stage QR model
specifies the conditional quantile function of $X$ given $Z$ to
be linear in $Z$, the DR model (\ref{eq:DR})
specifies the conditional distribution of $X$ given $Z$ to be generalized linear in
$Z$, i.e. linear after applying the link function $\Gamma$.

The second stage baseline has a reduced form: 
\begin{equation}
F_{Y}(Y \mid X,V) =  \Gamma( \beta_1(Y) + \beta_2(Y)X+\beta_3(Y)\Phi^{-1}(V)+\beta_4(Y)\Phi^{-1}(V)X).\label{eq:DR2} 
\end{equation}
When $Y$ is continuous,  an example of a structural model that has reduced form (\ref{eq:DR2})  is the latent random coefficient model 
\begin{equation}
\xi= \varepsilon_{1}+\varepsilon_{2}\Phi^{-1}(V),\quad\xi \mid X,V\sim\Gamma,\label{eq:DR2rc}
\end{equation}
with the restrictions 
\begin{equation*}
\varepsilon_{j}=\theta_{j}(Y)+\gamma_{j}(Y)X,\quad j\in\{1,2\},\label{eq:eps_j}
\end{equation*}
such that the mapping $y\mapsto\theta_{j}(y)+\gamma_{j}(y)x$ is strictly
increasing, and the following conditional independence property is
satisfied: 
\begin{equation}
F_{\varepsilon_{j}}(\varepsilon_{j} \mid V)=F_{\varepsilon_{j}}(\varepsilon_{j} \mid X,V), \quad j\in\{1,2\}.\label{eq:CI}
\end{equation}
Substituting the expression for $\varepsilon_{1}$ and $\varepsilon_{2}$
in \eqref{eq:DR2rc} yields 
\begin{equation*}
\xi=\theta_{1}(Y)+\gamma_{1}(Y)X+\theta_{2}(Y)\Phi^{-1}(V)+\gamma_{2}(Y)\Phi^{-1}(V)X,\label{eq:RFDR}
\end{equation*}
which has a reduced form for the distribution of $Y$ conditional on 
$(X,V)$ as in \eqref{eq:DR2}.

As in the quantile baseline, the specification \eqref{eq:DR2} can be used as starting point 
for a more general series approximation to the  distribution of $Y$ conditional on $X$ and $V$ 
based on including additional functions of $X$ and $\Phi^{-1}(V)$. All the parameters of model \eqref{eq:DR}-\eqref{eq:DR2} can be estimated by DR.\par\end{sloppy}

For the DR baseline, the QSF is the solution to 
\begin{equation*}
Q(\tau ,x)=G^{\leftarrow }(\tau ,x),\ \ G(y,x)=\int_{0}^{1}\Gamma (\beta
_{1}(y)+\beta _{2}(y)x+\beta _{3}(y)\Phi ^{-1}(v)+\beta _{4}(y)\Phi
^{-1}(v)x)dv.
\end{equation*}%
Compared to the QR baseline model, the ASF cannot be obtained as a linear
projection but it can be conveniently expressed as a linear functional of $%
G(y,x)$. Let $\mathcal{Y}$ denote the support of $Y$, $\mathcal{Y}^{+}=%
\mathcal{Y}\cap \lbrack 0,\infty )$ and $\mathcal{Y}^{-}=\mathcal{Y}\cap
(-\infty ,0)$. 
The ASF can be characterized as 
\begin{equation}
\mu (x)=\int_{0}^{1}E[Y\mid X=x, V=v]dv=\int_{\mathcal{Y}^{+}}[1-G(y,x)]\nu
(dy)-\int_{\mathcal{Y}^{-}}G(y,x)\nu (dy),  \label{eq:ASF_DR}
\end{equation}%
where $\nu $ is either the counting measure when $\mathcal{Y}$ is countable
or the Lebesgue measure otherwise, and we exploit the linear relationship
between the expected value and the distribution of a random variable. This
characterization simplifies both the computation and theoretical treatment
of the DR-based estimator for the ASF. It also applies to the QR
specification upon using the corresponding expression for $G(y,x)$.

Section \ref{sec:num} provides an example of a special case of the DR model.

\subsection{Identification}

\begin{sloppy}
The most general specifications that we consider include several exclusion
restrictions, covariates and transformations of the regressors in both
stages. For $d_{z_{1}}:=\textrm{dim}(Z_{1})$ and $r_{1}(Z_{1}):=r_{11}(Z_{11}) 
\otimes \dots \otimes r_{1L}(Z_{1d_{z_{1}}})$, let 
\[
R:=r(Z) \text{ and } W:=w(X,Z_{1},V) := p(X) \otimes r_{1}(Z_{1}) \otimes q(V)
\]
denote
the sets of regressors in the first and second stages, where $r$, $r_{1}$, $p$ and $q$
are vectors of transformations such as powers, b-splines and interactions,
and $\otimes$ denotes the Kronecker product. The simplest case is when $%
r(Z)=(1,Z)^{\prime }$, $r_{1}(Z_{1})=(1,Z_{1})^{\prime }$, $p(X)=(1,X)^{\prime }$ and $%
q(V)=(1,\Phi^{-1}(V))^{\prime }$, so that 
$w(X,Z_{1},V) =
(1,\Phi^{-1}(V),X,X \Phi^{-1}(V),Z_{1},Z_{1} \Phi^{-1}(V),X Z_{1}, X Z_{1}  \Phi^{-1}(V))^{\prime}$%
. The following assumption gathers the baseline specifications for the first
and second stages. 
\par\end{sloppy}

\begin{condition}
{[}Baseline Models{]} 
The outcome $Y$ has a conditional density function $y\mapsto f_{Y}(y\mid
X,Z_{1},V)$ with respect to some measure that is a.s. bounded away from zero
uniformly in $\mathcal{Y}$; and (a) $X$ conditional on $Z$ follows the QR
model 
\begin{equation*}
X=Q_{X}(V \mid Z) = R^{\prime }\pi(V),\ \ V \mid Z \sim U(0,1), 
\end{equation*}
and $Y$ conditional on $(X,Z_{1},V)$ follows the QR model 
\begin{equation*}
Y=Q_{Y}(U\mid X,Z_{1},V)=W^{\prime }\beta(U),\ \ V=F_{X}(X\mid Z),\ \ U\mid
X,Z_{1},V\sim U(0,1); 
\end{equation*}
or (b) $X$ conditional on $Z$ follows the DR model 
\begin{equation*}
V = \Lambda(R^{\prime }\pi(X)), \ \ V \mid Z \sim U(0,1), 
\end{equation*}
and $Y$ conditional on $(X,Z_{1},V)$ follows the DR model, 
\begin{equation*}
U = \Gamma(W^{\prime}\beta(Y)),\ \ V=F_{X}(X\mid Z), \ \ U\mid X,Z_{1},V\sim
U(0,1), 
\end{equation*}
where $\Gamma$ is either the standard normal or logistic CDF.\label%
{ass:Model assumptions}
\end{condition}

The structural functions of the baseline models involve quantile and
distribution regressions on the same set of regressors. A sufficient
condition for identification of the coefficients of these regressions is
that the second moment matrix of those regressors is nonsingular. The
regressors have a Kronecker product form $p(X) \otimes r_{1}(Z_{1}) \otimes q(V)$. 
The second moment matrix for these regressors will be nonsingular if the
joint distribution dominates a distribution where $X$, $Z_{1}$ and $V$ are
independent and the second moment matrices of $X$, $Z_{1}$ and $V$ are
positive definite\footnote{This condition is sufficient for identification and is in principle testable. However, in practice 
it will be easier to check directly if the sample second moment matrix for the regressors is of full rank.}. 
Define the product probability measure $\varsigma(z_{1}):=\times_{l=1}^{d_{z_{1}}} \varsigma_{l}(z_{1l})$.

\begin{sloppy}
\begin{condition}
The joint probability distribution of $X$, $Z_{1}$ and $V$ dominates a
product probability measure $\mu(x) \times \varsigma(z_{1})\times \rho(v)$ such that $%
E_{\mu}[p(X)p(X)^{\prime }]$, $E_{\varsigma_l}[r_{1l}(Z_{1l})r_{1l}(Z_{1l})^{\prime }]$, $l=1,\dots,d_{z_{1}}$, and $E_{\rho}[q(V)q(V)^{\prime }]$
are positive definite.
\label{ass:Identification}
\end{condition}

When $p(X)=(1,X)^{\prime }$, $r_{1l}(Z_{1l})=(1,Z_{1l})^{\prime }$, $l=1,\dots,d_{z_{1}}$, and $q(V)=(1,\Phi ^{-1}(V))^{\prime }$,
Assumption \ref{ass:Identification} simplifies to the requirement that the
joint distribution of $X$, $Z_{1}$ and $V$ be dominating one such that $\text{Var}%
_{\mu }(X)>0$, $\text{Var}%
_{\varsigma_l}(Z_{1l})>0$, $l=1,\dots,d_{z_{1}}$, and $\text{Var}_{\rho }(\Phi ^{-1}(V))>0$. For general
specifications where the regressors are higher order power series, it is
sufficient for Assumption \ref{ass:Identification} that the
joint distribution of $X$, $Z_{1}$ and $V$ be dominating one that has density
bounded away from zero on a hypercube. That will mean that the joint
distribution dominates a uniform distribution on that hypercube, and for a
uniform distribution on a hypercube 
$E[w(X,Z_{1},V) w(X,Z_{1},V)^{\prime}]$ is
nonsingular.
\par\end{sloppy}


\begin{lemma}
\label{Lemma:Nonsingular}If Assumption \ref{ass:Identification} holds, then $%
E[w(X,Z_{1},V) w(X,Z_{1},V)^{\prime}]$ 
is nonsingular.
\end{lemma}
Assumptions \ref{ass:Model assumptions}-\ref{ass:Identification} are
sufficient conditions for the map $y\mapsto F_{Y}(y\mid x,z_{1},v)$ to be
well-defined for all $(x,z_{1},v)$, and therefore for identification of the
structural functions.
\begin{theorem}
If Assumptions \ref{ass:Model assumptions} and \ref{ass:Identification} hold,
then the DSF, QSF and ASF are identified.\label{thm:IdentificationSF}
\end{theorem}
Given the semiparametric specifications in Assumption \ref{ass:Model
assumptions}, identification of structural functions does not restrict
the support of $Z$ to be continuous, and the full support assumption of Imbens
and Newey (2009) need not be satisfied. When $q(V)=(1,\Phi ^{-1}(V))^{\prime }$, 
for the second moment matrix of regressors to be nonsingular 
only requires the control function to have strictly positive variance across the support of $X$, which can be satisfied 
even if the support of $Z$ is binary or discrete. This is in sharp contrast with 
nonparametric identification which requires full support of the control variable 
at each value of $X$.
Theorem \ref{thm:IdentificationSF}
thus illustrates the identifying power of semiparametric restrictions and
the trade-off between these restrictions and the full support condition for
identification of structural functions.

\section{Estimation and Inference Methods\label{sec:Estimation}}

The QR and DR baselines
of the previous section lead to 
three-stage analog estimation and inference
methods for the DSF, QSF and ASF. 
The first
stage estimates the control function $V=F_{X}(X \mid Z)$. The second stage estimates the conditional distribution function $F_{Y}(y \mid X,Z_{1},V)$, replacing $V$ by the estimator from the first stage. The  third stage obtains estimators of the structural
functions, which are functionals of the first and second
stages building blocks. We provide a detailed description of the implementation
of each step for both QR and DR methods.
We also describe a weighted bootstrap procedure to perform uniform
inference on all structural functions considered. Detailed implementation algorithms are given in Appendix \ref{app:impl}.

We  assume that we observe a sample of $n$ independent
and identically distributed realizations $\{(Y_{i},X_{i},Z_{i})\}_{i=1}^{n}$
of the random vector $(Y,X,Z)$, and that $\textrm{dim}(X)=1$.
Calligraphic letters such as $\mathcal{Y}$
and $\mathcal{X}$ denote the supports of $Y$ and $X$; and $\mathcal{YX}$
denotes the joint support of $(Y,X)$. 
The description of all the stages includes individual
weights $e_{i}$ which are set to $1$ for the estimators,
or drawn from a distribution that satisfies Assumption \ref{ass:sampling}
in Section \ref{sec:Asymptotic-Theory} for the weighted bootstrap version of the estimators.

\subsection{First Stage: Estimation of Control Function\label%
{sub:Control-Function-Estimation}}

The first stage estimates the $n$ target values of the control function, $%
V_{i}=F_{X}(X_{i}\mid Z_{i})$, $i=1,\ldots ,n$. We estimate the conditional
distribution of $X$ in a trimmed support $\overline{\mathcal{X}}$ that
excludes extreme values. The purpose of the trimming is to avoid the far
tails. We consider a fixed trimming rule, which greatly simplifies the
derivation of the asymptotic properties. In our numerical and empirical
examples we find that the results are not sensitive to the trimming rule and
the choice of $\overline{\mathcal{X}}$ as the observed support of $X,$ i.e.
no trimming, works well. We use bars to denote trimmed supports with respect
to $X$, e.g., $\overline{\mathcal{X}\mathcal{Z}}=\{(x,z)\in \mathcal{X}%
\mathcal{Z}:x\in \overline{\mathcal{X}}\}$. A subscript in a set denotes a
finite grid covering the set, where the subscript is the number of grid
points. Unless otherwise specified, the points of the grid are sample
quantiles of the corresponding variable at equidistant probabilities in $%
[0,1]$. For example, $\mathcal{X}_{5}$ denotes a grid of $5$ points covering 
$\mathcal{X}$ located at the $0$, $1/4$, $1/2$, $3/4$ and $1$ sample
quantiles of $X$.

Denoting the usual check function by $\rho _{v}(z)=(v-1(z<0))z$, the first
stage in the QR baseline is 
\begin{eqnarray}
\widehat{F}_{X}^{e}(x\mid z) &=&\epsilon +\int_{\epsilon }^{1-\epsilon
}1\{R^{\prime }\widehat{\pi }^{e}(v)\leq x\}dv,\quad R=r(z),\quad (x,z)\in 
\overline{\mathcal{X}\mathcal{Z}},  \label{eq:CF_QR} \\
\widehat{\pi }^{e}(v) &\in &\arg \min_{\pi \in \mathbb{R}^{\dim
(R)}}\sum_{i=1}^{n}e_{i}\rho _{v}(X_{i}-R_{i}^{\prime }\pi ),
\label{eq:QR FS}
\end{eqnarray}%
for some small constant $\epsilon >0$. The adjustment in the limits of the
integral in (\ref{eq:CF_QR}) avoids tail estimation of quantiles.\footnote{%
Chernozhukov, Fernandez-Val and Melly (2013) provide conditions under which
this adjustment does not introduce bias.} 
The first stage in the DR baseline is, 
\begin{align}
&\widehat{F}_{X}^{e}(x\mid z) = \Gamma (R^{\prime }\widehat{\pi }%
^{e}(x)),\quad R=r(z),\quad (x,z)\in \overline{\mathcal{X}\mathcal{Z}},
\label{eq:CF_DR} \\
&\widehat{\pi }^{e}(x) \in \arg \min_{\pi \in \mathbb{R}^{\dim
(R)}}\sum_{i=1}^{n}e_{i}\left[ 1\left( X_{i}\leq x\right) \log \Gamma
(R_{i}^{\prime }\pi )\right. \label{eq:DR FS} \\
&
\quad \quad \quad \left. + 1\left( X_{i}>x\right) \log \left( 1-\Gamma
(R_{i}^{\prime }\pi )\right) \right]. \nonumber
\end{align}
When $e_{i}=1$ for all $i=1,\dots,n$, expressions (\ref{eq:CF_QR})-(\ref{eq:QR FS}) and (\ref%
{eq:CF_DR})-(\ref{eq:DR FS}) define $\widehat{F}_{X}$, the QR and DR 
estimators of $F_{X}$. 
For $(X_{i},Z_{i})\in \overline{\mathcal{XZ}}$, the estimator and weighted
bootstrap version of the control function are then $\widehat{V}_{i}=\widehat{%
F}_{X}(X_{i}\mid Z_{i})$ and $\widehat{V}_{i}^{e}=\widehat{F}%
_{X}^{e}(X_{i}\mid Z_{i})$, respectively, and we set $\widehat{V}_{i}=%
\widehat{V}_{i}^{e}=0$ otherwise.

\begin{rem}
For DR, the estimation of $\pi(x)$ at each $x=X_{i}$ can be computationally
expensive. 
Substantial gains in computational speed is achieved by first estimating $%
\pi(x)$ in a grid $\overline{\mathcal{X}}_M$, and then obtaining $\widehat{%
\pi}(x)$ at each $x=X_{i}$ by interpolation.
\end{rem}


\subsection{Second Stage: Estimation of $F_Y(\cdot \mid X,Z_1,V)$}

With the estimated control function in hand, the second building block
required for the estimation of structural functions is an estimate of the
reduced form CDF of $Y$ given $(X,Z_{1},V)$. The baseline models provide
direct estimation procedures based on QR and DR.

Let $T := 1(X \in \overline{\mathcal{X}})$ be a trimming indicator, which is
formally defined in Assumption \ref{ass:first} of Section \ref%
{sec:Asymptotic-Theory}. The estimator of $F_{Y}$ in the QR baseline is 
\begin{eqnarray}
&&\widehat{F}_{Y}^{e}(y\mid x,z_{1},v)=\epsilon +\int_{\epsilon
}^{1-\epsilon }1\{w(x,z_{1},v)^{\prime }\widehat{\beta}^{e}(u)\leq y\}du,\ \
(y,x,z_{1},v)\in \mathcal{Y}\overline{\mathcal{X}\mathcal{Z}_{1}\mathcal{V}},
\label{eq:FYXV_QR} \\
&&\widehat{\beta}^{e}(u)\in \arg \min_{\beta \in \mathbb{R}^{\text{dim}%
(W)}}\sum_{i=1}^{n}e_{i}T_{i}\rho _{u}(Y_{i}-\widehat{W}_{i}^{e\prime }\beta
),\ \ \widehat{W}_{i}^{e}=w(X_{i},Z_{1i},\widehat{V}_{i}^{e}),
\label{eq:QR SS}
\end{eqnarray}%
As for the first stage, the adjustment in the limits of the integral in (\ref%
{eq:FYXV_QR}) avoids tail estimation of quantiles. The estimator of $F_{Y}$
in the DR baseline is 
\begin{align}
&\widehat{F}_{Y}^{e}(y\mid x,z_{1},v) 
=\Gamma (w(x,z_{1},v)^{\prime }%
\widehat{\beta}^{e}(y)),\ \ (y,x,z_{1},v)\in \mathcal{Y}\overline{\mathcal{X}%
\mathcal{Z}_{1}\mathcal{V}},\label{eq:FYXV_DR} \\
&\widehat{\beta}^{e}(y) \in 
 \arg \min_{\beta \in \mathbb{R}^{\text{dim}%
(W)}} \sum_{i=1}^{n}e_{i}T_{i}\left[ 1\left( Y_{i}\leq y\right) \log \Gamma(%
\widehat{W}_{i}^{e\prime }\beta ) \right.\label{eq:DR SS}\\
&
\quad \quad \quad \left. + 1\left( Y_{i}>y\right) \log \left(
1-\Gamma (\widehat{W}_{i}^{e\prime }\beta )\right) \right] \nonumber.
\end{align}
When $e_{i}=1$ for all $i=1,\dots,n$, expressions (\ref{eq:FYXV_QR})-(\ref{eq:QR SS}) and (\ref%
{eq:FYXV_DR})-(\ref{eq:DR SS}) define $\widehat{F}_{Y}$, the quantile and
distribution regression estimators of $F_{Y}$, respectively. 

\subsection{Third Stage: Estimation of Structural Functions\label%
{sub:Estimation-of-Structural}}

\begin{sloppy}Given the estimators $(\{\widehat{V}_{i}\}_{i=1}^{n},\widehat{F}_{Y})$
and their bootstrap draws $(\{\widehat{V}_{i}^{e}\}_{i=1}^{n},\widehat{F}_{Y}^{e})$, we can form estimators of the
 structural functions as functionals of these building blocks. 

The estimator and bootstrap draw of the DSF are
\begin{equation}
\widehat{G}(y,x) = \frac{1}{n_T}\sum_{i=1}^{n}\widehat{F}_{Y}(y \mid x,Z_{1i},\widehat{V}_{i})T_{i},\label{eq:DSF est}
\end{equation}
where $n_T = \sum_{i=1}^n T_i$, and
\begin{equation}
\widehat{G}^{e}(y,x)=\frac{1}{n_T^e}\sum_{i=1}^{n}e_{i}\widehat{F}_{Y}^{e}(y \mid x,Z_{1i},\widehat{V}_{i}^{e})T_{i},\label{eq:DSF boot}
\end{equation}
where $n^e_T = \sum_{i=1}^n e_i T_i$.  
For the DR estimator,  $y\mapsto\widehat{G}(y,x)$
may not be monotonic. This can be addressed by applying the rearrangement
method of Chernozhukov, Fernandez-Val and Galichon (2010).

Given the DSF estimate and bootstrap draw, $\widehat{G}(y,x)$ and $\widehat{G}^{e}(y,x)$, the estimator and bootstrap draw of the QSF are
\begin{equation}
\widehat{Q}(\tau,x)=\int_{\Y^{+}}1\{\widehat{G}(y,x)\leq\tau\}\nu(dy)-\int_{\Y^{-}}1\{\widehat{G}(y,x)\geq\tau\}\nu(dy),\label{eq:QSF est}
\end{equation}
and
\begin{equation}
\widehat{Q}^{e}(\tau,x)=\int_{\Y^{+}}1\{\widehat{G}^{e}(y,x)\leq\tau\}\nu(dy)-\int_{\Y^{-}}1\{\widehat{G}^{e}(y,x)\geq\tau\}\nu(dy),\label{eq:QSF boot}
\end{equation}
respectively. Finally, the estimator and bootstrap draw of the ASF are
\begin{equation}
\widehat{\mu}(x)=\int_{\Y^{+}}[1-\widehat{G}(y,x)]\nu(dy)-\int_{\Y^{-}}\widehat{G}(y,x)\nu(dy),\label{eq:ASF est}
\end{equation}
and
\begin{equation}
\widehat{\mu}^{e}(x)=\int_{\Y^{+}}[1-\widehat{G}^{e}(y,x)]\nu(dy)-\int_{\Y^{-}}\widehat{G}^{e}(y,x)\nu(dy),\label{eq:ASF boot}
\end{equation}
respectively. 
When the set $\Y$ is uncountable, we approximate the previous integrals by sums over a fine mesh of equidistant points $\Y_{S}:=\{\inf[y\in\Y]=y_{1}<\cdots<y_{S}=\sup[y\in\Y]\}$
with mesh width $\delta$ such that $\delta\sqrt{n}\to0$. For example, (\ref{eq:QSF boot}) and (\ref{eq:ASF boot})  are approximated by
\begin{equation}
\widehat{Q}_{S}^{e}(\tau,x)=\delta\sum_{s=1}^{S}\left[1(y_{s}\geq0)-1\{\widehat{G}^{e}(y_{s},x)\geq\tau\}\right],\label{eq:QSF est2}
\end{equation}
and
\begin{equation}
\widehat{\mu}_{S}^{e}(x)=\delta\sum_{s=1}^{S}\left[1(y_{s}\geq0)-\widehat{G}^{e}(y_{s},x)\right].\label{eq:ASF est2}
\end{equation}

\subsection{Weighted Bootstrap Inference on Structural Functions}
We consider inference uniform over regions of values of $(y,x,\tau)$.   We denote the region of interest as $\mathcal{I}_G$ for the DSF, $\mathcal{I}_Q$ for the QSF, and $\mathcal{I}_{\mu}$ for the ASF. Examples include:
\begin{enumerate}
\item The DSF, $y\mapsto\widehat{G}^{e}(y,x)$, for fixed $x$
and over $y \in  \widetilde \Y \subset \Y$, by setting $\mathcal{I}_G=\widetilde \Y \times\{x\}$.
\item The QSF, $\tau \mapsto\widehat{Q}^{e}(\tau,x)$  for fixed $x$ and over $\tau \in \widetilde \T \subset (0,1)$, by setting $\mathcal{I}_Q= \widetilde \T  \times \{x\}$,
\item The ASF, $\widehat{\mu}^{e}(x)$, over $x \in \widetilde \X \subset \overline \X$,
by setting $\mathcal{I}_{\mu}=\widetilde \X$.
\end{enumerate}
When the region of interest is not a finite set, we approximate it by a finite grid.  
All the details of the procedure we implement are summarized in Algorithm \ref{%
alg:SFs_Est} in Appendix \ref{app:impl}.
\par\end{sloppy}


The weighted bootstrap versions of the DSF, QSF and ASF estimators are
obtained by rerunning the estimation procedure introduced in Section \ref%
{sub:Estimation-of-Structural} with sampling weights drawn from a
distribution that satisfies Assumption \ref{ass:sampling} in Section \ref%
{sec:Asymptotic-Theory}; see Algorithm \ref{alg:Weighted-bootstrap} in
Appendix \ref{app:impl} for details. 
They can then be used to perform uniform inference over the region of interest. 

For instance, a $(1-\alpha )$-confidence band for the DSF over the region $%
\mathcal{I}_G$ can be constructed as 
\begin{equation}
\left[ \widehat{G}(y,x)\pm \widehat{k}_{G}(1-\alpha )\widehat{\sigma}%
_{G}(y,x),(y,x)\in \mathcal{I}_G\right] ,  \label{eq:CI DSF}
\end{equation}%
where $\widehat{\sigma}_{G}(y,x)$ is an estimator of $\sigma _{G}(y,x),$ the
asymptotic standard deviation of $\widehat{G}(y,x)$, such as the rescaled
weighted bootstrap interquartile range\footnote{An alternative is to use the bootstrap standard deviation, 
but its validity requires  convergence of bootstrap moments in addition 
to convergence of the bootstrap distribution; cf. Remark 3.2 in Chernozhukov et al. (2013) for a discussion.}
\begin{equation}
\widehat{\sigma}_{G}(y,x)=\text{IQR}\left[ \widehat{G}^{e}(y,x) \right]
/1.349,  \label{eq:variance DSF est}
\end{equation}%
and $\widehat{k}_{G}(1-\alpha )$ denote a consistent estimator of the $%
(1-\alpha )$-quantile of the maximal $t$-statistic 
\begin{equation*}
\left\Vert t_{G}(y,x)\right\Vert _{\mathcal{I}_G}=\sup_{(y,x)\in \mathcal{I}%
_G}\left\vert \frac{\widehat{G}(y,x)-G(y,x)}{\sigma _{G}(y,x)}\right\vert ,
\end{equation*}%
such as the $(1-\alpha )$-quantile of the bootstrap draw of the maximal $t$%
-statistic 
\begin{equation}
\left\Vert t_{G}^{e}(y,x)\right\Vert _{\mathcal{I}_G}=\sup_{(y,x)\in 
\mathcal{I}_G}\left\vert \frac{\widehat{G}^{e}(y,x)-\widehat{G}(y,x)}{%
\widehat{\sigma}_{G}(y,x)}\right\vert .  \label{eq:max t stat DSF}
\end{equation}

Confidence bands for the ASF can be constructed by a similar procedure, 
using the bootstrap draws of the ASF estimator. 
For
the QSF, we can either use the same procedure based on the bootstrap draws
of the QSF, or invert the confidence bands for the DSF following the generic
method of Chernozhukov et al (2016). The first possibility works only when $Y
$ is continuous, whereas the second method is more generally applicable. We
provide algorithms for the construction of the bands in Appendix \ref%
{app:impl}.


\section{Asymptotic Theory\label{sec:Asymptotic-Theory}}

We derive asymptotic theory for the estimators of the ASF, DSF and QSF where
both the first and second stages are based on DR. The theory for the
estimators based on QR can be derived using similar arguments.

\begin{sloppy}In what follows, we shall use the following notation.
We let the random vector $A=(Y,X,Z,W,V)$ live on some probability
space $(\Omega_{0},\mathcal{F}_{0},P)$. Thus, the probability measure
$P$ determines the law of $A$ or any of its elements. We also let
$A_{1},...,A_{n}$, i.i.d. copies of $A$, live on the complete probability
space $(\Omega,\mathcal{F},\Pr)$, which contains the infinite product
of $(\Omega_{0},\mathcal{F}_{0},P)$. Moreover, this probability space
can be suitably enriched to carry also the random weights that appear
in the weighted bootstrap. The distinction between the two laws $P$
and $\Pr$ is helpful to simplify the notation in the proofs and in
the analysis. Unless explicitly mentioned, all functions appearing
in the statements are assumed to be measurable.\par\end{sloppy}

\begin{sloppy}
We now state formally the assumptions. The first assumption is about
sampling and the bootstrap weights.

\begin{condition}
{[}Sampling and Bootstrap Weights{]}\label{ass:sampling} (a) Sampling: the
data $\{Y_{i},X_{i},Z_{i}\}_{i=1}^{n}$ are a sample of size $n$ of
independent and identically distributed observations from the random vector $%
(Y,X,Z).$ (b) Bootstrap weights: $(e_{1},...,e_{n})$ are i.i.d. draws from a
random variable $e\geq0$, with ${\mathrm{E}}_{P}[e]=1$, $\mathrm{Var}%
_{P}[e]=1,$ and ${\mathrm{E}}_{P}|e|^{2+\delta}<\infty$ for some $\delta>0$;
live on the probability space $(\Omega,\mathcal{F},\Pr)$; and are
independent of the data $\{Y_{i},X_{i},Z_{i}\}_{i=1}^{n}$ for all $n$.
\end{condition}
\par\end{sloppy}

The second assumption is about the first stage where we estimate the control
function $(x,z) \mapsto \vartheta_{0}(x,z)$ defined as 
\begin{equation*}
\vartheta_{0}(x,z):=F_{X}(x\mid z),
\end{equation*}
with trimmed support $\overline{\mathcal{V}}=\{\vartheta_{0}(x,z):(x,z)\in%
\overline{\mathcal{X}\mathcal{Z}}\}$. We assume a logistic DR model for the
conditional distribution of $X$ in 
the trimmed support $\overline{\mathcal{X}}$. 

\begin{condition}
{[}First Stage{]}\label{ass:first} (a) Trimming: we consider a trimming rule
defined by the tail indicator 
\begin{equation*}
T=1(X\in\overline{\mathcal{X}}),
\end{equation*}
where $\overline{\mathcal{X}}=[\underline{x},\overline{x}]$ for some $%
-\infty<\underline{x}<\overline{x}<\infty$, such that $P(T=1)>0$. (b) Model:
the distribution of $X$ conditional on $Z$ follows Assumption \ref{ass:Model
assumptions}(b)  with $\Gamma = \Lambda$ in the trimmed support, where $%
\Lambda$ is the logit link function; 
the coefficients $x\mapsto\pi_{0}(x)$ are three times continuously
differentiable with uniformly bounded derivatives; $\overline{\mathcal{R}}$
is compact; and the minimum eigenvalue of ${\mathrm{E}}_{P}\left[%
\Lambda(R^{\prime }\pi_{0}(x))[1-\Lambda(R^{\prime }\pi_{0}(x))]RR^{\prime }%
\right]$ is bounded away from zero uniformly over $x\in\overline{\mathcal{X}}
$.
\end{condition}

For $x\in\overline{\mathcal{X}}$, let 
\begin{equation*}
\widehat{\pi}^{e}(x)\in\arg\min_{\pi\in\mathbb{R}^{\dim(R)}}\frac{1}{n}%
\sum_{i=1}^{n}e_{i}\{1(X_{i}\leq x)\log\Lambda(R_{i}^{\prime }\pi)+1(X_{i}>x)%
\log[1-\Lambda(R_{i}^{\prime }\pi)]\},
\end{equation*}
and set 
\begin{equation*}
\vartheta_{0}(x,r)=\Lambda(r^{\prime }\pi_{0}(x));\ \widehat{\vartheta}%
^{e}(x,r)=\Lambda(r^{\prime }\widehat{\pi}^{e}(x)),
\end{equation*}
if $(x,r)\in\overline{\mathcal{X}\mathcal{R}},$ and $\vartheta_{0}(x,r)=%
\widehat{\vartheta}^{e}(x,r)=0$ otherwise.

Theorem 4 of Chernozhukov, Fernandez-Val and Kowalski (2015) established the
asymptotic properties of the DR estimator of the control function. We repeat
the result here as a lemma for completeness and to introduce notation that
will be used in the results below. Let $T(x):=1(x\in\overline{\mathcal{X}})$%
, $\|f\|_{T,\infty}:=\sup_{a\in\mathcal{A}}|T(x)f(a)|$ for any function $f:%
\mathcal{A}\mapsto\mathbb{R}$, $\lambda=\Lambda(1-\Lambda)$, the density
of the logistic distribution.

\begin{lemma}
{[}First Stage{]}\label{lemma:first} Suppose that Assumptions \ref%
{ass:sampling} and \ref{ass:first} hold. Then, (1) 
\begin{eqnarray*}
\sqrt{n}(\widehat{\vartheta}^{e}(x,r)-\vartheta_{0}(x,r)) & = & \frac{1}{%
\sqrt{n}}\sum_{i=1}^{n}e_{i}\ell(A_{i},x,r)+o_{\Pr}(1)\rightsquigarrow%
\Delta^{e}(x,r)\text{ in }\ell^{\infty}(\overline{\mathcal{X}\mathcal{R}}),
\\
\ell(A,x,r) & := & \lambda(r^{\prime }\pi_{0}(x))[1\{X\leq
x\}-\Lambda(R^{\prime }\pi_{0}(x))]\times \\
& & \times r^{\prime }{\mathrm{E}}_{P}\left\{ \Lambda(R^{\prime
}\pi_{0}(x))[1-\Lambda(R^{\prime }\pi_{0}(x))]RR^{\prime }\right\} ^{-1}R, \\
{\mathrm{E}}_{P}[\ell(A,x,r)] & = & 0,{\mathrm{E}}_{P}[T\ell(A,X,R)^{2}]<%
\infty,
\end{eqnarray*}
where $(x,r)\mapsto\Delta^{e}(x,r)$ is a Gaussian process with uniformly
continuous sample paths and covariance function given by ${\mathrm{E}}%
_{P}[\ell(A,x,r)\ell(A,\tilde{x},\tilde{r})^{\prime }]$. (2) There exists $%
\widetilde{\vartheta}^{e}:\overline{\mathcal{XR}}\mapsto[0,1]$ that obeys
the same first order representation uniformly over $\overline{\mathcal{XR}}$%
, is close to $\widehat{\vartheta}^{e}$ in the sense that $\|\widetilde{%
\vartheta}^{e}-\widehat{\vartheta}^{e}\|_{T,\infty}=o_{\Pr}(1/\sqrt{n})$
and, with probability approaching one, belongs to a bounded function class $%
\Upsilon$ such that 
\begin{equation*}
\log N(\epsilon,\Upsilon,\|\cdot\|_{T,\infty})\lesssim\epsilon^{-1/2},\ \
0<\epsilon<1.
\end{equation*}
\end{lemma}

The next assumptions are about the second stage. We assume a logistic DR
model for the conditional distribution of $Y$ given $(X,Z_{1},V)$, impose
compactness and smoothness conditions, and provide sufficient conditions for
identification of the parameters. Compactness is imposed over the trimmed
supports and can be relaxed at the cost of more complicated and cumbersome
proofs. The smoothness conditions are fairly tight. The assumptions on $%
\mathcal{Y}$ cover continuous, discrete and mixed outcomes in the second
stage. We denote partial derivatives as $\partial_x f(x,y) := \partial f(x,y)/\partial x.$

\begin{condition}
{[}Second Stage{]}\label{ass:second} (a) Model: the distribution of $Y$
conditional on $(X,Z_1,V)$ follows Assumption \ref{ass:Model assumptions}(b)
 with $\Gamma = \Lambda$. 
(b) Compactness and smoothness: the set $\overline{\mathcal{X}\mathcal{Z}%
\mathcal{W}}$ is compact; the set $\mathcal{Y}$ is either a compact interval
in $\mathbb{R}$ or a finite subset of $\mathbb{R}$; $X$ has a continuous
conditional density function $x\mapsto f_{X}(x\mid z)$ that is bounded above
by a constant uniformly in $z\in\overline{\mathcal{Z}}$; if $\mathcal{Y}$ is
an interval, then $Y$ has a conditional density function $y\mapsto
f_{Y}(y\mid x,z)$ that is uniformly continuous in $y\in\mathcal{Y}$
uniformly in $(x,z)\in\overline{\mathcal{X}\mathcal{Z}}$, and bounded above
by a constant uniformly in $(x,z)\in\overline{\mathcal{X}\mathcal{Z}}$; the
derivative vector $\partial_{v}w(x,z_{1},v)$ exists and its components are
uniformly continuous in $v\in\overline{\mathcal{V}}$ uniformly in $%
(x,z_{1})\in\overline{\mathcal{X}\mathcal{Z}_{1}}$, and are bounded in
absolute value by a constant, uniformly in $(x,w,v)\in\overline{\mathcal{X}%
\mathcal{Z}_{1}\mathcal{V}}$; and for all $y\in\mathcal{Y}$, $\beta_{0}(y)\in%
\mathcal{B}$, where $\mathcal{B}$ is a compact subset of $\mathbb{R}%
^{\dim(W)}$. (c) Identification and nondegeneracy: 
Assumption \ref{ass:Identification} holds conditional on $T=1$, and the
matrix $C(y,v):=\mathrm{Cov}_{P}[f_{y}(A)+g_{y}(A),f_{v}(A)+g_{v}(A)\ ]$ is
finite and is of full rank uniformly in $y,v\in\mathcal{Y}$, where 
\begin{equation*}
f_{y}(A):=\{\Lambda(W^{\prime}\beta_{0}(y))-1(Y\leq y)\}WT,
\end{equation*}
and, for $\dot{W}=\partial_{v}w(X,Z_{1},v)|_{v=V}$, 
\begin{equation*}
g_{y}(A):={\mathrm{E}}_{P}[\{[\Lambda(W^{\prime }\beta_{0}(y))-1(Y\leq y)]%
\dot{W}+\lambda(W^{\prime }\beta_{0}(y))\dot{W}^{\prime
}\beta_{0}(y)W\}T\ell(a,X,R)]\big|_{a=A}.
\end{equation*}
\end{condition}

For $y\in\mathcal{Y}$, let 
\begin{equation*}
\widehat{\beta}(y)=\arg\min_{\beta\in\mathbb{R}^{\dim(W)}}\frac{1}{n}%
\sum_{i=1}^{n}T_{i}\rho_{y}(Y_{i},\beta^{\prime }\widehat{W}_{i}),\ \ 
\widehat{W}_{i}=w(X_{i},Z_{1i},\widehat{V}_{i}),\ \ \widehat{V}_{i}=\widehat{%
\vartheta}(X_{i},R_{i}),
\end{equation*}
where 
\begin{equation*}
\rho_{y}(Y,B):=-\{1(Y\leq y)\log\Lambda(B)+1(Y>y)\log[1-\Lambda(B)]\},
\end{equation*}
and $\widehat{\vartheta}$ is the estimator of the control function in the
unweighted sample; and 
\begin{equation*}
\widehat{\beta}^{e}(y)=\arg\min_{\beta\in\mathbb{R}^{\dim(W)}}\frac{1}{n}%
\sum_{i=1}^{n}e_{i}T_{i}\rho_{y}(Y_{i},\beta^{\prime }\widehat{W}_{i}^{e}),\
\ \widehat{W}_{i}^{e}=w(X_{i},Z_{1i},\widehat{V}_{i}^{e}),\ \ \widehat{V}%
_{i}^{e}=\widehat{\vartheta}^{e}(X_{i},R_{i}),
\end{equation*}
where $\widehat{\vartheta}^{e}$ is the estimator of the control function in
the weighted sample.

The following lemma establishes a functional central limit theorem and a
functional central limit theorem for the bootstrap for the estimator of the
DR coefficients in the second stage. Let $d_{w}:=\dim(W)$, and $%
\ell^{\infty}(\mathcal{Y})$ be the set of all uniformly bounded real
functions on $\mathcal{Y}$, and define the matrix $J(y) := {\mathrm{E}}_{P}[\lambda(W'\beta_{0}(y))WW' T]$ for $y\in \mathcal{Y}$. We use $\rightsquigarrow_{\Pr}$ to denote
bootstrap consistency, i.e. weak convergence conditional on the data in
probability, which is formally defined in Appendix \ref{app:notation}.

\begin{lemma}
{[}FCLT and Bootstrap FCLT for $\widehat{\beta}(y)${]}\label{thm:fclt} Under
Assumptions \ref{ass:Model assumptions}--\ref{ass:second}, in $\ell^{\infty}(%
\mathcal{Y})^{d_{w}}$, 
\begin{equation*}
\sqrt{n}(\widehat{\beta}(y)-\beta_{0}(y))\rightsquigarrow J(y)^{-1}G(y),%
\text{\ \ and \ \ }\sqrt{n}(\widehat{\beta}^{e}(y)-\widehat{\beta}%
(y))\rightsquigarrow_{\Pr}J(y)^{-1}G(y),
\end{equation*}
where $y\mapsto G(y)$ is a $d_{w}$-dimensional zero-mean Gaussian process
with uniformly continuous sample paths and covariance function 
\begin{equation*}
{\mathrm{E}}_{P}[G(y)G(v)^{\prime }]=C(y,v),\ \ y,v\in\mathcal{Y}.
\end{equation*}
\end{lemma}

We consider now the estimators of the main quantities of interest -- the
structural functions. Let $W_{x}:=w(x,Z_{1},V)$, $\widehat{W}_{x}:=w(x,Z_{1},%
\widehat{V})$, and $\widehat{W}_{x}^{e}:=w(x,Z_{1},\widehat{V}^{e})$. The DR
estimator and bootstrap draw of the DSF in the trimmed support, $G_T(y,x)={%
\mathrm{E}}_{P}\{\Lambda[\beta_{0}(y)^{\prime }W_{x}] \mid T = 1\}$, are $%
\widehat{G}(y,x)=\sum_{i=1}^{n}\Lambda[\widehat{\beta}(y)^{\prime }\widehat{W%
}_{xi}]T_{i}/n_T$, and $\widehat{G}^{e}(y,x)=\sum_{i=1}^{n}e_{i}\Lambda[%
\widehat{\beta}^{e}(y)^{\prime }\widehat{W}_{xi}^{e}]T_{i}/n_T^e$. 
Let $p_T := P(T=1)$. The next result gives large sample theory for these
estimators.

\begin{theorem}[FCLT and Bootstrap FCLT for DSF]
\label{fclt:sdf} Under Assumptions \ref{ass:Model assumptions}--\ref%
{ass:second}, in $\ell^{\infty}(\mathcal{Y}\overline{\mathcal{X}})$, 
\begin{equation*}
\sqrt{np_T}(\widehat{G}(y,x)-G_T(y,x))\rightsquigarrow Z(y,x)\text{ and }%
\sqrt{np_T}(\widehat{G}^{e}(y,x)-\widehat{G}(y,x))\rightsquigarrow_{%
\Pr}Z(y,x),
\end{equation*}
where $(y,x)\mapsto Z(y,x)$ is a zero-mean Gaussian process with covariance
function 
\begin{equation*}
\mathrm{Cov}_{P}[\Lambda[W_{x}^{\prime }\beta_{0}(y)] +h_{y,x}(A),\Lambda[%
W_{u}^{\prime }\beta_{0}(v)] +h_{v,u}(A) \mid T = 1],
\end{equation*}
with 
\begin{multline*}
h_{y,x}(A)={\mathrm{E}}_{P}\{\lambda[W_{x}^{\prime }\beta_{0}(y)]%
W_{x}T\}^{\prime -1}[f_{y}(A)+g_{y}(A)]+ \\
{\mathrm{E}}_{P}\{\lambda[W_{x}^{\prime }\beta_{0}(y)]\dot{W}_{x}^{\prime
}\beta_{0}(y)T\ell(a,X,R)\}\big|_{a=A}.
\end{multline*}
\end{theorem}


When $Y$ is continuous and $y\mapsto G_T(y,x)$ is strictly increasing, we
can also characterize the asymptotic distribution of $\widehat{Q}(\tau,x)$,
the estimator of the QSF in the trimmed support. Let $g_T(y,x)$ be the
density of $y\mapsto G_T(y,x)$
, $\overline{\mathcal{T}}:=\{\tau\in(0,1):Q(\tau,x)\in \mathcal{Y},
g_T(Q(\tau,x),x)>\epsilon, x \in \overline{\mathcal{X}} \}$ for fixed $%
\epsilon>0$, and $Q_{T}(\tau,x)$ the QSF in the trimmed support $\overline{%
\mathcal{TX}}$ defined as 
\begin{equation*}
Q_{T}(\tau,x)=\int_{\mathcal{Y}^{+}}1\{G_{T}(y,x)\leq\tau\}dy-\int_{\mathcal{%
Y}^{-}}1\{G_{T}(y,x)\geq\tau\}dy. 
\end{equation*}
The estimator and its bootstrap draw given in \eqref{eq:QSF est}-%
\eqref{eq:QSF boot} follow the functional central limit theorem:

\begin{theorem}[FCLT and Bootstrap FCLT for QSF]
\label{fclt:sqf} Assume that $y\mapsto G_T(y,x)$ is strictly increasing in $%
\overline{\mathcal{Y}}$ and $(y,x)\mapsto G_T(y,x)$ is continuously
differentiable in $\overline{\mathcal{Y}\mathcal{X}}$. Under Assumptions \ref%
{ass:Model assumptions}--\ref{ass:second}, in $\ell^{\infty}(\overline{%
\mathcal{T}\mathcal{X}})$, 
\begin{multline*}
\sqrt{np_T}(\widehat{Q}(\tau,x)-Q_{T}(\tau,x))\rightsquigarrow-\frac{%
Z(Q(\tau,x),x)}{g_T(Q(\tau,x),x)}\text{ and } \\ \sqrt{np_T}(\widehat{Q}%
^{e}(\tau,x)-\widehat{Q}(\tau,x))\rightsquigarrow_{\Pr}-\frac{Z(Q(\tau,x),x)%
}{g_T(Q(\tau,x),x)},
\end{multline*}
where $(y,x)\mapsto Z(y,x)$ is the same Gaussian process as in Theorem \ref%
{fclt:sdf}.
\end{theorem}



Finally, we consider the ASF in the trimmed support 
\begin{equation*}
\mu_T(x)=\int_{\mathcal{Y}^{+}}[1-G_T(y,x)]\nu(dy)-\int_{\mathcal{Y}%
^{-}}G_T(y,x)\nu(dy).
\end{equation*}
The estimator and its bootstrap draw given in \eqref{eq:ASF est}-%
\eqref{eq:ASF boot} follow the functional central limit theorem:


\begin{theorem}[FCLT and Bootstrap FCLT for ASF]
\label{fclt:asf} Under Assumptions \ref{ass:Model assumptions}--\ref%
{ass:second}, in $\ell^{\infty}(\overline{\mathcal{X}})$, 
\begin{multline*}
\sqrt{np_T}(\widehat{\mu}(x)-\mu_T(x))\rightsquigarrow-\int_{\mathcal{Y}%
}Z(y,x)\nu(dy)\text{ and } \\ \sqrt{np_T}(\widehat{\mu}^{e}(x)-\widehat{\mu}%
(x))\rightsquigarrow_{\Pr}-\int_{\mathcal{Y}}Z(y,x)\nu(dy),
\end{multline*}
where $(y,x)\mapsto Z(y,x)$ is the same Gaussian process as in Theorem \ref%
{fclt:sdf}.
\end{theorem}

\section{Empirical Application: Engel Curves for Food and Leisure
Expenditure}

\label{sec:num}

In this section we apply our methods to the estimation of a semiparametric
nonseparable triangular model for Engel curves. We focus on the structural
relationship between household's total expenditure and household's demand
for two goods: food and leisure. We take the outcome $Y$ to be the
expenditure share on either food or leisure, and $X$ the logarithm of total
expenditure. Following Blundell, Chen and Kristensen (2007) we use as an
exclusion restriction the logarithm of gross earnings of the head of household. We also
include an additional binary covariate $Z_{1}$ accounting for the presence
of children in the household.

There is an extensive literature on Engel curve estimation (e.g., see Lewbel
(2006) for a review), and the use of nonseparable triangular models for the
identification and estimation of Engel curves has been considered in the
recent literature. Blundell, Chen and Kristensen (2007) estimate
semi-nonparametrically Engel curves for several categories of expenditure,
Imbens and Newey (2009) estimate the QSF nonparametrically for food and
leisure, and Chernozhukov, Fernandez-Val and Kowalski (2015) estimate Engel
curves for alcohol accounting for censoring. For comparison purposes we use
the same dataset as these papers, the 1995 U.K. Family Expenditure Survey.
We restrict the sample to 1,655 married or cohabiting couples with two or
fewer children, in which the head of the household is employed and between
the ages of 20 and 55 years. For this sample we estimate the DSF, QSF and
ASF for both goods. Unlike Imbens and Newey (2009) we also account for the
presence of children in the household and we impose semiparametric
restrictions through our baseline models. In contrast to Chernozhukov,
Fernandez-Val and Kowalski (2015), we do not impose separability between the
control function and other regressors, and we estimate the structural
functions.

\begin{figure}[tbp]
\subfloat[Food.]{\includegraphics[width=7.9cm,height=7cm]{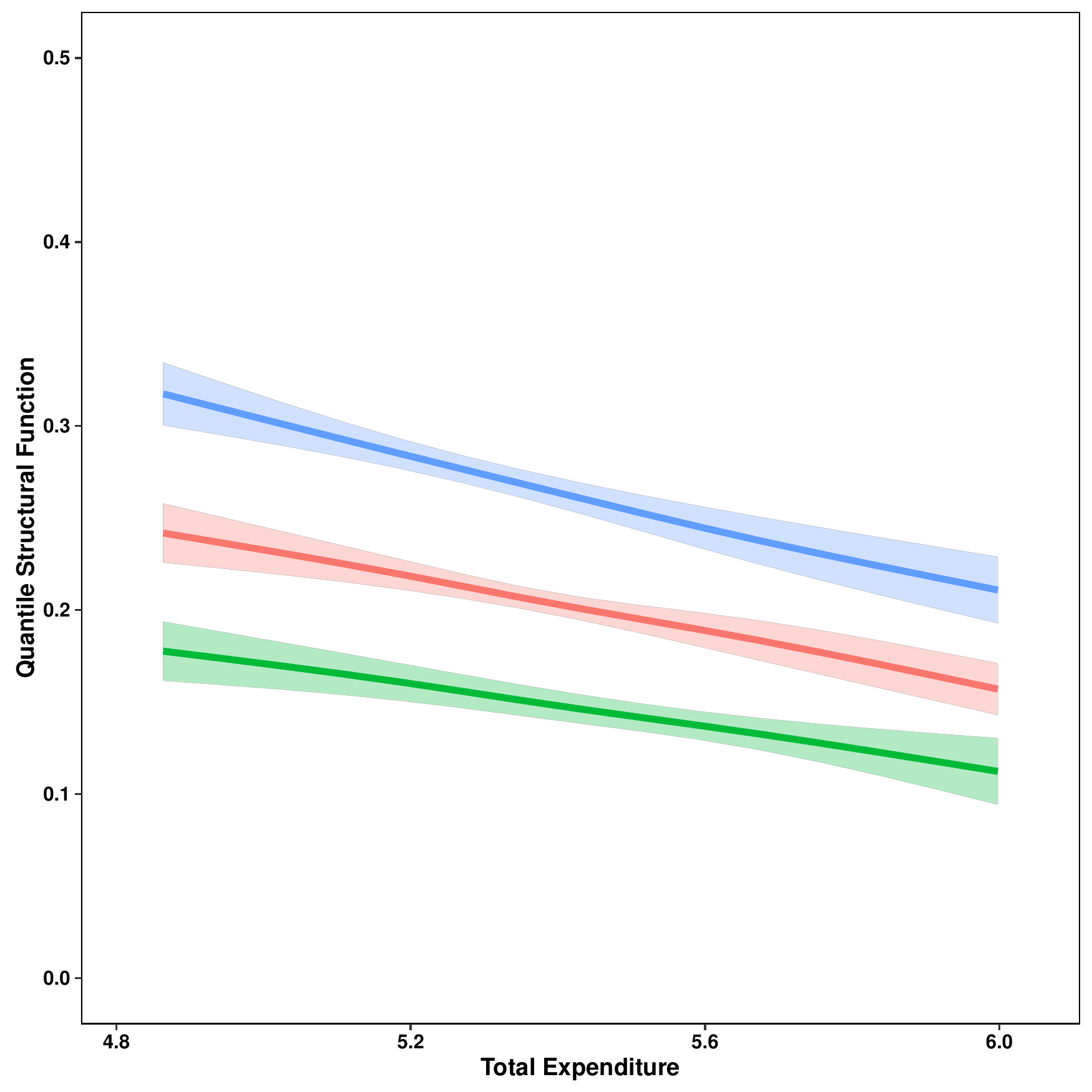}\includegraphics[width=7.9cm,height=7cm]{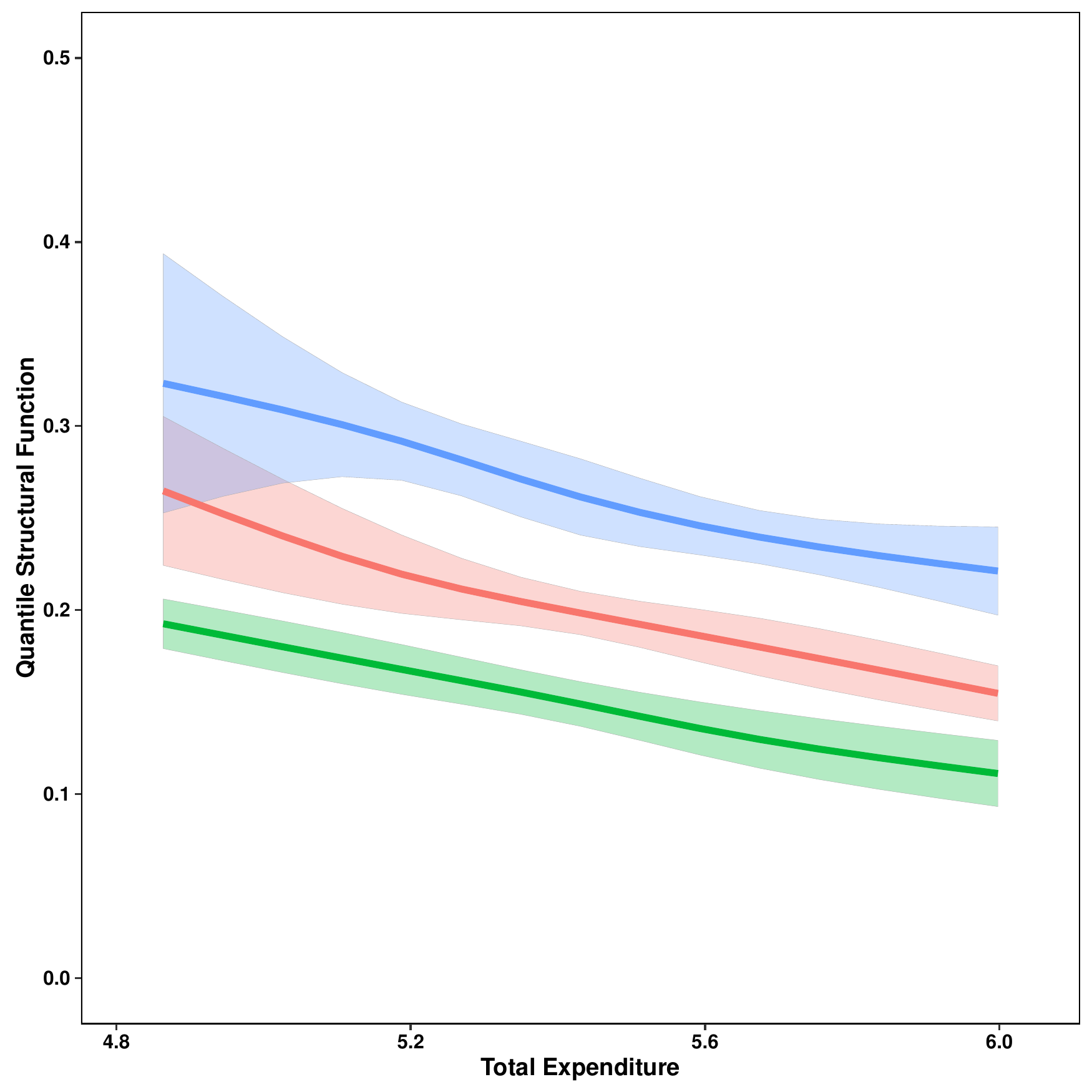}
}
\par
\subfloat[Leisure.]{\includegraphics[width=7.9cm,height=7cm]{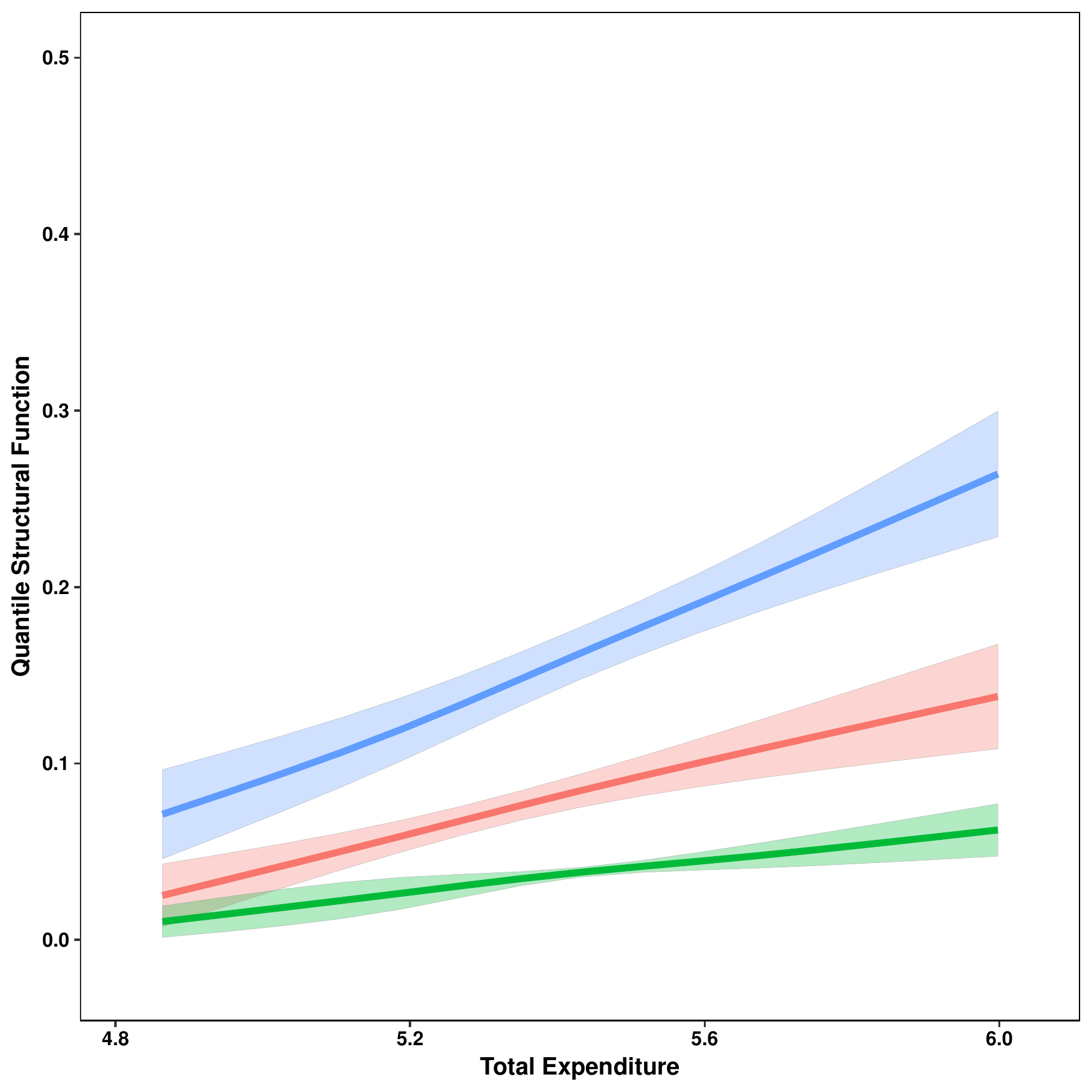}\includegraphics[width=7.9cm,height=7cm]{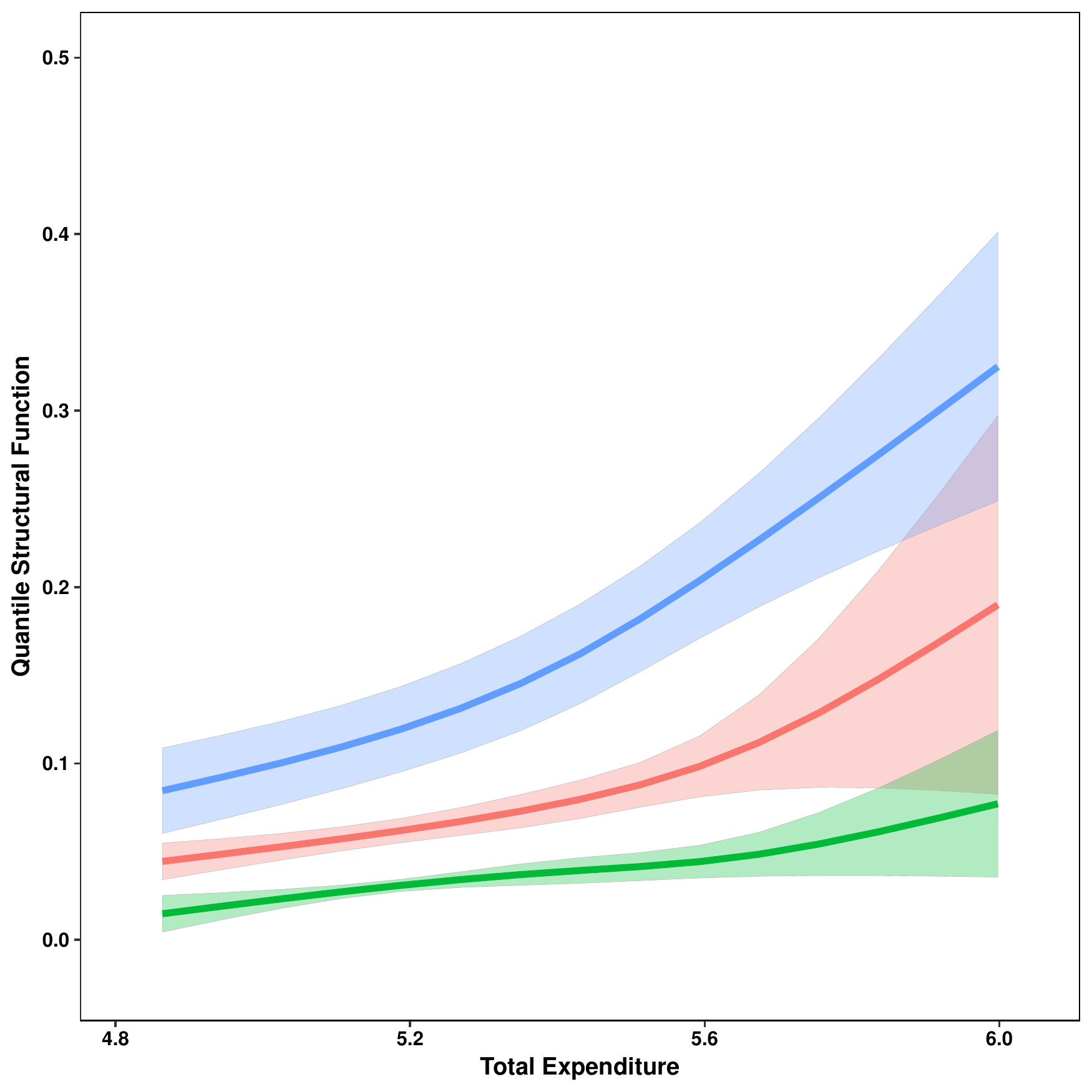}
}
\caption{QSF. Quantile (left) and distribution regression (right).}
\label{fig:QSF}
\end{figure}

\begin{figure}[tbp]
\includegraphics[width=7.9cm,height=7cm]{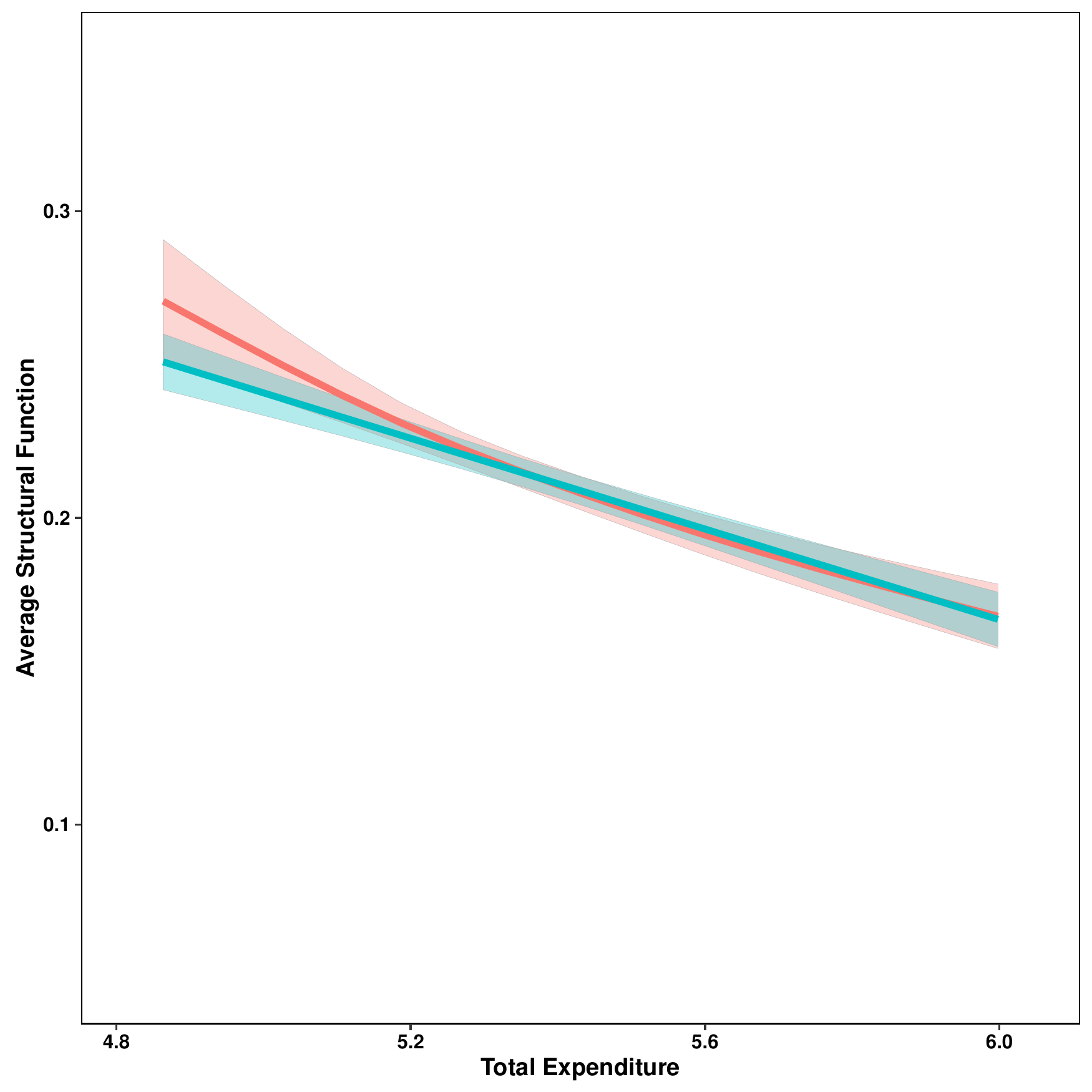}%
\includegraphics[width=7.9cm,height=7cm]{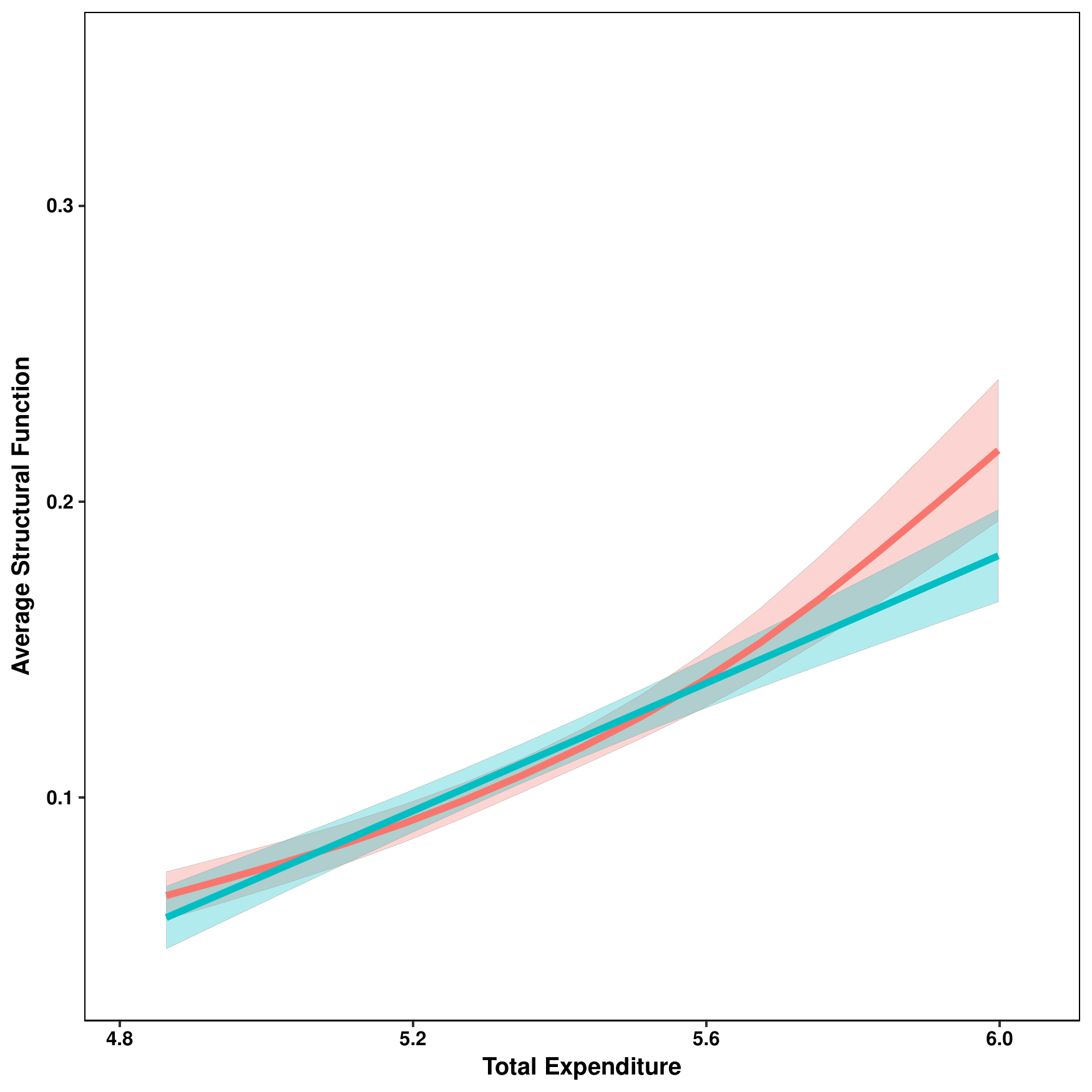}
\caption{ASF for food (left) and leisure (right). Quantile (blue) and
distribution regression (red). }
\label{fig:ASF}
\end{figure}

\begin{figure}[tbp]
\subfloat[Food.]{%
\includegraphics[width=7.9cm,height=7cm]{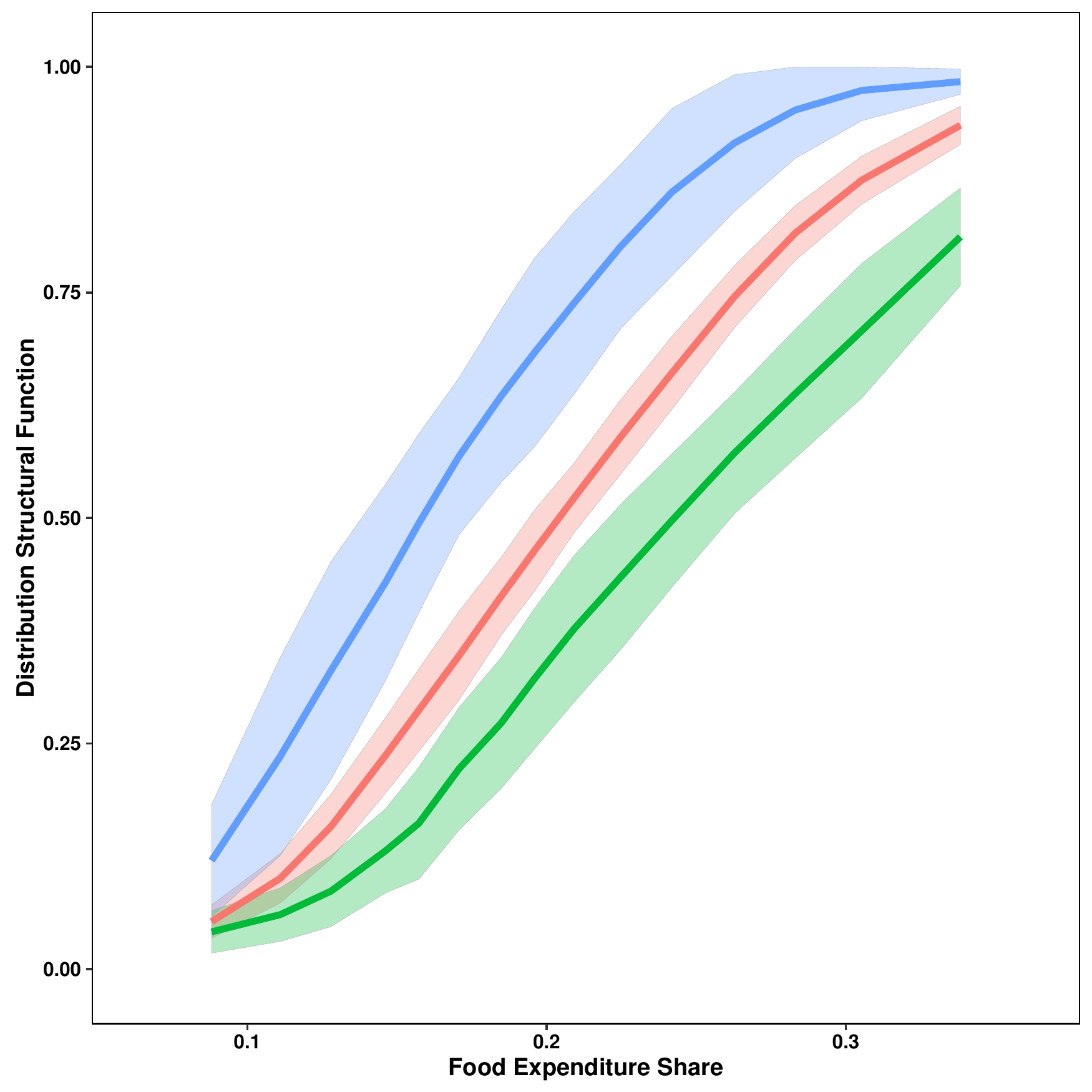}
\includegraphics[width=7.9cm,height=7cm]{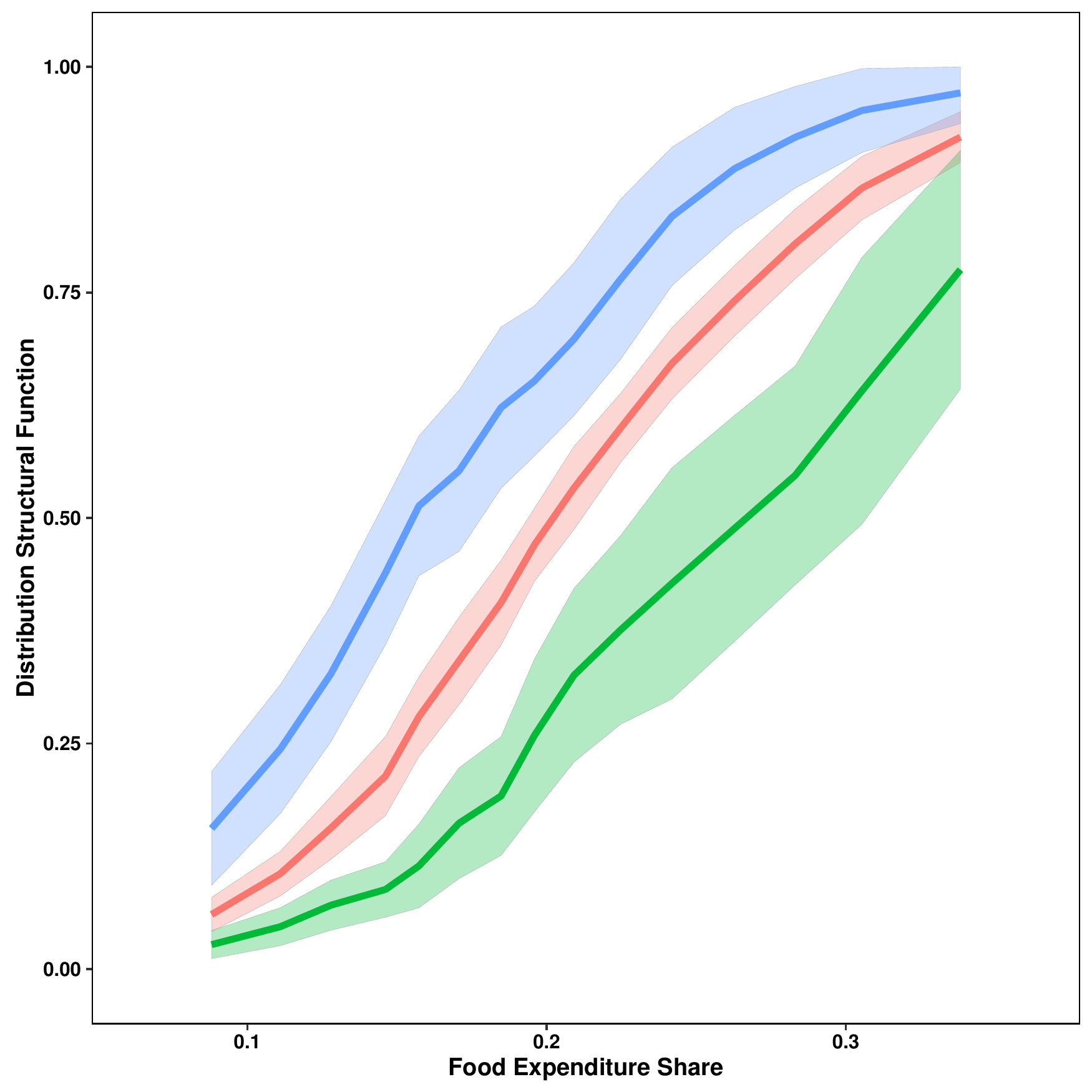}}
\par
\subfloat[Leisure.]{\includegraphics[width=7.9cm,height=7cm]{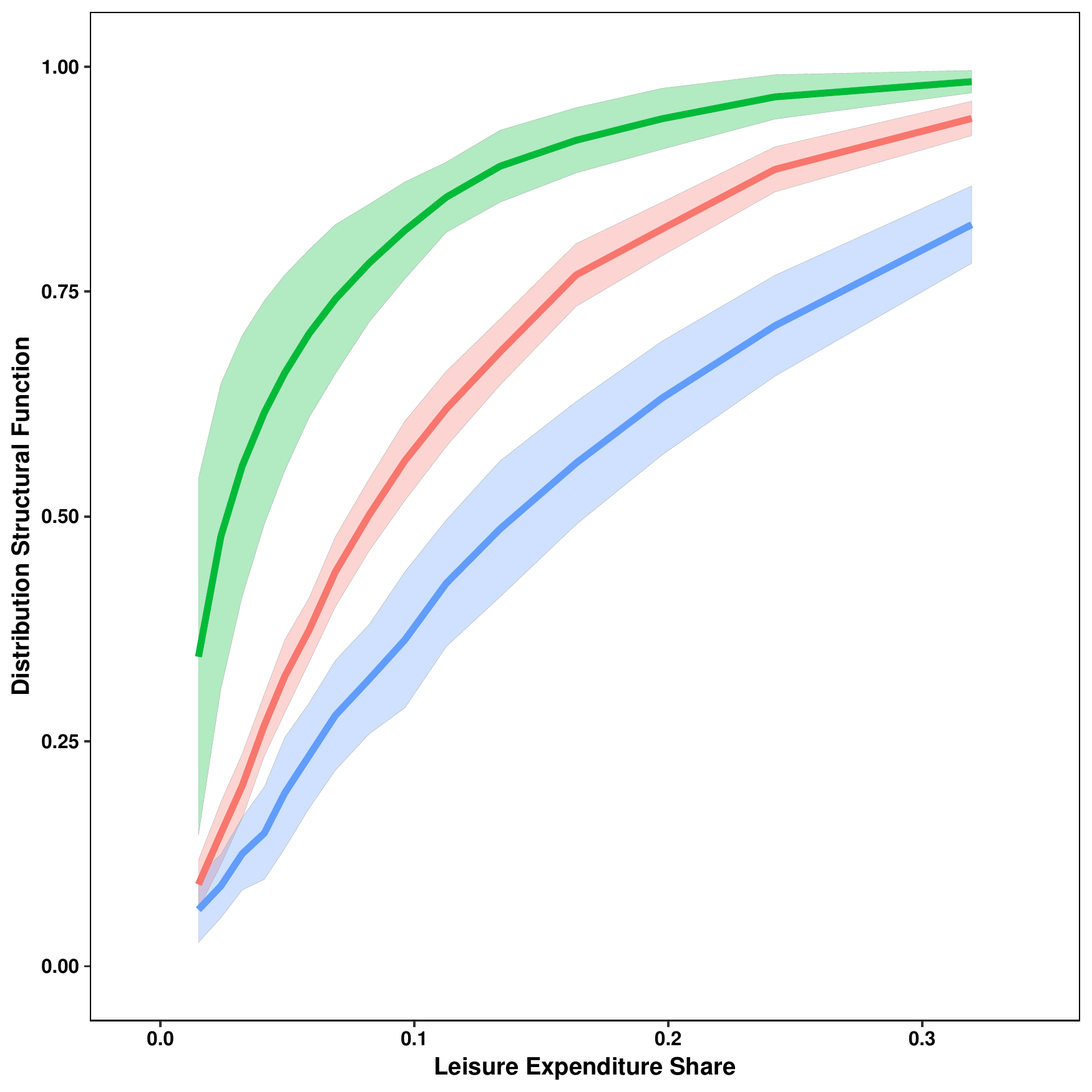}\includegraphics[width=7.9cm,height=7cm]{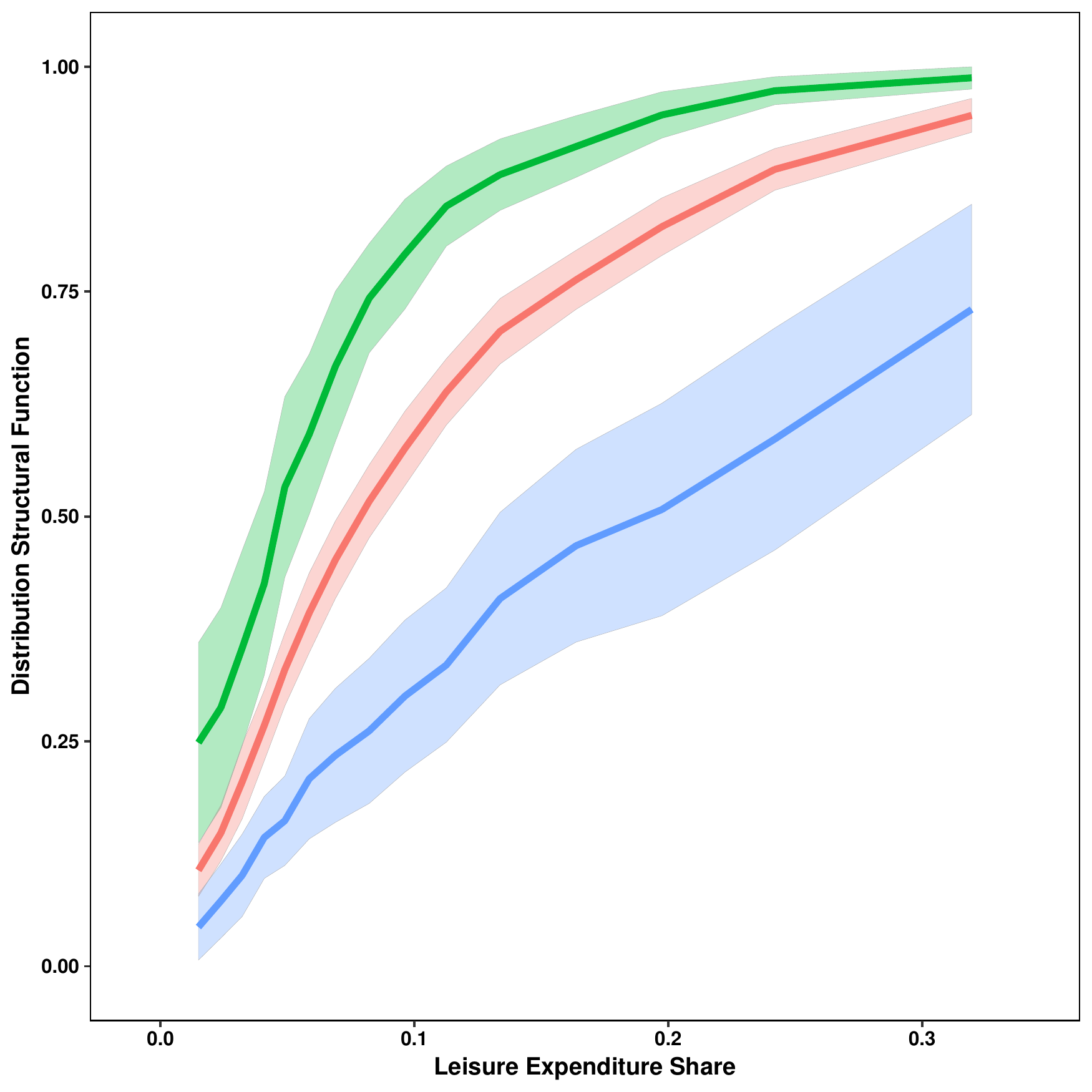}

}
\caption{DSF. Quantile (left) and distribution regression (right).}
\label{fig:DSF}
\end{figure}

All structural functions are estimated by both QR and DR methods, following
exactly the description of the implementation presented in Section \ref%
{sec:Estimation} with the specifications $r(Z) = (1,Z)^{\prime}$, $r_{1}(Z_{1})=(1,Z_{1})^{\prime}$, $p(X)
= (1,X)^{\prime}$,  and $q(V) = (1,\Phi^{-1}(V))^{\prime}$. We set $M=599$ and $%
\epsilon=0.01$ in Algorithm \ref{alg:SFs_Est}, approximate the integrals
using $S=599$ points, and run $B=199$ bootstrap replications in Algorithm %
\ref{alg:Weighted-bootstrap} for both methods. 
The regions of interest are $\widetilde{\mathcal{X}}=[\widehat
Q_{X}(0.1),\widehat Q_{X}(0.9)]$ and $\widetilde{\mathcal{Y}}=[\widehat
Q_{Y}(0.1),\widehat Q_{Y}(0.9)]$, where $\widehat Q_{X}(u)$ and $\widehat
Q_{Y}(u)$ are the sample $u$-quantiles of $X$ and $Y$. We approximate $%
\widetilde{\mathcal{X}}$ by a grid $\widetilde{\mathcal{X}}_{K}$ with $K =
3,5$, and $\widetilde{\mathcal{Y}}$ by a grid $\widetilde{\mathcal{Y}}_{15}$%
. 
We estimate the structural functions and perform uniform inference over the
following regions:

\begin{enumerate}
\item For the QSF, $\widehat{Q}(\tau,x)$, we take $\widetilde{\mathcal{T}}%
=\{0.25,0.5,0.75\}$, and then set: $\mathcal{I}_Q= \widetilde{\mathcal{T}}%
\widetilde{\mathcal{X}}_{5}$.


\item For the DSF, $\widehat{G}(y,x)$, we set: $\mathcal{I}_G=\widetilde{%
\mathcal{Y}}_{15}\widetilde{\mathcal{X}}_{3}$.

\item For the ASF, $\widehat{\mu}(x)$, we set: $\mathcal{I}_{\mu}=\widetilde{%
\mathcal{X}}_{5}$.
\end{enumerate}

We implement the DR estimator using the logit
 link function. Since the
estimated DSF may be non-monotonic in $y$, we apply rearrangement to $%
y\mapsto \widehat{G}(y,x)$ at each value of $x$ in $\mathcal{I}_G$. None of
the methods uses trimming, that is we set $T=1$ a.s.

Figures \ref{fig:QSF}-\ref{fig:DSF} show the QSF, ASF and DSF for both goods%
\footnote{%
For graphical representation the QSF and ASF are interpolated by splines
over $\overline{\mathcal{X}}$ and the DSF over $\overline{\mathcal{Y}}$.}.
For each structural function, we report weighted bootstrap 90\%-confidence
bands that are uniform over the corresponding region specified above. Our
empirical results illustrate that QR and DR specifications are able to
capture different features of structural functions, and are therefore
complementary. For food, both estimation methods deliver very similar QSF estimates, 
close to being linear, although linearity is not
imposed in the estimation procedure. For leisure, the QSF and ASF estimated by DR are
able to capture some nonlinearity which is absent from those obtained by QR.
For QR, this reflects the specified linear structure of the ASF which also 
constrains the shape of the QSF. In addition, some degree of
heteroskedasticity appears to be a feature of the structural model for both
goods, although much more markedly for leisure, so our methods are
well-suited for this problem. Increased dispersion across quantile levels in
Figure \ref{fig:QSF} is reflected by the increasing spread across
probability levels between the two extreme DSF estimates in Figure \ref%
{fig:DSF}. Finally, our semiparametric specifications are able to capture
the asymmetry across leisure expenditure shares, an important feature of the
data highlighted in Imbens and Newey (2009).

\begin{figure}
\subfloat[Food.]{\includegraphics[width=8.5cm,height=8.5cm]{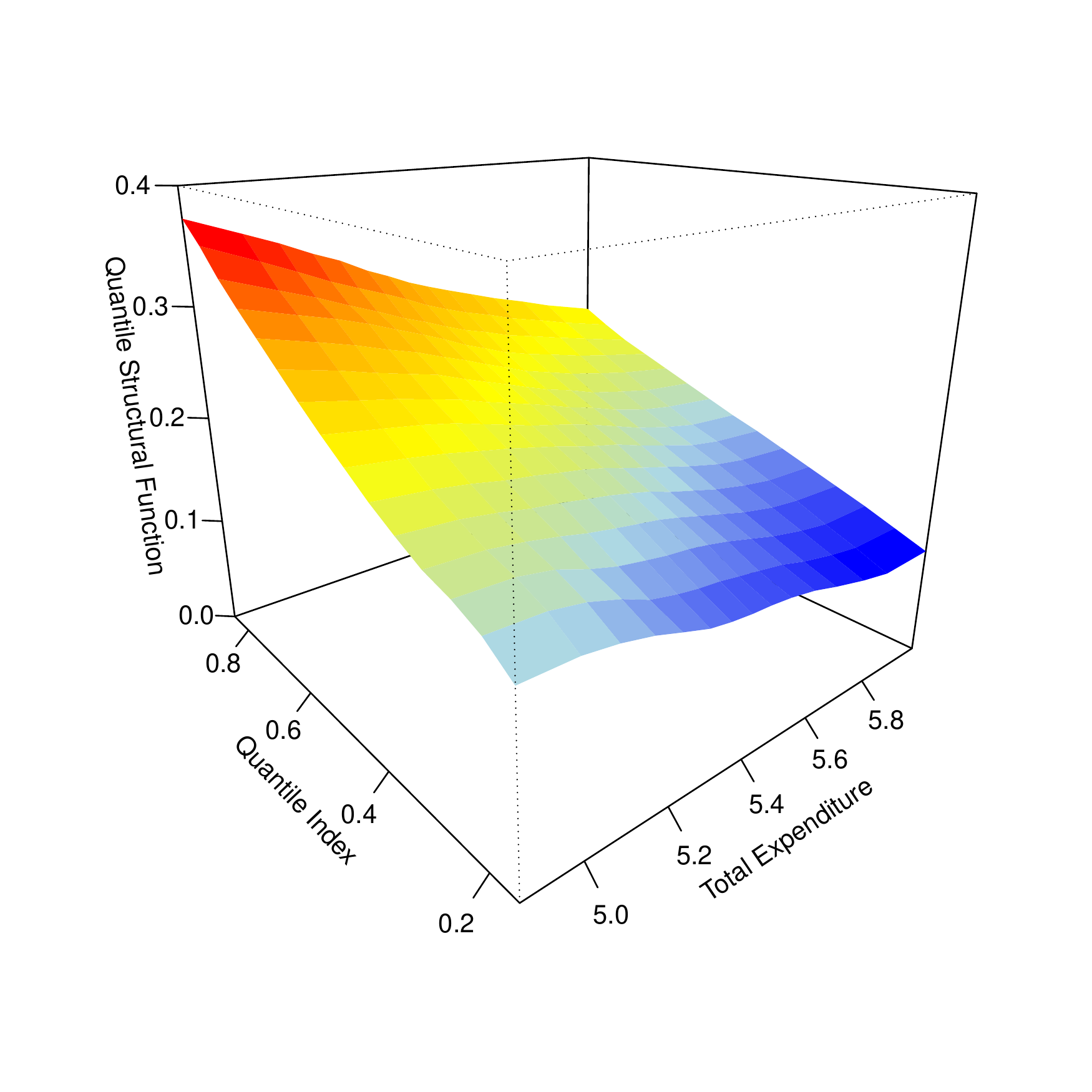} 
\includegraphics[width=8.5cm,height=8.5cm]{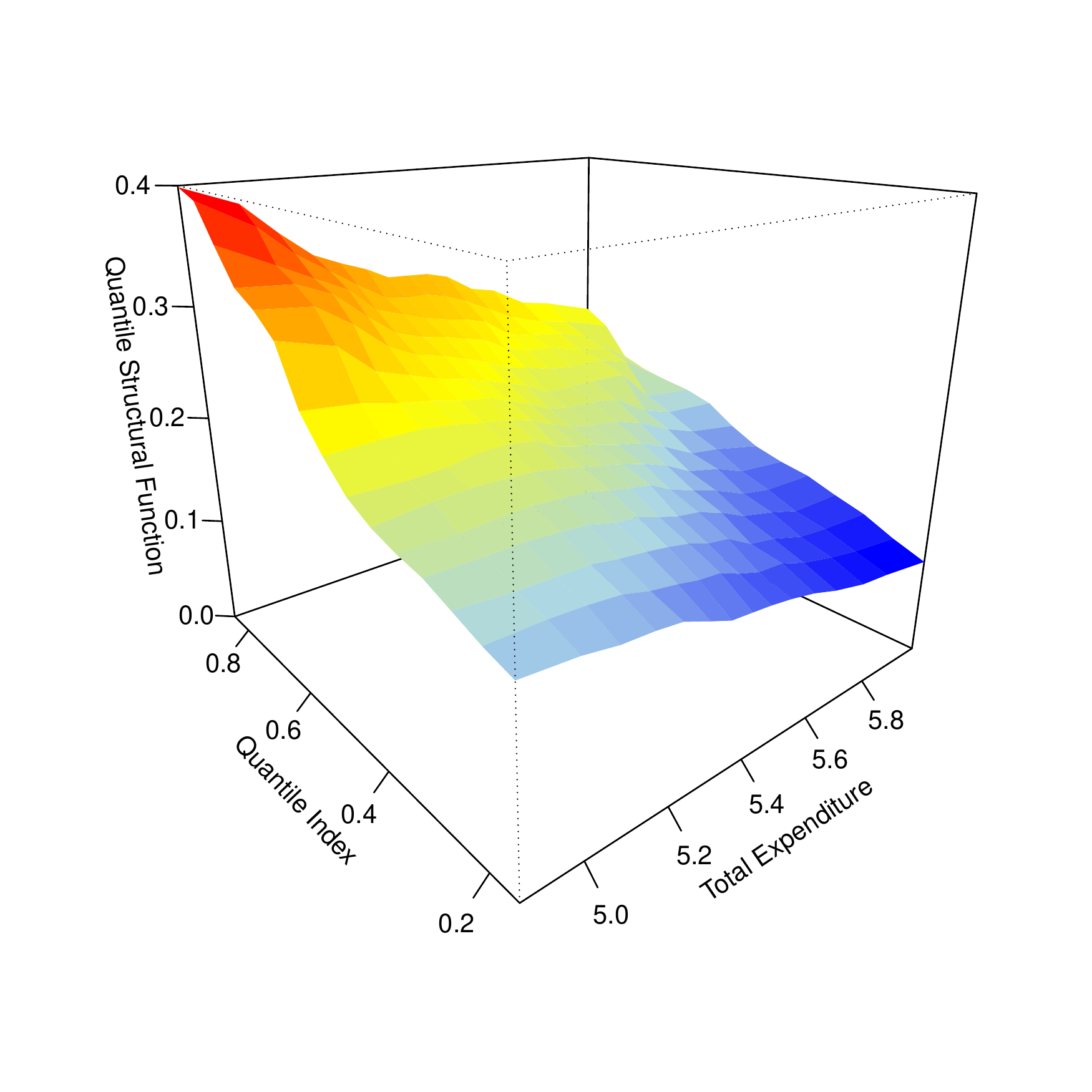}
}

\subfloat[Leisure.]{\includegraphics[width=8.5cm,height=8.5cm]{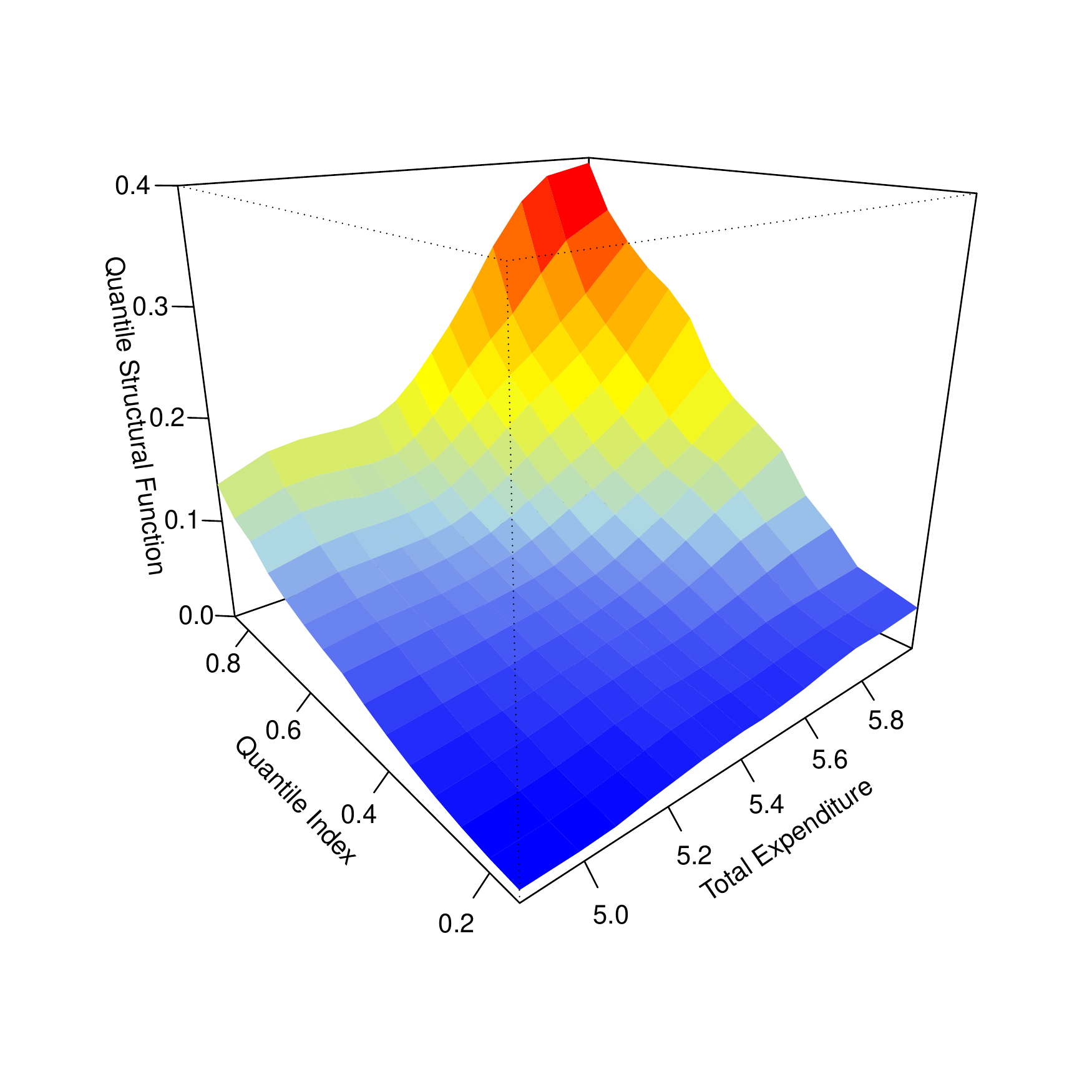} 
\includegraphics[width=8.5cm,height=8.5cm]{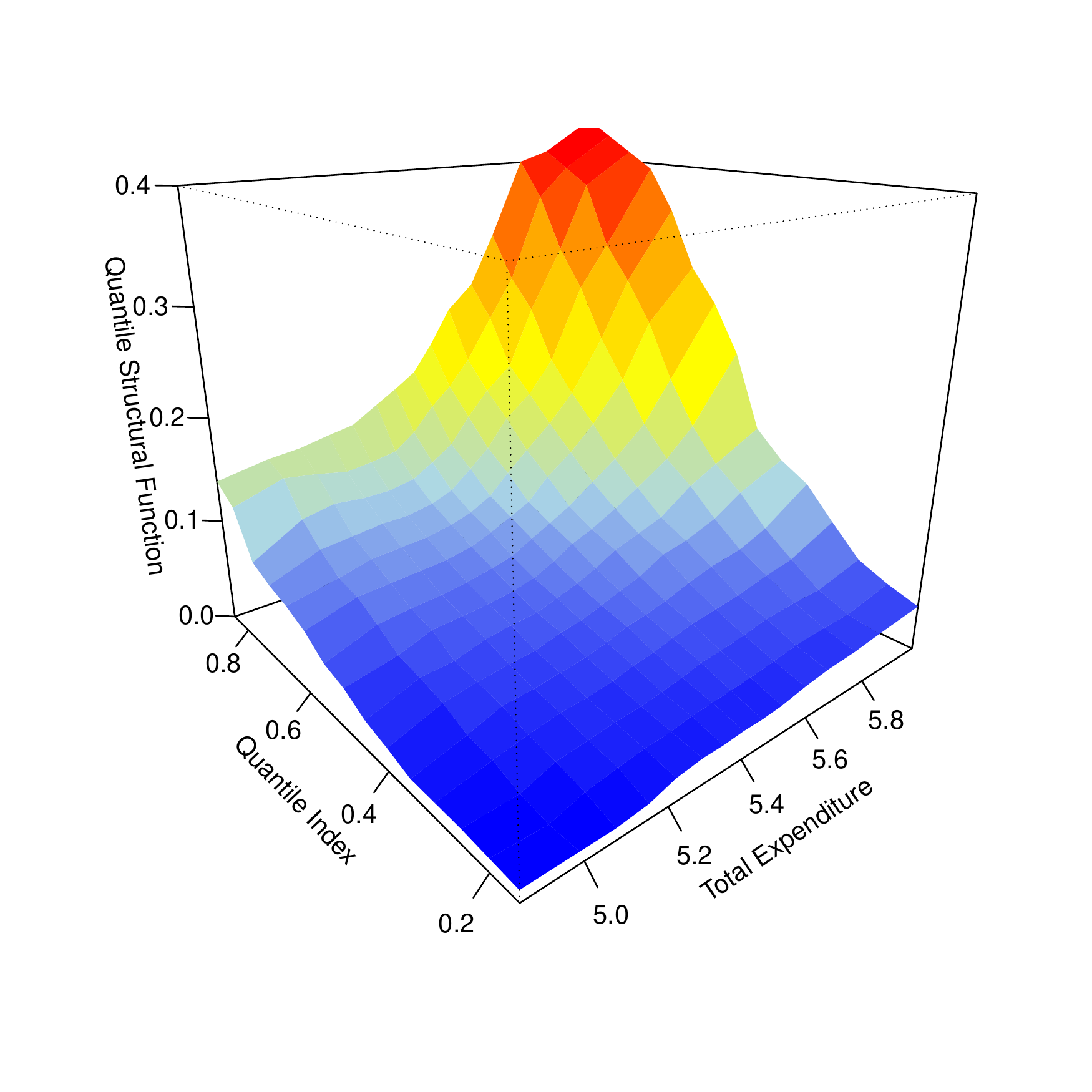}
}

\caption{Flexible QSF specification. QR (left) and DR (right) .\label{fig:FlexSpec}}
\end{figure}

Our baseline models naturally allow for the inclusion of transformations of
covariates - for instance spline transformations - in order to account for
potential nonlinearities in data. In practice, these augmented specifications 
are useful to verify the robustness of the baseline specifications empirical findings. 
In order to illustrate nonlinear
implementations of our approach and robustness of our baseline estimates, the QSF for food and
leisure obtained by taking cubic B-splines transformations with 4 knots of
log-total expenditure are shown in Figure \ref{fig:FlexSpec}, for both DR and QR methods. A complete
description of the structural stochastic relationship between total
expenditure and food and leisure shares is then obtained, and  confirms the
essentially linear form of the QSF for food, as well as the nonlinearity
already detected by DR for leisure in the empirical application - 
without the inclusion of nonlinear transformations of log-total expenditure.

Compared to existing studies of this dataset, the empirical results 
presented for the DSF are new. Our semiparametric estimates of the ASF and QSF capture
the main features displayed by the nonparametric estimates of 
Imbens and Newey (2009), or those we obtain with more flexible specifications in Figure  \ref{fig:FlexSpec}. 
Moreover, our results and methods 
further make it possible to construct uniform confidence regions for structural functions, thereby 
providing applied researchers with useful inferential tools.
These empirical results thus illustrate that our parsimonious models are able to capture complex features of the data, such as 
asymmetric distributions and nonlinear structural relationships, while leading to relatively easy-to-implement estimators and inferential methods 
that can be augmented straightforwardly for robustness checks and additional flexibility. 
This is demonstrated further in the Supplementary Material where we perform a thorough sensitivity analysis
which further shows that our empirical results are robust to the modelling,
estimation and integration choices.


\appendix

\section{Implementation Algorithms}

\label{app:impl} This section gathers the algorithms for the three-stage
estimation procedure, weighted bootstrap, and the constructions of uniform
bands for the structural functions.

\begin{algorithm}[H]
\noindent \begin{raggedright}
\textit{For $i=1,\ldots,n$, set $e_{i}=1$.} 

\par\end{raggedright}

\noindent 
\begin{raggedright}
\textbf{\textit{First Stage. }}\textit{{[}Control
function estimation{]}}
\par\end{raggedright}
\begin{enumerate}
\item
\noindent \begin{raggedright}
\textit{(QR) For $\epsilon$ in $(0,0.5)$
(e.g., $\epsilon = .01$)  and a
fine mesh of $M$ values $\{\epsilon=v_{1} <
\cdots < v_{M}=1-\epsilon\}$,
estimate $\{\widehat{\pi}^{e}(v_{m})%
\}_{m=1}^{M}$ by solving (\ref{eq:QR FS}).
Then set $\widehat{V}_{i}^{e}=%
\widehat{F}_{X}^{e}(X_{i} \mid Z_{i})$, $i=1,\ldots,n$,
 as in (\ref{%
eq:CF_QR}).}
\par\end{raggedright}
\item \noindent \begin{raggedright}
\textit{(DR) Estimate $\{\widehat{\pi}(X_{i})\}_{i=1}^{n}$
by solving (\ref{%
eq:DR FS}). Then set $\widehat{V}_{i}^{e}=\widehat{F}_{X}^{e}(X_{i} \mid
Z_{i})$,
$i=1,\ldots,n$, as in (\ref{eq:CF_DR}).}\medskip{}

\par\end{%
raggedright}
\end{enumerate}
\noindent \begin{raggedright}
\textbf{%
\textit{Second Stage. }}\textit{[Reduced-form CDF estimation]}
\par\end{raggedright}
\begin{enumerate}
\item \begin{raggedright}
\textit{(QR) (a) For $\epsilon$ in $(0,0.5)$ (e.g., $\epsilon=.01$) and
a
fine mesh of $M$ values $\{\epsilon=u_{1},\ldots,u_{M}=1-\epsilon\}$%
,
estimate $\{\widehat{\beta}^{e}(u_{m})\}_{m=1}^{M}$ by solving (\ref{%
eq:QR SS}).
(b) 
Obtain} $\widehat{F}_{Y}^{e}(y \mid x,Z_{1i},\widehat{V}_{i}^{e})$
\textit{as in (\ref{eq:FYXV_QR}) }
\par\end{raggedright}
\item \noindent 
\begin{raggedright}
\textit{(DR) (a) For each $y_{m}\in \Y_{M}$,
estimate $%
\{\widehat{\beta}(y_{m})\}_{m=1}^{M}$ by solving (\ref{eq:DR SS}).
(b)
Obtain }$%
\widehat{F}_{Y}^{e}(y \mid x,Z_{1i},\widehat{V}_{i}^{e})$
\textit{as in (%
\ref{eq:FYXV_DR}).}\medskip{}

\par\end{raggedright}
\end{enumerate}%

\noindent \raggedright{}\textbf{\textit{Third Stage.}}\textit{
{[}Structural
functions estimation{]} 
Compute $\widehat{G}^{e}(y,x)$, $%
\widehat{Q}_{S}^{e}(\tau,x)$ and $\widehat{\mu}_{S}^{e}(x)$ using (\ref{%
eq:DSF boot}), (\ref{eq:QSF est2}) and (\ref{eq:ASF est2}).
\caption{%
\textbf{\textit{Three-Stage Estimation Procedure.\label{alg:SFs_Est}}}}
}
\end{algorithm}


\begin{rem}
The size of the grids $M$ 
can differ across stages and methods. For our empirical application, we have
found that the estimates are not very sensitive to $M$.
\end{rem}

\begin{rem}
All the estimation steps can also be implemented keeping $Z_{1}$, or some
component of $Z_{1}$, fixed as a conditioning variable. The estimated
structural functions are then evaluated at values of the conditioning
variable(s) of interest. Denoting the DSF estimator and bootstrap draw by $%
\widehat{G}(y,x,z_{1})=\sum_{i=1}^{n}\widehat{F}_{Y}(y \mid x,z_{1},\widehat{%
V}_{i})T_{i}/n_{T}$ and $\widehat{G}^{e}(y,x,z_{1})=\sum_{i=1}^{n}e_{i}%
\widehat{F}_{Y}^{e}(y \mid x,z_{1},\widehat{V}_{i}^{e})T_{i}/n_{T}^{e}$, the
corresponding QSF and ASF estimators and bootstrap draws obtain upon
substituting $\widehat{G}(y,x,z_{1})$ and $\widehat{G}^{e}(y,x,z_{1})$ for $%
\widehat{G}(y,x)$ and $\widehat{G}^{e}(y,x)$ in (\ref{eq:DSF est})-(\ref%
{eq:DSF boot}).
\end{rem}

\begin{rem}
\label{remark: ASF-LS} For the QR specification, the estimator of the ASF in
the second and third stages can be replaced by $\widehat{\mu}(x) = w(x,\bar{Z%
}_{1},0)^{\prime }\widehat \beta,$ where $\bar{Z}_{1}=\sum_{i=1}^{n}Z_{1i}/n$
and $\widehat \beta$ the least squares estimator of the linear regression of 
$Y$ on $\widehat{W}_{i}^{e}$. Our numerical implementation in the
Supplementary Material  shows that estimates thus obtained are very similar
to those formed according to (\ref{eq:ASF est2}).
\end{rem}

\begin{algorithm}[H]
\medskip{}

\noindent \begin{raggedright}
\textit{For $b=1,\ldots,B$, repeat the following steps: }
\par\end{raggedright}

\noindent \begin{raggedright}
\textbf{\textit{Step 0.}}\textit{ Draw $e_{b}:=\{e_{ib}\}_{i=1}^{n}$
$i.i.d.$ from a random variable that satisfies Assumption \ref{ass:sampling} (e.g., the standard exponential distribution). }
\par\end{raggedright}

\noindent \begin{raggedright}
\textbf{\textit{Step 1. }}\textit{Reestimate the control function
$\widehat{V}_{ib}^{e}=\widehat{F}_{X,b}^{e}(X_{i} \mid Z_{i})$ in the weighted
sample, according to (\ref{eq:CF_QR})-(\ref{eq:QR FS}) or (\ref{eq:CF_DR})-(\ref{eq:DR FS}).}
\par\end{raggedright}

\noindent \begin{raggedright}
\textbf{\textit{Step 2. }}\textit{Reestimate the reduced form CDF
$\widehat{F}_{Y,b}^{e}$ in the weighted sample according to (\ref{eq:FYXV_QR})-(\ref{eq:QR SS})
or (\ref{eq:FYXV_DR})-(\ref{eq:DR SS}).}
\par\end{%
raggedright}%

\noindent \raggedright{}\textbf{\textit{Step 3. }}\textit{%
For $n_{Tb}^e =
\sum_{i=1}^n
e_{ib}T_i$, compute
$
\widehat{G}_{b}^{e}(y,x)=\sum_{i=1}^{n}e_{ib}%
\widehat{%
F}_{Y,b}^{e}(y \mid x,Z_{1i},\widehat{V}%
_{ib}^{e})T_{i}/n_{Tb}^e,
$
\ $
\widehat{Q}_{b}^{e}(\tau,x)=\delta\sum_{s=1}^{S}%
\left[%
1(y_{s}\geq0)-1\{\widehat{G}_{b}^{e}(y_{s},x)\geq\tau\}\right],
$
and
$%
\widehat{\mu}%
_{b}^{e}(x)=\delta\sum_{s=1}^{S}\left[1(y_{s}\geq0)-\widehat{%
G}%
_{b}^{e}(y_{s},x)\right],
$
\caption{\textbf{\textit{Weighted
Bootstrap.\label{%
alg:Weighted-bootstrap}}}}
}
\end{algorithm}

\begin{algorithm}[H]
\medskip{}

\noindent \begin{raggedright}
\textbf{\textit{Step 1. }}\textit{Given B bootstrap draws $\left\{ (\widehat{G}_{b}^{e}(y,x),\widehat{\mu}_{b}^{e}(x)\right\} _{b=1}^{B}$,
compute the standard errors of $\widehat{G}(y,x)$ and $\widehat{\mu}(x)$ as  
\[
\widehat{\sigma}_G(y,x)=\text{IQR}\left[\left\{ \widehat{G}_{b}^{e}(y,x)\right\} _{b=1}^{B}\right]/1.349,\quad\widehat{\sigma}_{\mu}(x)= \text{IQR}\left[\left\{ \widehat{\mu}_{b}^{e}(x)\right\} _{b=1}^{B}\right]/1.349.
\]
}\textbf{\textit{Step 2. }}\textit{For $b= 1,\ldots,B,$ compute the bootstrap draws of the
maximal $t$-statistics for the DSF and ASF as
\[
\left\Vert t_{G,b}^{e}(y,x)\right\Vert _{\mathcal{I}_G}=\sup_{(y,x)\in\mathcal{I}_G}\left|\frac{\widehat{G}_b^{e}(y,x)-\widehat{G}(y,x)}{\widehat{\sigma}_{G}(y,x)}\right|,
\quad\left\Vert t_{\mu,b}^{e}(x)\right\Vert _{\mathcal{I}_{\mu}}=\sup_{x\in\mathcal{I}_{\mu}}\left|\frac{\widehat{\mu}_b^{e}(x)-\widehat{\mu}(x)}{\widehat{\sigma}_{\mu}(x)}\right|.
\]}
\par\end{raggedright}

\noindent \begin{raggedright}
\textbf{\textit{Step 3. }}\textit{Form $(1-\alpha)$-confidence
bands for the DSF and ASF as
\[
\left\{ \widehat{G}(y,x)\pm \widehat k_{G}(1-\alpha)\widehat{\sigma}_G(y,x):(y,x)\in\mathcal{I}_G\right\} ,\quad 
\left\{ \widehat{\mu}(x)\pm \widehat k_{\mu}(1-\alpha)\widehat{\sigma}_{\mu}(x):x\in\mathcal{I}_{\mu}\right\},
\]
where  $\widehat k_{G}(1-\alpha)$ is the sample $(1-\alpha)$-quantile of
$\left\{ \left\Vert t_{G,b}^{e}(y,x)\right\Vert _{\mathcal{I}_G}:1\leq b\leq B\right\} $, and $\widehat k_{\mu}(1-\alpha)$ is the sample $(1-\alpha)$-quantile of
$\left\{ \left\Vert t_{\mu,b}^{e}(x)\right\Vert _{\mathcal{I}_{\mu}}:1\leq b\leq B\right\} $.}
\par\end{raggedright}

\textit{\caption{\textbf{\textit{Uniform Inference for DSF and ASF.\label{alg:Unif Inference}}}}
}
\end{algorithm}

\begin{algorithm}[H]
\medskip{}

\noindent \begin{raggedright}
\textbf{\textit{Step 1. }}\textit{Given B bootstrap draws $\left\{ (\widehat{G}_{b}^{e}(y,x),\widehat{Q}_{b}^{e}(\tau,x))\right\} _{b=1}^{B}$,
compute the standard errors of  $\widehat{G}(y,x)$ and $\widehat{Q}(\tau,x)$ as  
\[
\widehat{\sigma}_G(y,x)=\text{IQR}\left[\left\{ \widehat{G}_{b}^{e}(y,x)\right\} _{b=1}^{B}\right]/1.349,\quad\widehat{\sigma}_Q(\tau,x)=\text{IQR}\left[\left\{ \widehat{Q}_{b}^{e}(\tau,x)\right\} _{b=1}^{B}\right]/1.349.
\]
}
\textbf{\textit{Step 2. }}\textit{For $b= 1,\ldots,B,$ compute the bootstrap draws of the
maximal $t$-statistics for the DSF and ASF as
\[
\left\Vert t_{G,b}^{e}(\tau,x)\right\Vert _{\mathcal{I}_G}=\sup_{(y,x)\in\mathcal{I}_G}\left|\frac{\widehat{G}_b^{e}(y,x)-\widehat{G}(y,x)}{\widehat{\sigma}_{G}(y,x)}\right|,\quad \left\Vert t_{Q,b}^{e}(\tau,x)\right\Vert _{\mathcal{I}_Q}=\sup_{(\tau,x)\in\mathcal{I}_Q}\left|\frac{\widehat{Q}_b^{e}(\tau,x)-\widehat{Q}(\tau,x)}{\widehat{\sigma}_{Q}(\tau,x)}\right|.
\]}
\par\end{raggedright}

\noindent \begin{raggedright}
\textbf{\textit{Step 3. }}\textit{If $Y$ is continuous, form a $(1-\alpha)$-confidence
band for the QSF as 
\[
\left\{ \widehat{Q}(\tau,x)\pm \widehat k_{Q}(1-\alpha)\widehat{\sigma}_Q(\tau,x):(\tau,x)\in\mathcal{I}_Q\right\},
\]
where  $\widehat k_{Q}(1-\alpha)$ is the sample $(1-\alpha)$-quantile of
$\left\{ \left\Vert t_{Q,b}^{e}(\tau,x)\right\Vert _{\mathcal{I}_Q}:1\leq b\leq B\right\} $. Otherwise, form a $(1-\alpha)$-confidence
band for the QSF as
\[
\ensuremath{\left\{ \left[\widehat{G}_{U}^{\leftarrow}(\tau,x),\widehat{G}_{L}^{\leftarrow}(\tau,x)\right]:(\tau,x)\in\mathcal{I}_G^{\leftarrow} \right\} ,}
\]
where $\mathcal{I}_G^{\leftarrow} = \{(\tau,x) : \widehat{G}_{L}(y,x) = \tau, (y,x) \in \mathcal{I}_G \} \cap \{(\tau,x) : \widehat{G}_{U}(y,x) = \tau, (y,x) \in \mathcal{I}_G \}$, 
$$
\widehat{G}_{L}(y,x) = \widehat{G}(y,x) - \widehat k_{G}(1-\alpha)\widehat{\sigma}_G(y,x), \quad \widehat{G}_{U}(y,x) = \widehat{G}(y,x) + \widehat k_{G}(1-\alpha)\widehat{\sigma}_G(y,x),
$$
and $\widehat k_{G}(1-\alpha)$ is the sample $(1-\alpha)$-quantile of
$\left\{ \left\Vert t_{G,b}^{e}(y,x)\right\Vert _{\mathcal{I}_G}:1\leq b\leq B\right\} $.
}
\par\end{raggedright}

\textit{\caption{\textbf{\textit{Uniform Inference for QSF.\label{alg:Unif Inference}}}}
}
\end{algorithm}

\section{Identification}

\subsection{Proof of Lemma \protect\ref{Lemma:Nonsingular}}

\begin{sloppy}
By Assumption \ref{ass:Identification} $E_{\mu}[p(X)p(X)^{%
\prime}]$, $E_{\varsigma_{l}}[r_{1l}(Z_{1l})r_{1l}(Z_{1l})^{%
\prime}]$, $l=1,\dots,d_{z_{1}}$, and $E_{\rho}[q(V)q(V)^{\prime}]$ are positive definite. Also, with $W=w(X,Z_{1},V)$, 
there is a positive constant $C$ such that 
\begin{align*}
E[w(X,Z_{1},V) w(X,Z_{1},V)^{\prime}] & \geq C\int w(x,z_{1},v) w(x,z_{1},v)^{\prime}[\mu(dx) \times \varsigma(dz_{1}) \times \rho(dv)] \\
&
=C\int\{p(x)p(x)^{\prime}\} \otimes \{r_{11}(z_{11})r_{11}(z_{11})^{\prime}\} \otimes \cdots \\ 
& \quad \otimes \{r_{1d_{z_{1}}}(z_{1d_{z_{1}}})r_{1d_{z_{1}}}(z_{1d_{z_{1}}})^{\prime}\} \otimes \{q(v)q(v)^{\prime}\}
[\mu(dx) \times \varsigma(dz_{1})\times\rho(dv)] \\
& =CE_{\mu}[p(X)p(X)^{\prime}]\otimes
E_{\varsigma_{1}}[r_{11}(Z_{11})r_{11}(Z_{11})^{\prime}]\otimes \cdots \\
& \quad \otimes E_{\varsigma_{d_{z_{1}}}}[r_{1d_{z_{1}}}(Z_{1d_{z_{1}}})r_{1d_{z_{1}}}(Z_{1d_{z_{1}}})^{\prime}]\otimes
E_{\rho}[q(V)q(V)^{\prime}].
\end{align*}
where the inequality means no less than in the usual partial ordering for
positive semi-definite matrices. The conclusion then follows by the
matrices following the last equality being positive definite. \qed
\par\end{sloppy}

\subsection{Proof of Theorem \protect\ref{thm:IdentificationSF}}

\begin{sloppy}Under Assumption \ref{ass:Identification}, Lemma \ref{Lemma:Nonsingular}
implies that the QR coefficients $\beta(U)$ and DR coefficients $\beta(Y)$
are unique. For the QR specification, suppose there exists $\tilde{\beta}(U)$
such that $\beta(U)^{\prime} w(X,Z_{1},V) =\tilde{\beta}(U)^{\prime} w(X,Z_{1},V) $.
Then $\{\beta(U)-\tilde{\beta}(U)\}^{\prime}  w(X,Z_{1},V) =0$,
and after applying iterated expectations, independence of $U$ and
$(X,Z_{1},V)$ implies 
\begin{eqnarray*}
0 & = & E[(\beta(U)-\tilde{\beta}(U))'  \left\{w(X,Z_{1},V)  w(X,Z_{1},V) '\right\}(\beta(U)-\tilde{\beta}(U))]\\
 & = & E[(\beta(U)-\tilde{\beta}(U))'E[  w(X,Z_{1},V)  w(X,Z_{1},V)'\mid U](\beta(U)-\tilde{\beta}(U))]\\
 & \geq & CE[||\beta(U)-\tilde{\beta}(U)||^{2}]
\end{eqnarray*}
for some positive constant $C$, by positive definiteness of $E[  w(X,Z_{1},V)  w(X,Z_{1},V)']$.
Therefore, the map $u\mapsto Q_{Y}(u\mid x,v)$ is well-defined for
all $(x,z_{1},v)\in\mathcal{X}\mathcal{Z}_{1}\mathcal{V}$ under Assumption
\ref{ass:Model assumptions}(a). Strict monotonicity of $u\mapsto Q_{Y}(u\mid x,z_{1},v)$
for all $(x,z_{1},v)\in\mathcal{X}\mathcal{Z}_{1}\mathcal{V}$ then
implies that the inverse map $y\mapsto F_{Y}(y\mid x,z_{1},v)=Q_{Y}^{-1}(y\mid x,z_{1},v)$
is well-defined for all $(x,z_{1},v)\in\mathcal{X}\mathcal{Z}_{1}\mathcal{V}$.
For the DR specification, positive definiteness of $E[ w(X,Z_{1},V)  w(X,Z_{1},V)']$
is also sufficient for uniqueness of DR coefficients by standard identification
results for Logit and Probit models, e.g.,  see Example 1.2 in Newey and
McFadden (1994). Therefore, the map $y\mapsto F_{Y}(y\mid x,z_{1},v)$
is well-defined for all $(x,z_{1},v)\in\mathcal{X}\mathcal{Z}_{1}\mathcal{V}$
under Assumption \ref{ass:Model assumptions}(b). For both specifications
the result now follows from the definitions of structural functions
in 
Section \ref{sec:Model}.\qed \par\end{sloppy}

\section{Asymptotic Theory}

\subsection{Notation}

\label{app:notation} In what follows $\vartheta$ denotes a generic value for
the control function. It is convenient also to introduce some additional
notation, which will be extensively used in the proofs. Let $%
V_{i}(\vartheta):=\vartheta(X_{i},Z_{i})$, $W_{i}(%
\vartheta):=w(X_{i},Z_{1i},V_{i}(\vartheta))$, and $\dot{W}%
_{i}(\vartheta):=\partial_{v}w(X_{i},Z_{1i},v)|_{v=V_{i}(\vartheta)}$. When
the previous functions are evaluated at the true values we use $%
V_{i}=V_{i}(\vartheta_{0}),$ $W_{i}=W_{i}(\vartheta_{0})$, and $\dot{W}_{i}=%
\dot{W}_{i}(\vartheta_{0})$. Also, let $\rho_{y}(u,v):=-1(u\leq
y)\log\Lambda(v)-1(u>y)\log\Lambda(-v)$. Recall that $A:=(Y,X,Z,W,V)$, $%
T(x)=1(x\in\overline{\mathcal{X}}),$ and $T=T(X)$. For a function $f:%
\mathcal{A}\mapsto\mathbb{R}$, we use $\|f\|_{T,\infty}=\sup_{a\in\mathcal{A}%
}|T(x)f(a)|$; for a $K$-vector of functions $f:\mathcal{A}\mapsto\mathbb{R}%
^{K}$, we use $\|f\|_{T,\infty}=\sup_{a\in\mathcal{A}}\|T(x)f(a)\|_{2}$. We
make functions in $\Upsilon$ as well as estimators $\widehat{\vartheta}$ to
take values in $[0,1]$, the support of the control function $V$. This allows
us to simplify notation in what follows.

We adopt the standard notation in the empirical process literature (see,
e.g., van der Vaart, 1998), 
\begin{equation*}
{\mathbb{E}_n}[f]={\mathbb{E}_n}[f(A)]=n^{-1}\sum_{i=1}^{n}f(A_{i}),
\end{equation*}
and 
\begin{equation*}
\mathbb{G}_n[f]=\mathbb{G}_n[f(A)]=n^{-1/2}\sum_{i=1}^{n}(f(A_{i})-{\mathrm{E%
}}_{P}[f(A)]).
\end{equation*}
When the function $\widehat{f}$ is estimated, the notation should
interpreted as: 
\begin{equation*}
\mathbb{G}_n[\widehat{f}\ ]=\mathbb{G}_n[f]\mid_{f=\widehat{f}}\text{\ and \ 
}{\mathrm{E}}_{P}[\widehat{f}\ ]={\mathrm{E}}_{P}[f]\mid_{f=\widehat{f}}.
\end{equation*}
We also use the concepts of covering entropy and bracketing entropy in the
proofs. The covering entropy $\log N(\epsilon,\mathcal{F},\|\cdot\|)$ is the
logarithm of the minimal number of $\|\cdot\|$-balls of radius $\epsilon$
needed to cover the set of functions $\mathcal{F}$. The bracketing entropy $%
\log N_{[]}(\epsilon,\mathcal{F},\|\cdot\|)$ is the logarithm of the minimal
number of $\epsilon$-brackets in $\|\cdot\|$ needed to cover the set of
functions $\mathcal{F}$. An $\epsilon$-bracket $[\ell,u]$ in $\|\cdot\|$ is
the set of functions $f$ with $\ell\leq f\leq u$ and $\|u-\ell\|<\epsilon$.

For a sequence of random functions $y\mapsto f_{n}(y)$ and a deterministic
sequence $a_{n}$, we use $f_{n}(y)=\bar{o}_{\Pr }(a_{n})$ and $f_{n}(y)=\bar{%
O}_{\Pr }(a_{n})$ to denote uniform in $y\in \mathcal{Y}$ orders in
probability, i.e. $\sup_{y\in \mathcal{Y}}f_{n}(y)=o_{\Pr }(a_{n})$ and $%
\sup_{y\in \mathcal{Y}}f_{n}(y)=O_{\Pr }(a_{n})$, respectively. The uniform
in $y\in \mathcal{Y}$ deterministic orders $\bar{o}(a_{n})$ and $\bar{O}%
(a_{n})$ are defined analogously suppressing the $\Pr $ subscripts.

We follow the notation and definitions in van der Vaart and Wellner (1996)
of bootstrap consistency. Let $D_{n}$ denote the data vector and $E_{n}$ be
the vector of bootstrap weights. Consider the random element $%
Z_{n}^{e}=Z_{n}(D_{n},E_{n})$ in a normed space $\mathbb{Z}$. We say that
the bootstrap law of $Z_{n}^{e}$ consistently estimates the law of some
tight random element $Z$ and write $Z_{n}^{e}\rightsquigarrow_{\Pr}Z$ in $%
\mathbb{Z}$ if 
\begin{equation}
\begin{array}{r}
\sup_{h\in\text{BL}_{1}(\mathbb{Z})}\left|{\mathrm{E}}_{P}^{e}h%
\left(Z_{n}^{e}\right)-{\mathrm{E}}_{P}h(Z)\right|\rightarrow_{\Pr^{*}}0,%
\end{array}
\label{boot1}
\end{equation}
where $\text{BL}_{1}(\mathbb{Z})$ denotes the space of functions with
Lipschitz norm at most 1, ${\mathrm{E}}_{P}^{e}$ denotes the conditional
expectation with respect to $E_{n}$ given the data $D_{n}$, and $%
\rightarrow_{\Pr^{*}}$ denotes convergence in (outer) probability.

\subsection{Proof of Lemma \protect\ref{thm:fclt}}

We only consider the case where $\mathcal{Y}$ is a compact interval of $%
\mathbb{R}$. The case where $\mathcal{Y}$ is finite is simpler and follows
similarly.

\subsubsection{Auxiliary Lemmas}

We start with 2 results on stochastic equicontinuity and a local expansion
for the second stage estimators that will be used in the proof of Lemma \ref%
{thm:fclt}.

\begin{lemma}
{[}Stochastic equicontinuity{]}\label{lemma SE} Let $e\geq0$ be a positive
random variable with ${\mathrm{E}}_{P}[e]=1$, $\mathrm{Var}_{P}[e]=1,$ and ${%
\mathrm{E}}_{P}|e|^{2+\delta}<\infty$ for some $\delta>0$, that is
independent of $(Y,X,Z,W,V)$, including as a special case $e=1$, and set,
for $A=(e,Y,X,Z,W,V)$, 
\begin{equation*}
f_{y}(A,\vartheta,\beta):=e\cdot[\Lambda(W(\vartheta)^{\prime
}\beta)-1(Y\leq y)]\cdot W(\vartheta)\cdot T.
\end{equation*}
Under Assumptions \ref{ass:sampling}--\ref{ass:second} the following
relations are true.

\begin{itemize}
\item[(a)] Consider the set of functions 
\begin{equation*}
\mathcal{F}=\{f_{y}(A,\vartheta,\beta)^{\prime
}\alpha:(\vartheta,\beta,y)\in\Upsilon_{0}\times\mathcal{B}\times\mathcal{Y}%
,\alpha\in\mathbb{R}^{\dim(W)},\|\alpha\|_{2}\leq1\},
\end{equation*}
where $\mathcal{Y}$ is a compact subset of $\mathbb{R}$, $\mathcal{B}$ is a
compact set under the $\|\cdot\|_{2}$ metric containing $\beta_{0}(y)$ for
all $y\in\mathcal{Y}$, $\Upsilon_{0}$ is the intersection of $\Upsilon$,
defined in Lemma \ref{lemma:first}, with a neighborhood of $\vartheta_{0}$
under the $\|\cdot\|_{T,\infty}$ metric. This class is $P$-Donsker with a
square integrable envelope of the form $e$ times a constant.

\item[(b)] Moreover, if $(\vartheta,\beta(y))\to(\vartheta_{0},\beta_{0}(y))$
in the $\|\cdot\|_{T,\infty}\vee\|\cdot\|_{2}$ metric uniformly in $y\in%
\mathcal{Y}$, then 
\begin{equation*}
\sup_{y\in\mathcal{Y}}\|f_{y}(A,\vartheta,\beta(y))-f_{y}(A,\vartheta_{0},%
\beta_{0}(y))\|_{P,2}\to0.
\end{equation*}

\item[(c)] Hence for any $(\widetilde{\vartheta},\widetilde{\beta}%
(y))\to_{\Pr}(\vartheta_{0},\beta_{0}(y))$ in the $\|\cdot\|_{T,\infty}\vee%
\|\cdot\|_{2}$ metric uniformly in $y\in\mathcal{Y}$ such that $\widetilde{%
\vartheta}\in\Upsilon_{0}$, 
\begin{equation*}
\sup_{y\in\mathcal{Y}}\|\mathbb{G}_n f_{y}(A,\widetilde{\vartheta},%
\widetilde{\beta}(y))-\mathbb{G}_n
f_{y}(A,\vartheta_{0},\beta_{0}(y))\|_{2}\to_{\Pr}0.
\end{equation*}

\item[(d)] For any $(\widehat{\vartheta},\widetilde{\beta}%
(y))\to_{\Pr}(\vartheta_{0},\beta_{0}(y))$ in the $\|\cdot\|_{T,\infty}\vee%
\|\cdot\|_{2}$ metric uniformly in $y\in\mathcal{Y}$, so that 
\begin{equation*}
\|\widehat{\vartheta}-\widetilde{\vartheta}\|_{T,\infty}=o_{\Pr}(1/\sqrt{n}),%
\text{ where }\widetilde{\vartheta}\in\Upsilon_{0},
\end{equation*}
we have that 
\begin{equation*}
\sup_{y\in\mathcal{Y}}\|\mathbb{G}_n f_{y}(A,\widehat{\vartheta},\widetilde{%
\beta}(y))-\mathbb{G}_n f_{y}(A,\vartheta_{0},\beta_{0}(y))\|_{2}\to_{\Pr}0.
\end{equation*}
\end{itemize}
\end{lemma}

\textbf{Proof of Lemma \ref{lemma SE}.} The proof is divided in subproofs of
each of the claims.

Proof of Claim (a). The proof proceeds in several steps.

Step 1. Here we bound the bracketing entropy for 
\begin{equation*}
\mathcal{I}_{1}=\{[\Lambda(W(\vartheta)^{\prime }\beta)-1(Y\leq y)]T:\beta\in%
\mathcal{B},\vartheta\in\Upsilon_{0},y\in\mathcal{Y}\}.
\end{equation*}
For this purpose consider a mesh $\{\vartheta_{k}\}$ over $\Upsilon_{0}$ of $%
\|\cdot\|_{T,\infty}$ width $\delta$, a mesh $\{\beta_{l}\}$ over $\mathcal{B%
}$ of $\|\cdot\|_{2}$ width $\delta$, and a mesh $\{y_{j}\}$ over $\mathcal{Y%
}$ of $\|\cdot\|_{2}$ width $\delta$. A generic bracket over $\mathcal{I}%
_{1} $ takes the form 
\begin{equation*}
[i_{1}^{0},i_{1}^{1}]=[\{\Lambda(W(\vartheta_{k})^{\prime
}\beta_{l}-\kappa\delta)-1(Y\leq
y_j-\delta)\}T,\{\Lambda(W(\vartheta_{k})^{\prime
}\beta_{l}+\kappa\delta)-1(Y\leq y_j+\delta)\}T],
\end{equation*}
where $\kappa=L_{W}\max_{\beta\in\mathcal{B}}\|\beta\|_{2}+L_{W}$, and $%
L_{W}:=\|\partial_{v}w\|_{T,\infty}\vee\|w\|_{T,\infty}.$

Note that this is a valid bracket for all elements of $\mathcal{I}_{1}$
because for any $\vartheta$ located within $\delta$ from $\vartheta_{k}$ and
any $\beta$ located within $\delta$ from $\beta_{l}$, 
\begin{eqnarray}
|W(\vartheta)^{\prime }\beta-W(\vartheta_{k})^{\prime }\beta_{l}|T & \leq &
|(W(\vartheta)-W(\vartheta_{k}))^{\prime }\beta|T+|W(\vartheta_{k})^{\prime
}(\beta-\beta_{l})|T  \notag \\
& \leq & L_{W}\delta\max_{\beta\in\mathcal{B}}\|\beta\|_{2}+L_{W}\delta\leq%
\kappa\delta,  \label{eq: converge 1}
\end{eqnarray}
and the $\|\cdot\|_{P,2}$-size of this bracket is given by 
\begin{eqnarray*}
\|i_{1}^{0}-i_{1}^{1}\|_{P,2} & \leq & \sqrt{{\mathrm{E}}_{P}[P\{Y\in[%
y\pm\delta]\mid X,Z\}T]} \\
& + & \sqrt{{\mathrm{E}}_{P}[\{\Lambda(W(\vartheta_{k})^{\prime
}\beta_{l}+\kappa\delta)-\Lambda(W(\vartheta_{k})^{\prime
}\beta_{l}-\kappa\delta)\}^{2}T]} \\
& \leq &
\sqrt{ \| f_{Y}(\cdot \mid \cdot)\|_{T,\infty} 2 \delta} + \kappa \delta/2,
\end{eqnarray*}
because $\|\lambda(\cdot)\|_{T,\infty}\leq1/4$, where $\lambda=\Lambda(1-%
\Lambda)$ is the derivative of $\Lambda$.

Hence, counting the number of brackets induced by the mesh created above, we
arrive at the following relationship between the bracketing entropy of $%
\mathcal{I}_{1}$ and the covering entropies of $\Upsilon_{0}$, $\mathcal{B}$%
, and $\mathcal{Y}$, 
\begin{multline*}
\log N_{[]}(\epsilon,\mathcal{I}_{1},\|\cdot\|_{P,2})\lesssim\log
N(\epsilon^{2},\Upsilon_{0},\|\cdot\|_{T,\infty})+\log N(\epsilon^{2},%
\mathcal{B},\|\cdot\|_{2})+\log N(\epsilon^{2},\mathcal{Y},\|\cdot\|_{2}) \\
\lesssim1/(\epsilon^{2}\log^{4}\epsilon)+\log(1/\epsilon)+\log(1/\epsilon),
\end{multline*}
and so $\mathcal{I}_{1}$ is $P$-Donsker with a constant envelope.

Step 2. Similarly to Step 1, it follows that 
\begin{equation*}
\mathcal{I}_{2}=\{W(\vartheta)^{\prime }\alpha
T:\vartheta\in\Upsilon_{0},\alpha\in\mathbb{R}^{\dim(W)},\|\alpha\|_{2}\leq1%
\}
\end{equation*}
also obeys a similar bracketing entropy bound 
\begin{equation*}
\log N_{[]}(\epsilon,\mathcal{I}_{2},\|\cdot\|_{P,2})\lesssim1/(\epsilon^{2}%
\log^{4}\epsilon)+\log(1/\epsilon)
\end{equation*}
with a generic bracket taking the form $[i_{2}^{0},i_{2}^{1}]=[\{W(%
\vartheta_{k})^{\prime }\beta_{l}-\kappa\delta\}T,\{W(\vartheta_{k})^{\prime
}\beta_{l}+\kappa\delta\}T]$. Hence, this class is also $P$-Donsker with a
constant envelope.

Step 3. In this step we verify the claim (a). Note that $\mathcal{F}=e\cdot%
\mathcal{I}_{1}\cdot\mathcal{I}_{2}.$ This class has a square-integrable
envelope under P. The class $\mathcal{F}$ is $P$-Donsker by the following
argument. Note that the product $\mathcal{I}_{1}\cdot\mathcal{I}_{2}$ of
uniformly bounded classes is $P$-Donsker, e.g., by Theorem 2.10.6 of van der
Vaart and Wellner (1996). Under the stated assumption the final product of
the random variable $e$ with the $P$-Donsker class remains to be $P$-Donsker
by the Multiplier Donsker Theorem, namely Theorem 2.9.2 in van der Vaart and
Wellner (1996).

Proof of Claim (b). The claim follows by the Dominated Convergence Theorem,
since any $f\in\mathcal{F}$ is dominated by a square-integrable envelope
under $P$, and, uniformly in $y\in\mathcal{Y}$, $\Lambda[W(\vartheta)^{%
\prime }\beta(y)]T\to\Lambda[W^{\prime }\beta_{0}(y)]T$ and $%
|W(\vartheta)^{\prime }\beta(y)T-W^{\prime }\beta_{0}(y)T|\to0$ in view of
the relation such as (\ref{eq: converge 1}).

Proof of Claim (c). This claim follows from the asymptotic equicontinuity of
the empirical process $(\mathbb{G}_n[f_{y}],f_{y}\in\mathcal{F})$ under the $%
L_{2}(P)$ metric, and hence also with respect to the $\|\cdot\|_{T,\infty}%
\vee\|\cdot\|_{2}$ metric uniformly in $y\in\mathcal{Y}$ in view of Claim
(b).

Proof of Claim (d). It is convenient to set $\widehat{f}_{y}:=f_{y}(A,%
\widehat{\vartheta},\widetilde{\beta}(y))$ and $\widetilde{f}_{y}:=f_{y}(A,%
\widetilde{\vartheta},\widetilde{\beta}(y)).$ Note that 
\begin{eqnarray*}
\max_{1\leq j \leq \dim W} |\mathbb{G}_n[\widehat{f}_{y}-\widetilde{f}_{y}]|_j & \leq & \max_{1\leq j \leq \dim W}|\sqrt{n}{\mathbb{%
E}_n}[\widehat{f}_{y}-\widetilde{f}_{y}]|_j+ \max_{1\leq j \leq \dim W}|\sqrt{n}{\mathrm{E}}_{P}(\widehat{%
f}_{y}-\widetilde{f}_{y})|_j \\
& \lesssim & \sqrt{n}{\mathbb{E}_n}[\widehat{\zeta}\ ]+\sqrt{n}{\mathrm{E}}%
_{P}[\widehat{\zeta}\ ] 
 \lesssim  \mathbb{G}_n[\widehat{\zeta}\ ]+2\sqrt{n}{\mathrm{E}}_{P}[%
\widehat{\zeta}\ ],
\end{eqnarray*}
where $|f_{y}|_j$ denotes the $j$th element of an application of absolute value to each element of
the vector $f_{y}$, and $\widehat{\zeta}$ is defined by the following
relationship, which holds with probability approaching one uniformly in $y\in%
\mathcal{Y}$, 
\begin{multline*}
\max_{1\leq j \leq \dim W}|\widehat{f}_{y}-\widetilde{f}_{y}|_j\lesssim|e|\cdot\{\|W(\widehat{\vartheta}%
)-W(\widetilde{\vartheta})\|_{2}+|\Lambda[W(\widehat{\vartheta})^{\prime }%
\widetilde{\beta}(y)]-\Lambda[W(\widetilde{\vartheta})^{\prime }\widetilde{%
\beta}(y)]|\}\cdot T \\ \lesssim\widehat{\zeta}:=e\cdot\kappa\Delta_{n}, 
\end{multline*}
where  $\kappa=L_{W}\max_{\beta\in\mathcal{B}}\|\beta\|_{2}+L_{W}$, $%
L_{W}=\|\partial_{v}w\|_{T,\infty}\vee\|w\|_{T,\infty}$, and $\Delta_{n}=o(1/%
\sqrt{n})$ is a deterministic sequence such that 
\begin{equation*}
\Delta_{n}\geq\|\widehat{\vartheta}-\widetilde{\vartheta}\|_{T,\infty}. 
\end{equation*}

By part (c) the result follows from 
\begin{eqnarray*}
\mathbb{G}_n[\widehat{\zeta}\ ]=\bar{o}_{\Pr}(1),\ \ \ \ \sqrt{n}{\mathrm{E}}%
_{P}[\widehat{\zeta}\ ]=\bar{o}_{\Pr}(1).
\end{eqnarray*}
Indeed, 
\begin{equation*}
\|e\cdot\kappa\Delta_{n}\|_{P,2}=\bar{o}(1)\Rightarrow\mathbb{G}_n[\widehat{%
\zeta}\ ]=\bar{o}_{\Pr}(1), 
\end{equation*}
and 
\begin{equation*}
\|e\cdot\kappa\Delta_{n}\|_{P,1}\leq{\mathrm{E}}_{P}|e|\cdot\kappa\Delta_{n}=%
\bar{o}(1/\sqrt{n})\Rightarrow{\mathrm{E}}_{P}|\widehat{\zeta}|=\bar{o}%
_{\Pr}(1/\sqrt{n}), 
\end{equation*}
since $\Delta_{n}=o(1/\sqrt{n})$.

\begin{lemma}
{[}Local expansion{]} \label{lemma Expand} Under Assumptions \ref%
{ass:sampling}--\ref{ass:second}, for 
\begin{eqnarray*}
& & \widehat{\delta}(y)=\sqrt{n}(\widetilde{\beta}(y)-\beta_{0}(y))=\bar{O}%
_{\Pr}(1); \\
& & \widehat{\Delta}(x,r)=\sqrt{n}(\widehat{\vartheta}(x,r)-%
\vartheta_{0}(x,r))=\sqrt{n}\ {\mathbb{E}_n}[\ell(A,x,r)]+o_{\Pr}(1)\text{
in }\ell^{\infty}(\overline{\mathcal{XR}}), \\
& & \|\sqrt{n}\ {\mathbb{E}_n}[\ell(A,\cdot)]\|_{T,\infty}=O_{\Pr}(1),
\end{eqnarray*}
we have that 
\begin{eqnarray*}
& & \sqrt{n}\ {\mathrm{E}}_{P}[\{\Lambda[W(\widehat{\vartheta})^{\prime }%
\widetilde{\beta}(y)]-1(Y\leq y)\}W(\widehat{\vartheta})T]=J(y)\widehat{%
\delta}(y)+\sqrt{n}\ {\mathbb{E}_n}\left[g_{y}(A)\right]+\bar{o}_{\Pr}(1),
\end{eqnarray*}
where 
\begin{equation*}
g_{y}(a)={\mathrm{E}}_{P}\{[\Lambda(W^{\prime }\beta_{0}(y))-1(Y\leq y)]\dot{%
W}+\lambda(W^{\prime }\beta_{0}(y))W\dot{W}^{\prime
}\beta_{0}(y)\}T\ell(a,X,R).
\end{equation*}
\end{lemma}

\textbf{Proof of Lemma \ref{lemma Expand}.}

Uniformly in $\xi:=(X,Z)\in\overline{\mathcal{X}\mathcal{Z}}$ and $y\in%
\mathcal{Y}$, 
\begin{eqnarray*}
& & \sqrt{n}{\mathrm{E}}_{P}\{\Lambda[W(\widehat{\vartheta})^{\prime }%
\widetilde{\beta}(y)]-1(Y\leq y)\mid X,Z\}T \\
& & =\sqrt{n}{\mathrm{E}}_{P}\{\Lambda[W^{\prime }\beta_{0}(y)]-1(Y\leq
y)\mid X,Z\}T \\
& & +\lambda[W(\bar{\vartheta}_{\xi})^{\prime }\bar{\beta}_{\xi}(y)]\{W(\bar{%
\vartheta}_{\xi})^{\prime }\widehat{\delta}(y)+\dot{W}(\bar{\vartheta}%
_{\xi})^{\prime }\bar{\beta}_{\xi}\widehat{\Delta}(X,R)\}T \\
& & =\sqrt{n}{\mathrm{E}}_{P}\{\Lambda[W^{\prime }\beta_{0}(y)]-1(Y\leq
y)\mid X,Z\}T \\
& & +\lambda[W^{\prime }\beta_{0}(y)]\{W^{\prime }\widehat{\delta}(y)+\dot{W}%
^{\prime }\beta_{0}(y)\widehat{\Delta}(X,R)\}T+R_{\xi}(y), 
\end{eqnarray*}
and
$$
\bar R(y) = \sup_{\{\xi \in \overline{\X\Z}\}} |R_{\xi}(y)| =
\bar o_{\Pr}(1)
$$
where $\bar{\vartheta}_{\xi}$ is on the line connecting $\vartheta_{0}$ and $%
\widehat{\vartheta}$ and $\bar{\beta}_{\xi}(y)$ is on the line connecting $%
\beta_{0}(y)$ and $\widetilde{\beta}(y)$. The first equality follows by the
mean value expansion. The second equality follows by uniform continuity of $%
\lambda(\cdot)$, uniform continuity of $W(\cdot)$ and $\dot{W}(\cdot)$, and
by $\|\widehat{\vartheta}-\vartheta_{0}\|_{T,\infty}\to_{\Pr}0$ and $%
\sup_{y\in\mathcal{Y}}\|\widetilde{\beta}(y)-\beta_{0}(y)\|_{2}\to_{\Pr}0$.

Since $\lambda(\cdot)$ and the entries of $W$ and $\dot{W}$ are bounded, $%
\widehat{\delta}(y)=\bar{O}_{\Pr}(1),$ and $\|\widehat{\Delta}%
\|_{T,\infty}=O_{\Pr}(1)$, with probability approaching one uniformly in $%
y\in\mathcal{Y}$, 
\begin{multline*}
\sqrt{n}{\mathrm{E}}_{P}\{\Lambda[W(\widehat{\vartheta})^{\prime }\widetilde{%
\beta}(y)]-1(Y\leq y)\}W(\widehat{\vartheta})T={\mathrm{E}}%
_{P}\{\Lambda(W^{\prime }\beta_{0}(y))-1(Y\leq y)\}\dot{W}T\widehat{\Delta}%
(X,R) \\
+{\mathrm{E}}_{P}\{\lambda[W^{\prime }\beta_{0}(y)]WW^{\prime }T\}\widehat{%
\delta}(y)+{\mathrm{E}}_{P}\{\lambda[W^{\prime }\beta_{0}(y)]W\dot{W}%
^{\prime }\beta_{0}(y)T\widehat{\Delta}(X,R)\}+O_{\Pr}(\bar{R}(y)) \\
=J(y)\widehat{\delta}(y)+{\mathrm{E}}_{P}[\{\Lambda(W^{\prime
}\beta_{0}(y))-1(Y\leq y)\}\dot{W}+\lambda[W^{\prime }\beta_{0}(y)]W\dot{W}%
^{\prime }\beta_{0}(y)]T\widehat{\Delta}(X,R)+o_{\Pr}(1).
\end{multline*}

Substituting in $\widehat{\Delta}(x,r)=\sqrt{n}\ {\mathbb{E}_n}%
[\ell(A,x,r)]+o_{\Pr}(1)$ and interchanging ${\mathrm{E}}_{P}$ and ${\mathbb{%
E}_n}$, we obtain 
\begin{equation*}
{\mathrm{E}}_{P}[\{\Lambda(W^{\prime }\beta_{0}(y))-1(Y\leq y)\}\dot{W}+%
\lambda[W^{\prime }\beta_{0}(y)]W\dot{W}^{\prime }\beta_{0}(y)]T\widehat{%
\Delta}(X,R)=\sqrt{n}\ {\mathbb{E}_n}[g_{y}(A)]+\bar{o}_{\Pr}(1),
\end{equation*}
since $[\{\Lambda(W^{\prime }\beta_{0}(y))-1(Y\leq y)\}\dot{W}+\lambda[%
W^{\prime }\beta_{0}(y)]W\dot{W}^{\prime }\beta_{0}(y)]T$ is bounded
uniformly in $y\in\mathcal{Y}$. The claim of the lemma follows. \qed

\subsubsection{Proof of Lemma \protect\ref{thm:fclt}.}

The proof is divided in two parts corresponding to the FCLT and bootstrap
FCLT.

\textbf{Part 1: FCLT}

In this part we show $\sqrt{n}(\widehat{\beta}(y)-\beta_{0}(y))%
\rightsquigarrow J(y)^{-1}G(y)$ in $\ell^{\infty}(\mathcal{Y})^{d_{w}}$.

Step 1. This step shows that $\sqrt{n}(\widehat{\beta}(y)-\beta_{0}(y))=\bar{%
O}_{\Pr}(1)$.

Recall that 
\begin{equation*}
\widehat{\beta}(y)=\arg\min_{\beta\in\mathbb{R}^{\dim(W)}}\mathbb{E}%
_{n}[\rho_{y}(Y,W(\widehat{\vartheta})^{\prime }\beta)T].
\end{equation*}
Due to convexity of the objective function, it suffices to show that for any 
$\epsilon>0$ there exists a finite positive constant $B_{\epsilon}$ such
that uniformly in $y\in\mathcal{Y}$, 
\begin{eqnarray}
\liminf_{n\to\infty}\Pr\left(\inf_{\|\eta\|_{2}=1}\sqrt{n}\eta^{\prime }{%
\mathbb{E}_n}\Big[\widehat{f}_{\eta,B_{\epsilon},y}\Big]>0\right)\geq1-%
\epsilon,  \label{eq: prob0}
\end{eqnarray}
where 
\begin{equation*}
\widehat{f}_{\eta,B_{\epsilon},y}(A):=\left\{ \Lambda[W(\widehat{\vartheta}%
)^{\prime }(\beta_{0}(y)+B_{\epsilon}\eta/\sqrt{n})]-1(Y\leq y)\right\} W(%
\widehat{\vartheta})T.
\end{equation*}
Let 
\begin{equation*}
f_{y}(A):=\left\{ \Lambda[W^{\prime }\beta_{0}(y)]-1(Y\leq y)\right\} WT.
\end{equation*}
Then uniformly in $\|\eta\|_{2}=1$, 
\begin{eqnarray*}
\sqrt{n}\eta^{\prime }{\mathbb{E}_n}[\widehat{f}_{\eta,B_{\epsilon},y}] & =
& \eta^{\prime }\mathbb{G}_n[\widehat{f}_{\eta,B_{\epsilon},y}]+\sqrt{n}%
\eta^{\prime }{\mathrm{E}}_{P}[\widehat{f}_{\eta,B_{\epsilon},y}] \\
& =_{(1)} & \eta^{\prime }\mathbb{G}_n[f_{y}]+\bar{o}_{\Pr}(1)+\eta^{\prime }%
\sqrt{n}{\mathrm{E}}_{P}[\widehat{f}_{\eta,B_{\epsilon},y}] \\
& =_{(2)} & \eta^{\prime }\mathbb{G}_n[f_{y}]+\bar{o}_{\Pr}(1)+\eta^{\prime
}J(y)\eta B_{\epsilon}+\eta^{\prime }\mathbb{G}_n[g_{y}]+\bar{o}_{\Pr}(1) \\
& =_{(3)} & \bar{O}_{\Pr}(1)+\bar{o}_{\Pr}(1)+\eta^{\prime }J(y)\eta
B_{\epsilon}+\bar{O}_{\Pr}(1)+\bar{o}_{\Pr}(1),
\end{eqnarray*}
where relations (1) and (2) follow by Lemma \ref{lemma SE} and Lemma \ref%
{lemma Expand} with $\widetilde{\beta}(y)=\beta_{0}(y)+B_{\epsilon}\eta/%
\sqrt{n}$, respectively, using that $\|\widehat{\vartheta}-\widetilde{%
\vartheta}\|_{T,\infty}=o_{\Pr}(1/\sqrt{n})$, $\widetilde{\vartheta}%
\in\Upsilon$, $\|\widetilde{\vartheta}-\vartheta_{0}\|_{T,\infty}=O_{\Pr}(1/%
\sqrt{n})$ and $\|\beta_{0}(y)+B_{\epsilon}\eta/\sqrt{n}-\beta_{0}(y)\|_{2}=%
\bar{O}(1/\sqrt{n})$; relation (3) holds because $f_{y}$ and $g_{y}$ are $P$%
-Donsker by step-2 below. Since uniformly in $y\in\mathcal{Y}$, $J(y)$ is
positive definite, with minimal eigenvalue bounded away from zero, the
inequality (\ref{eq: prob0}) follows by choosing $B_{\epsilon}$ as a
sufficiently large constant.

Step 2. In this step we show the main result. Let 
\begin{equation*}
\widehat{f}_{y}(A):=\left\{ \Lambda[W(\widehat{\vartheta})^{\prime }\widehat{%
\beta}(y)]-1(Y\leq y)\right\} W(\widehat{\vartheta})T.
\end{equation*}
From the first order conditions of the distribution regression problem, 
\begin{eqnarray*}
0=\sqrt{n}{\mathbb{E}_n}\left[\widehat{f}_{y}\right] & = & \mathbb{G}_n\left[%
\widehat{f}_{y}\right]+\sqrt{n}{\mathrm{E}}_{P}\left[\widehat{f}_{y}\right]
\\
& =_{(1)} & \mathbb{G}_n[f_{y}]+\bar{o}_{\Pr}(1)+\sqrt{n}{\mathrm{E}}_{P}%
\left[\widehat{f}_{y}\right] \\
& =_{(2)} & \mathbb{G}_n[f_{y}]+\bar{o}_{\Pr}(1)+J(y)\sqrt{n}(\widehat{\beta}%
(y)-\beta_{0}(y))+\mathbb{G}_n[g_{y}]+\bar{o}_{\Pr}(1),
\end{eqnarray*}
where relations (1) and (2) follow by Lemma \ref{lemma SE} and Lemma \ref%
{lemma Expand} with $\widetilde{\beta}(y)=\widehat{\beta}(y)$, respectively,
using that $\|\widehat{\vartheta}-\widetilde{\vartheta}\|_{T,\infty}=o_{%
\Pr}(1/\sqrt{n})$, $\widetilde{\vartheta}\in\Upsilon$, and $\|\widetilde{%
\vartheta}-\vartheta\|_{T,\infty}=O_{\Pr}(1/\sqrt{n})$ by Lemma \ref%
{lemma:first}, and $\|\widehat{\beta}(y)-\beta_{0}(y)\|_{2}=\bar{O}_{\Pr}(1/%
\sqrt{n})$.

Therefore by uniform invertibility of $J(y)$ in $y\in\mathcal{Y}$, 
\begin{equation*}
\sqrt{n}(\widehat{\beta}(y)-\beta_{0}(y))=-J(y)^{-1}\mathbb{G}%
_n(f_{y}+g_{y})+\bar{o}_{\Pr}(1).
\end{equation*}
The function $f_{y}$ is $P$-Donsker by standard argument for distribution
regression (e.g., step 3 in the proof of Theorem 5.2 of Chernozhukov,
Fernandez-Val and Melly, 2013). Similarly, $g_{y}$ is $P$-Donsker by Example
19.7 in van der Vaart (1998) because $g_{y}\in\{h_{y}(A):|h_{y}(A)-h_{v}(A)|%
\leq M(A)|y-v|;{\mathrm{E}}_{P}M(A)^{2}<\infty;y,v\in\mathcal{Y}\}$, since 
\begin{equation*}
|g_{y}-g_{v}|\leq L{\mathrm{E}}_{P}[T|\ell(a,X,R)|]\big|_{a=A}|y-v|,
\end{equation*}
with $L=2L_{W}+L_{W}^{2}\max_{\beta\in\mathcal{B}}\|\beta\|_{2}/4$, $%
L_{W}:=\|\partial_{v}w\|_{T,\infty}\vee\|w\|_{T,\infty}$, and ${\mathrm{E}}%
_{P}[T\ell(A,X,R)^{2}]<\infty$ by Lemma \ref{lemma:first}. Hence, by the
Functional Central Limit Theorem 
\begin{equation*}
\mathbb{G}_n(f_{y}+g_{y})\rightsquigarrow G(y)\text{ in }\ell^{\infty}(%
\mathcal{Y})^{d_{w}},
\end{equation*}
where $y\mapsto G(y)$ is a zero mean Gaussian process with uniformly
continuous sample paths and the covariance function $C(y,v)$ specified in
the lemma. Conclude that 
\begin{equation*}
\sqrt{n}(\widehat{\beta}(y)-\beta_{0}(y))\rightsquigarrow J(y)^{-1}G(y)\text{
in }\ell^{\infty}(\mathcal{Y})^{d_{w}}.
\end{equation*}
\qed

\textbf{Part 2: Bootstrap FCLT}

In this part we show $\sqrt{n}(\widehat{\beta}^{e}(y)-\widehat{\beta}%
(y))\rightsquigarrow_{\Pr}J(y)^{-1}G(y)$ in $\ell^{\infty}(\mathcal{Y}%
)^{d_{w}}$.

Step 1. This step shows that $\sqrt{n}(\widehat{\beta}^{e}(y)-\beta_{0}(y))=%
\bar{O}_{\Pr}(1)$ under the unconditional probability $\Pr$.

Recall that 
\begin{equation*}
\widehat{\beta}^{e}(y)=\arg\min_{\beta\in\mathbb{R}^{\dim(W)}}\mathbb{E}%
_{n}[e\rho_{y}(Y,W(\widehat{\vartheta}^{e})^{\prime }\beta)T],
\end{equation*}
where $e$ is the random variable used in the weighted bootstrap. Due to
convexity of the objective function, it suffices to show that for any $%
\epsilon>0$ there exists a finite positive constant $B_{\epsilon}$ such that
uniformly in $y\in\mathcal{Y}$, 
\begin{eqnarray}
\liminf_{n\to\infty}\Pr\left(\inf_{\|\eta\|_{2}=1}\sqrt{n}\eta^{\prime }{%
\mathbb{E}_n}\Big[\widehat{f}_{\eta,B_{\epsilon},y}^{e}\Big]%
>0\right)\geq1-\epsilon,  \label{eq: prob}
\end{eqnarray}
where 
\begin{equation*}
\widehat{f}_{\eta,B_{\epsilon},y}^{e}(A):=e\cdot\left\{ \Lambda[W(\widehat{%
\vartheta}^{e})^{\prime }(\beta_{0}(y)+B_{\epsilon}\eta/\sqrt{n})]-1(Y\leq
y)\right\} W(\widehat{\vartheta}^{e})T.
\end{equation*}
Let 
\begin{equation*}
f_{y}^{e}(A):=e\cdot\left\{ \Lambda[W^{\prime }\beta_{0}(y)]-1(Y\leq
y)\right\} WT.
\end{equation*}
Then uniformly in $\|\eta\|_{2}=1$, 
\begin{eqnarray*}
\sqrt{n}\eta^{\prime }{\mathbb{E}_n}[\widehat{f}_{\eta,B_{\epsilon},y}^{e}]
& = & \eta^{\prime }\mathbb{G}_n[\widehat{f}_{\eta,B_{\epsilon},y}^{e}]+%
\sqrt{n}\eta^{\prime }{\mathrm{E}}_{P}[\widehat{f}_{\eta,B_{\epsilon},y}^{e}]
\\
& =_{(1)} & \eta^{\prime }\mathbb{G}_n[f_{y}^{e}]+\bar{o}_{\Pr}(1)+\eta^{%
\prime }\sqrt{n}{\mathrm{E}}_{P}[\widehat{f}_{\eta,B_{\epsilon},y}^{e}] \\
& =_{(2)} & \eta^{\prime }\mathbb{G}_n[f_{y}^{e}]+\bar{o}_{\Pr}(1)+\eta^{%
\prime }J(y)\eta B_{\epsilon}+\eta^{\prime }\mathbb{G}_n[g_{y}^{e}]+\bar{o}%
_{\Pr}(1) \\
& =_{(3)} & \bar{O}_{\Pr}(1)+\bar{o}_{\Pr}(1)+\eta^{\prime }J(y)\eta
B_{\epsilon}+\bar{O}_{\Pr}(1)+\bar{o}_{\Pr}(1),
\end{eqnarray*}
where relations (1) and (2) follow by Lemma \ref{lemma SE} and Lemma \ref%
{lemma Expand} with $\widetilde{\beta}(y)=\beta_{0}(y)+B_{\epsilon}\eta/%
\sqrt{n}$, respectively, using that $\|\widehat{\vartheta}^{e}-\widetilde{%
\vartheta}^{e}\|_{T,\infty}=o_{\Pr}(1/\sqrt{n})$, $\widetilde{\vartheta}%
^{e}\in\Upsilon$ and $\|\widetilde{\vartheta}^{e}-\vartheta_{0}\|_{T,%
\infty}=O_{\Pr}(1/\sqrt{n})$ by Lemma \ref{lemma:first}, and $%
\|\beta_{0}(y)+B_{\epsilon}\eta/\sqrt{n}-\beta_{0}(y)\|_{2}=\bar{O}(1/\sqrt{n%
})$; relation (3) holds because $f_{y}^{e}=e\cdot f_{y}$ and $%
g_{y}^{e}=e\cdot g_{y}$, where $f_{y}$ and $g_{y}$ are $P$-Donsker by step-2
of the proof of Theorem \ref{thm:fclt} and ${\mathrm{E}}_{P}e^{2}<\infty$.
Since uniformly in $y\in\mathcal{Y}$, $J(y)$ is positive definite, with
minimal eigenvalue bounded away from zero, the inequality (\ref{eq: prob})
follows by choosing $B_{\epsilon}$ as a sufficiently large constant.

Step 2. In this step we show that $\sqrt{n}(\widehat{\beta}%
^{e}(y)-\beta_{0}(y))=-J(y)^{-1}\mathbb{G}_n(f_{y}^{e}+g_{y}^{e})+\bar{o}%
_{\Pr}(1)$ under the unconditional probability $\Pr$.

Let 
\begin{equation*}
\widehat{f}_{y}^{e}(A):=e\cdot\{\Lambda[W(\widehat{\vartheta}^{e})^{\prime }%
\widehat{\beta}^{e}(y)]-1(Y\leq y)\}W(\widehat{\vartheta}^{e})T.
\end{equation*}
From the first order conditions of the distribution regression problem in
the weighted sample, uniformly in $y\in\mathcal{Y}$, 
\begin{eqnarray*}
0=\sqrt{n}{\mathbb{E}_n}\left[\widehat{f}_{y}^{e}\right] & = & \mathbb{G}_n%
\left[\widehat{f}_{y}^{e}\right]+\sqrt{n}{\mathrm{E}}_{P}\left[\widehat{f}%
_{y}^{e}\right] \\
& =_{(1)} & \mathbb{G}_n[f_{y}^{e}]+\bar{o}_{\Pr}(1)+\sqrt{n}{\mathrm{E}}_{P}%
\left[\widehat{f}_{y}^{e}\right] \\
& =_{(2)} & \mathbb{G}_n[f_{y}^{e}]+\bar{o}_{\Pr}(1)+J(y)\sqrt{n}(\widehat{%
\beta}^{e}(y)-\beta_{0}(y))+\mathbb{G}_n[g_{y}^{e}]+\bar{o}_{\Pr}(1),
\end{eqnarray*}
where relations (1) and (2) follow by Lemma \ref{lemma SE} and Lemma \ref%
{lemma Expand} with $\widetilde{\beta}(y)=\widehat{\beta}^{e}(y)$,
respectively, using that $\|\widehat{\vartheta}^{e}-\widetilde{\vartheta}%
^{e}\|_{T,\infty}=o_{\Pr}(1/\sqrt{n})$, $\widetilde{\vartheta}%
^{e}\in\Upsilon $ and $\|\widetilde{\vartheta}^{e}-\vartheta_{0}\|_{T,%
\infty}=O_{\Pr}(1/\sqrt{n})$ by Lemma \ref{lemma:first}, and $\|\widehat{%
\beta}^{e}(y)-\beta_{0}(y)\|_{2}=\bar{O}_{\Pr}(1/\sqrt{n})$.

Therefore by uniform invertibility of $J(y)$ in $y\in\mathcal{Y}$, 
\begin{equation*}
\sqrt{n}(\widehat{\beta}^{e}(y)-\beta_{0}(y))=-J(y)^{-1}\mathbb{G}%
_n(f_{y}^{e}+g_{y}^{e})+\bar{o}_{\Pr}(1).
\end{equation*}

Step 3. In this final step we establish the behavior of $\sqrt{n}(\widehat{%
\beta}^{e}(y)-\widehat{\beta}(y))$ under $\Pr^{e}$. Note that $\Pr^{e}$
denotes the conditional probability measure, namely the probability measure
induced by draws of $e_{1},...,e_{n}$ conditional on the data $%
A_{1},...,A_{n}$. By Step 2 of the proof of Theorem 1 and Step 2 of this
proof, we have that under $\Pr$: 
\begin{eqnarray*}
&&\sqrt{n}(\widehat{\beta}^{e}(y)-\beta_{0}(y))=-J(y)^{-1}\mathbb{G}%
_n(f_{y}^{e}+g_{y}^{e})+\bar{o}_{\Pr}(1),\\
&& \sqrt{n}(\widehat{\beta}%
(y)-\beta_{0}(y))=-J(y)^{-1}\mathbb{G}_n(f_{y}+g_{y})+\bar{o}_{\Pr}(1).
\end{eqnarray*}
Hence, under $\Pr$ 
\begin{multline*}
\sqrt{n}(\widehat{\beta}^{e}(y)-\widehat{\beta}(y))=-J(y)^{-1}\mathbb{G}%
_n(f_{y}^{e}-f_{y}+g_{y}^{e}-g_{y})+r_{n}(y) \\ =-J(y)^{-1}\mathbb{G}%
_n((e-1)(f_{y}+g_{y}))+r_{n}(y),
\end{multline*}
where $r_{n}(y)=\bar{o}_{\Pr}(1)$. Note that it is also true that 
\begin{equation*}
r_{n}(y)=\bar{o}_{\Pr^{e}}(1)\text{ in $\Pr$-probability},
\end{equation*}
where the latter statement means that for every $\epsilon>0$, $%
\Pr^{e}(\|r_{n}(y)\|_{2}>\epsilon)=\bar{o}_{\Pr}(1).$ Indeed, this follows
from Markov inequality and by 
\begin{equation*}
{\mathrm{E}}_{\mathbb{P}}[\Pr^{e}(\|r_{n}(y)\|_{2}>\epsilon)]=\Pr(\|r_{n}(y)%
\|_{2}>\epsilon)=\bar{o}(1),
\end{equation*}
where the latter holds by the Law of Iterated Expectations and $r_{n}(y)=%
\bar{o}_{\Pr}(1)$.

Note that $f_{y}^{e}=e\cdot f_{y}$ and $g_{y}^{e}=e\cdot g_{y}$, where $%
f_{y} $ and $g_{y}$ are $P$-Donsker by step-2 of the proof of the first part
and ${\mathrm{E}}_{P}e^{2}<\infty$. Then, by the Conditional Multiplier
Functional Central Limit Theorem, e.g., Theorem 2.9.6 in van der Vaart and
Wellner (1996), 
\begin{equation*}
G_{n}^{e}(y):=\mathbb{G}_n((e-1)(f_{y}+g_{y}))\rightsquigarrow_{\Pr}G(y)%
\text{ in }\ell^{\infty}(\mathcal{Y})^{d_{w}}.
\end{equation*}
Conclude that 
\begin{equation*}
\sqrt{n}(\widehat{\beta}^{e}(y)-\widehat{\beta}(y))\rightsquigarrow_{%
\Pr}J(y)^{-1}G(y)\text{ in }\ell^{\infty}(\mathcal{Y})^{d_{w}}.
\end{equation*}
\qed

\subsection{Proof of Theorems \protect\ref{fclt:sdf}--\protect\ref{fclt:asf}}

\begin{sloppy}
In this section we use the notation $W_{x}(\vartheta)=w(x,Z_{1},V(%
\vartheta)) $ such that $W_{x}=w(x,Z_{1},V(\vartheta_{0}))$. Again we focus
on the case where $\mathcal{Y}$ is a compact interval of $\mathbb{R}$.
\par\end{sloppy}

\subsubsection{Proof of Theorem \protect\ref{fclt:sdf}}

The result follows by a similar argument to the proof of Lemma \ref{thm:fclt}
using Lemmas \ref{lemma SE2} and \ref{lemma Expand2} in place of Lemmas \ref%
{lemma SE} and \ref{lemma Expand}, and the delta method. For the sake of
brevity, here we just outline the proof of the FCLT.

Let $\psi_{x}(A,\vartheta,\beta):=\Lambda(W_{x}(\vartheta)^{\prime }\beta)T$
such that $G_T(y,x)={\mathrm{E}}_{P}\psi_{x}(A,\vartheta_{0},\beta_{0}(y))/{%
\mathrm{E}}_P T$ and $\widehat{G}(y,x)={\mathbb{E}_n}\psi_{x}(A,\widehat{%
\vartheta},\widehat{\beta}(y))/{\mathbb{E}_n} T$. Then, for $\widehat{\psi}%
_{y,x}:=\psi_{x}(A,\widehat{\vartheta},\widehat{\beta}(y))$ and $%
\psi_{y,x}:=\psi_{x}(A,\vartheta_{0},\beta_{0}(y))$, 
\begin{eqnarray*}
\sqrt{n}\left[{\mathbb{E}_n}\psi_{x}(A,\widehat{\vartheta},\widehat{\beta}%
(y))-{\mathrm{E}}_{P}\psi_{x}(A,\vartheta_{0},\beta_{0}(y))\right] & = & 
\mathbb{G}_n\left[\widehat{\psi}_{y,x}\right]+\sqrt{n}{\mathrm{E}}_{P}\left[%
\widehat{\psi}_{y,x}-\psi_{y,x}\right] \\
& =_{(1)} & \mathbb{G}_n[\psi_{y,x}]+\bar{o}_{\Pr}(1)+\sqrt{n}{\mathrm{E}}%
_{P}\left[\widehat{\psi}_{y,x}-\psi_{y,x}\right] \\
& =_{(2)} & \mathbb{G}_n[\psi_{y,x}]+\bar{o}_{\Pr}(1)+\mathbb{G}_n[h_{y,x}]+%
\bar{o}_{\Pr}(1),
\end{eqnarray*}
where relations (1) and (2) follow by Lemma \ref{lemma SE2} and Lemma \ref%
{lemma Expand2} with $\widetilde{\beta}(y)=\widehat{\beta}(y)$,
respectively, using that $\|\widehat{\vartheta}-\widetilde{\vartheta}%
\|_{T,\infty}=o_{\Pr}(1/\sqrt{n})$, $\widetilde{\vartheta}\in\Upsilon$, and $%
\|\widetilde{\vartheta}-\vartheta\|_{T,\infty}=O_{\Pr}(1/\sqrt{n})$ by Lemma %
\ref{lemma:first}, and $\sqrt{n}(\widehat{\beta}(y)-\beta_{0}(y))=-J(y)^{-1}%
\mathbb{G}_n(f_{y}+g_{y})+\bar{o}_{\Pr}(1)$ from step 2 of the proof of
Lemma \ref{thm:fclt}.

The functions $(y,x)\mapsto\psi_{y,x}$ and $(y,x)\mapsto h_{y,x}$ are $P$%
-Donsker by Example 19.7 in van der Vaart (1998) because they are Lipschitz
continuous on $\mathcal{Y}\overline{\mathcal{X}}$. Hence, by the Functional
Central Limit Theorem 
\begin{equation*}
\mathbb{G}_n(\psi_{y,x}+h_{y,x})\rightsquigarrow Z(y,x)\text{ in }%
\ell^{\infty}(\mathcal{Y}\overline{\mathcal{X}}),
\end{equation*}
where $(y,x)\mapsto Z(y,x)$ is a zero mean Gaussian process with uniformly
continuous sample paths and covariance function 
\begin{equation*}
\mathrm{Cov}_{P}[\psi_{y,x}+h_{y,x},\psi_{v,u}+h_{v,u}],\ \ (y,x),(v,u)\in%
\mathcal{Y}\overline{\mathcal{X}}.
\end{equation*}

\begin{sloppy}
The result follows by the functional delta method applied to the ratio of ${%
\mathbb{E}_n}\psi_{x}(A,\widehat{\vartheta},\widehat{\beta}(y))$ and ${%
\mathbb{E}_n} T$ using that 
\begin{equation*}
 \left( 
\begin{array}{c}
{\mathbb{G}_n}\psi_{x}(A,\widehat{\vartheta},\widehat{\beta}(y)) \\ 
{\mathbb{G}_n} T %
\end{array}
\right) \rightsquigarrow \left( 
\begin{array}{c}
Z(y,x) \\ 
Z_T%
\end{array}
\right),
\end{equation*}
where $Z_T \sim N(0, p_T(1-p_T))$, 
\begin{equation*}
\mathrm{Cov}_P(Z(y,x),Z_T) = G_T(y,x) p_T(1-p_T),
\end{equation*}
and 
\begin{multline*}
\mathrm{Cov}_{P}[\psi_{y,x}+h_{y,x},\psi_{v,u}+h_{v,u} \mid T = 1] \\
= \frac{\mathrm{Cov}_{P}[\psi_{y,x}+h_{y,x},\psi_{v,u}+h_{v,u}] - G_T(y,x)
G_T(v,u) p_T(1-p_T)}{p_T}.
\end{multline*}
\par\end{sloppy}
\qed

\begin{lemma}
{[}Stochastic equicontinuity{]}\label{lemma SE2} Let $e\geq0$ be a positive
random variable with ${\mathrm{E}}_{P}[e]=1$, $\mathrm{Var}_{P}[e]=1,$ and ${%
\mathrm{E}}_{P}|e|^{2+\delta}<\infty$ for some $\delta>0$, that is
independent of $(Y,X,Z,W,V)$, including as a special case $e=1$, and set,
for $A=(e,Y,X,Z,W,V)$, 
\begin{equation*}
\psi_{x}(A,\vartheta,\beta):=e\cdot\Lambda(W_{x}(\vartheta)^{\prime
}\beta)\cdot T.
\end{equation*}
Under Assumptions \ref{ass:sampling}--\ref{ass:second}, the following
relations are true.

\begin{itemize}
\item[(a)] Consider the set of functions 
\begin{equation*}
\mathcal{F}:=\{\psi_{x}(A,\vartheta,\beta):(\vartheta,\beta,x)\in%
\Upsilon_{0}\times\mathcal{B}\times\overline{\mathcal{X}}\},
\end{equation*}
where $\overline{\mathcal{X}}$ is a compact subset of $\mathbb{R}$, $%
\mathcal{B}$ is a compact set under the $\|\cdot\|_{2}$ metric containing $%
\beta_{0}(y)$ for all $y\in\mathcal{Y}$, $\Upsilon_{0}$ is the intersection
of $\Upsilon$, defined in Lemma \ref{lemma:first}, with a neighborhood of $%
\vartheta_{0}$ under the $\|\cdot\|_{T,\infty}$ metric. This class is $P$%
-Donsker with a square integrable envelope of the form $e$ times a constant.

\item[(b)] Moreover, if $(\vartheta,\beta(y))\to(\vartheta_{0},\beta_{0}(y))$
in the $\|\cdot\|_{T,\infty}\vee\|\cdot\|_{2}$ metric uniformly in $y\in%
\mathcal{Y}$, then 
\begin{equation*}
\sup_{(y,x)\in\mathcal{Y}\overline{\mathcal{X}}}\|\psi_{x}(A,\vartheta,%
\beta(y))-\psi_{x}(A,\vartheta_{0},\beta_{0}(y))\|_{P,2}\to0.
\end{equation*}

\item[(c)] Hence for any $(\widetilde{\vartheta},\widetilde{\beta}%
(y))\to_{\Pr}(\vartheta_{0},\beta_{0}(y))$ in the $\|\cdot\|_{T,\infty}\vee%
\|\cdot\|_{2}$ metric uniformly in $y\in\mathcal{Y}$ such that $\widetilde{%
\vartheta}\in\Upsilon_{0}$, 
\begin{equation*}
\sup_{(y,x)\in\mathcal{Y}\overline{\mathcal{X}}}\|\mathbb{G}_n\psi_{x}(A,%
\widetilde{\vartheta},\widetilde{\beta}(y))-\mathbb{G}_n\psi_{x}(A,%
\vartheta_{0},\beta_{0}(y))\|_{2}\to_{\Pr}0.
\end{equation*}

\item[(d)] For any $(\widehat{\vartheta},\widetilde{\beta}%
(y))\to_{\Pr}(\vartheta_{0},\beta_{0}(y))$ in the $\|\cdot\|_{T,\infty}\vee%
\|\cdot\|_{2}$ metric uniformly in $y\in\mathcal{Y}$, so that 
\begin{equation*}
\|\widehat{\vartheta}-\widetilde{\vartheta}\|_{T,\infty}=o_{\Pr}(1/\sqrt{n}),%
\text{ where }\widetilde{\vartheta}\in\Upsilon_{0},
\end{equation*}
we have that 
\begin{equation*}
\sup_{(y,x)\in\mathcal{Y}\overline{\mathcal{X}}}\|\mathbb{G}_n\psi_{x}(A,%
\widehat{\vartheta},\widetilde{\beta}(y))-\mathbb{G}_n\psi_{x}(A,%
\vartheta_{0},\beta_{0}(y))\|_{2}\to_{\Pr}0.
\end{equation*}
\end{itemize}
\end{lemma}

\textbf{Proof of Lemma \ref{lemma SE2}.} The proof is omitted because is
similar to the proof of Lemma \ref{lemma SE}. \qed

\begin{lemma}
{[}Local expansion{]} \label{lemma Expand2} Under Assumptions \ref%
{ass:sampling}--\ref{ass:second}, for 
\begin{eqnarray*}
& & \widehat{\delta}(y)=\sqrt{n}(\widetilde{\beta}(y)-\beta_{0}(y))=\bar{O}%
_{\Pr}(1); \\
& & \widehat{\Delta}(x,r)=\sqrt{n}(\widehat{\vartheta}(x,r)-%
\vartheta_{0}(x,r))=\sqrt{n}\ {\mathbb{E}_n}[\ell(A,x,r)]+o_{\Pr}(1)\text{
in }\ell^{\infty}(\overline{\mathcal{XR}}), \\
& & \|\sqrt{n}\ {\mathbb{E}_n}[\ell(A,\cdot)]\|_{T,\infty}=O_{\Pr}(1),
\end{eqnarray*}
we have that 
\begin{multline*}
\sqrt{n}\ \left\{ {\mathrm{E}}_{P}\Lambda[W_{x}(\widehat{\vartheta})^{\prime
}\widetilde{\beta}(y)]T-{\mathrm{E}}_{P}\Lambda[W_{x}^{\prime }\beta_{0}(y)]%
T\right\} ={\mathrm{E}}_{P}\{\lambda[W_{x}^{\prime }\beta_{0}(y)]%
W_{x}T\}^{\prime }\widehat{\delta}(y) \\
+{\mathrm{E}}_{P}\{\lambda[W_{x}^{\prime }\beta_{0}(y)]\dot{W}_{x}^{\prime
}\beta_{0}(y)T\ell(a,X,R)\}\big|_{a=A}+\bar{o}_{\Pr}(1),
\end{multline*}
where $\bar{o}_{\Pr}(1)$ denotes order in probability uniform in $(y,x)\in%
\mathcal{Y}\overline{\mathcal{X}}$.
\end{lemma}

\textbf{Proof of Lemma \ref{lemma Expand2}.} The proof is omitted because is
similar to the proof of Lemma \ref{lemma Expand}. \qed

\subsubsection{Proof of Theorem \protect\ref{fclt:sqf}}

The result follows from Theorem \ref{fclt:sdf} and the functional delta
method, because the map $\phi:H\mapsto\int_{\mathcal{Y}^{+}}1(H(y,x)\leq%
\tau)dy-\int_{\mathcal{Y}^{-}}1(H(y,x)\geq\tau)dy$ is Hadamard
differentiable at $H=G_T$ under the conditions of the theorem by Proposition
2 of Chernozhukov, Fernandez-Val and Galichon (2010) with derivative 
\begin{equation*}
\phi_{G_T}^{\prime }(h)=-\frac{h(\phi(\cdot,x),x)}{g_T(\phi(\cdot,x),x)}.
\end{equation*}

\subsubsection{Proof of Theorem \protect\ref{fclt:asf}}

The result follows from Theorem \ref{fclt:sdf} and the functional delta
method, because the map $\varphi:H\mapsto\int_{\mathcal{Y}%
}[1(y\geq0)-H(y,x)]dy$ is Hadamard differentiable at $H=G_T$ by Lemma \ref%
{lemma:hd} with derivative 
\begin{equation*}
\varphi_{G_T}^{\prime }(h)=-\int_{\mathcal{Y}}h(y,x)\nu(dy).
\end{equation*}

\begin{lemma}
{[}Hadamard Differentiability of ASF Map{]} \label{lemma:hd} The ASF map $%
\varphi:\ell^{\infty}(\mathcal{Y}\overline{\mathcal{X}})\to\ell^{\infty}(%
\overline{\mathcal{X}})$ defined by 
\begin{equation*}
H\mapsto\varphi(H):=\int_{\mathcal{Y}}[1(y\geq0)-H(y,x)]\nu(dy),
\end{equation*}
is Hadamard-differentiable at $H=G$, tangentially to the set of uniformly
continuous functions on $\mathcal{Y}\overline{\mathcal{X}}$, with derivative
map $h\mapsto\varphi^{\prime }_{G}(h)$ defined by 
\begin{equation*}
\varphi^{\prime }_{G}(h):=-\int_{\mathcal{Y}}h(y,x)\nu(dy),
\end{equation*}
where the derivative is defined and is continuous on $\ell^{\infty}(\mathcal{%
Y}\overline{\mathcal{X}})$.
\end{lemma}

\textbf{Proof of Lemma \ref{lemma:hd}.} Consider any sequence $%
H^{t}\in\ell^{\infty}(\mathcal{Y}\overline{\mathcal{X}})$ such that for $%
h^{t}:=(H^{t}-G)/t$, $h^{t}\to h$ in $\ell^{\infty}(\mathcal{Y}\overline{%
\mathcal{X}})$ as $t\searrow0$, where $h$ is a uniformly continuous function
on $\mathcal{Y}\overline{\mathcal{X}}$. We want to show that as $t\searrow0$%
, 
\begin{equation*}
\frac{\varphi(H^{t})-\varphi(G)}{t}-\varphi^{\prime }_{G}(h)\to0\text{ in }%
\ell^{\infty}(\mathcal{Y}\overline{\mathcal{X}}).
\end{equation*}
The result follows because by linearity of the map $\varphi$ 
\begin{equation*}
\frac{\varphi(H^{t})-\varphi(G)}{t}=-\int_{\mathcal{Y}}h^{t}(y,x)\nu(dy)\to-%
\int_{\mathcal{Y}}h(y,x)\nu(dy)=\varphi^{\prime }_{G}(h).
\end{equation*}
The derivative is well-defined over $\ell^{\infty}(\mathcal{Y}\overline{%
\mathcal{X}})$ and continuous with respect to the sup-norm on $\ell^{\infty}(%
\mathcal{Y}\overline{\mathcal{X}})$.


\end{document}